\pdfoutput=1 

\documentclass[a4paper,12pt,twoside,openright]{reportMultibib}

\usepackage[paper=a4paper, left=2cm,right=2cm,top=2cm,bottom=2cm]{geometry}
\usepackage[hang,splitrule]{footmisc}
\usepackage[normalem]{ulem}
\usepackage[
pdftitle={Digital Ecosystems}, 
pdfauthor={Gerard Briscoe},
pdfsubject={Ecosystem-Oriented Architectures},
colorlinks=true, linkcolor=black, citecolor=black, filecolor=black, urlcolor=black,
hyperfootnotes=false]{hyperref} 

\usepackage{graphicx, setspace, acronym, amsthm, tobiShell, ifthen, substr, multibib, color, amsmath, tocloft}

\newboolean{final}\setboolean{final}{true}

\ifthenelse{\boolean{final}}{}{\usepackage{showlabels}\usepackage{citeext}}



\input{blocked.sty}
\input{uhead.sty}
\input{icthesis.sty}

\newcites{pub}{publications}

\newcommand{\normallinespacing}{\renewcommand{\baselinestretch}{1.5} \normalsize}

\newcommand{\narrowlinespacing}{\renewcommand{\baselinestretch}{1.0} \normalsize}

\theoremstyle{plain}
\newtheorem{definition}{Definition}
\newtheorem{theorem}{Theorem}
\newcommand{\be}{\begin{equation}}
\newcommand{\eeq}[1]{\label{#1}\end{equation}}
\newcommand{\bx}{{\bf X}}
\newcommand{\by}{{\bf Y}}
\newcommand{\bone}{{\bf M_{max}}}
\newcommand{\bfif}{{\bf M_{half}}}

\newcommand{\bxi}{\mbox{\boldmath $\xi$}}
\newcommand{\ra}{1, \ldots, n}

\newcommand{\Circ}[1]{\textcircled{\scriptsize #1}}
\newcommand{\execute}[1]{\immediate\write18{#1}}

\definecolor{tred}{RGB}{255,0,0} 
\definecolor{tblue}{RGB}{102,204,255} 
\definecolor{red}{RGB}{128,0,0} 
\definecolor{blue}{RGB}{0,0,128} 
\definecolor{green}{RGB}{0,128,0} 
\definecolor{yellow}{RGB}{128,128,0} 
\definecolor{purple}{RGB}{128,0,128} 
\definecolor{turquoise}{RGB}{0,128,128} 
\definecolor{grey}{RGB}{76,76,76}
\definecolor{brown}{RGB}{128,64,0}

\newcommand{\bold}{\textbf}
\newcommand{\italics}{\textit}

\newcommand{\mfigure}[7]
	{
	\vspace{#5}
		\IfSubStringInString{!}{#7}
			{\begin{figure}[#7]}{\begin{figure}[!t]}
		#1
		\vspace{#6}
		\caption[#2]{\italics{#2: #3}}
		\label{#4} 
	\end{figure}
	}

\newcommand{\tfigure}[9]
	{
	\IfSubStringInString{!}{#7}{\begin{figure}[#7]}{\begin{figure}[!t]}
	\IfSubStringInString{mm}{#8}{\vspace{#8}}{}
	\centering
	
	\IfSubStringInString{pdf}{#3}
		{
		\execute{cd images; ln -s #2.pdf .#2.gdf}
		\href{file://localhost/Users/g/Desktop/PhDthesis/images/.#2.gdf}{\includegraphics[#1]{images/#2}}
		}
		{\IfSubStringInString{graph}{#3}
			{
			\execute{cd images; ./makeGraph.sh #2; ln -s #2.pdf .#2.gdf}
			\ifthenelse{\boolean{final}}
				{\includegraphics[#1]{images/#2}}
				{\href{file://localhost/Users/g/Desktop/PhDthesis/images/.#2.gdf}{\includegraphics[#1]{images/#2}}}
			}
			{
			\execute{cd images; ./pdfcrop.sh #2}
			\ifthenelse{\boolean{final}}
				{\includegraphics[#1]{images/#2-crop.pdf}}
				{\href{file://localhost/Users/g/Desktop/PhDthesis/images/#2.#3}{\includegraphics[#1]{images/#2-crop.pdf}}}
			}
		}
		
	\vspace{#6}
	\caption[#4]
		{
		\label{#2}
		\tcaption{#4}{#5}
		}
	\IfSubStringInString{mm}{#9}{\vspace{#9}}{}
	\end{figure}
	}

\newcommand{\square}[3]
	{
	\squaresub{#1}{#2}{#3}{-1pt}
	}
	
\newcommand{\squaresub}[4]
	{
	\immediate\write18{cd images; ./pdfcrop.sh square#2}
	\ifthenelse{\boolean{final}}
		{\hspace{#1}\raisebox{#4}{$\includegraphics[clip=true, trim=0mm 0.25mm 0.25mm 0mm]{images/square#2-crop.pdf}$}\hspace{#3}}
		{\href{file://localhost/Users/g/Desktop/PhDthesis/images/square#2.graffle}{\hspace{#1}\raisebox{#4}{$\includegraphics[clip=true, trim=0mm 0.25mm 0.25mm 0mm]{images/square#2-crop.pdf}$}\hspace{#3}}}
	}

\newcommand{\tcaption}[2]
	{
	\IfSubStringInString{:}{#2}{\italics{#1 #2}}{\italics{#1: #2}}
	}

\ifthenelse{\boolean{final}}
	{}
	{}
\newcommand{\red}[1]{\color{red}#1\normalcolor}
\newcommand{\blue}[1]{\color{blue}#1\normalcolor}
\newcommand{\green}[1]{\color{green}#1\normalcolor}
\newcommand{\yellow}[1]{\color{yellow}#1\normalcolor}
\newcommand{\purple}[1]{\color{purple}#1\normalcolor}
\newcommand{\turquoise}[1]{\color{turquoise}#1\normalcolor}
\newcommand{\grey}[1]{\color{grey}#1\normalcolor}
\newcommand{\brown}[1]{\color{brown}#1\normalcolor}
\newcommand{\white}[1]{\color{white}#1\normalcolor}

\newcommand{\setCap}[2]{#1\immediate\write18{./mkcaption.sh #2}}
\newcommand{\getCap}[1]{\acl*{#1}}

\begin{document}

\bShell
cd linkRefs
./linkrefs.sh &
\eShell

\bShell
./openMenu.sh &
\eShell

\ifthenelse{\boolean{final}}
	{}
	{\immediate\write18{./linkpdfs.sh &}}

\bShell
./version.sh
\eShell

\ifthenelse{\boolean{final}}
	{\newcommand{\highlight}{\normalcolor}}
		{\newcommand{\highlight}{\colorbox{tred}}}

\ifthenelse{\boolean{final}}
	{\newcommand{\darklight}{\normalcolor}}
		{\newcommand{\darklight}{\colorbox{tblue}}}

\def\cited#1{\IfSubStringInString{?}{#1}{\darklight{[?]}}{\cite{#1}}}

\acrodef{PCG}{Projected Conjugate Gradient} 
\acrodef{QP}{quadratic programming}
\acrodef{RBF}{Radial-Basis Function}
\acrodef{ABM}{Agent-Based Modelling}
\acrodef{AI}{Artificial Intelligence}
\acrodef{DAI}{Distributed Artificial Intelligence}
\acrodef{API}{Application Programming Interface}
\acrodef{ARF}{p14ARF human tumor-suppressor gene}
\acrodef{B2B}{business-to-business}
\acrodef{BDP}{Biological Design Pattern}
\acrodef{BGS}{Best Guess Solution}
\acrodef{BIC}{Biologically-Inspired Computing}
\acrodef{BML}{Business Modelling Language}
\acrodef{BPEL}{Business Process Execution Language}
\acrodef{BPMN}{Business Process Modelling Notation}
\acrodef{CAS}{Complex Adaptive Systems}
\acrodef{COBOL}{COmmon Business-Oriented Language}
\acrodef{DBE}{Digital Business Ecosystem}
\acrodef{DE}{Digital Ecosystem}
\acrodef{DEC}{distributed evolutionary computing}
\acrodef{DGA}{Distributed genetic algorithms}
\acrodef{DIS}{Distributed Intelligence System}
\acrodef{DNA}{Deoxyribose Nucleic Acid}
\acrodef{DOP}{DBE Open Protocol}
\acrodef{DSS}{Distributed Storage System}
\acrodef{EAP}{Evolving Agent Population}
\acrodef{ebXML}{e-business eXtensible Markup Language}
\acrodef{EC}{Evolutionary Computing}
\acrodef{ECJ}{Evolutionary Computing in Java}
\acrodef{EE}{Evolutionary Environment}
\acrodef{EFL}{Evolutionary Framework for Language}
\acrodef{FLE}{Framework for Language Ecosystems}
\acrodef{EOA}{Ecosystem-Oriented Architecture}
\acrodef{ESS}{evolutionary stable strategy}
\acrodef{EvE}{Evolutionary Environment}
\acrodef{ExE}{Execution Environment}
\acrodef{FCB}{Framework for Computational Biomimicry}
\acrodef{FFF}{Fitness Function Framework}
\acrodef{FL}{Fitness Landscape}
\acrodef{HWU}{Heriot-Watt University}
\acrodef{ICL}{Imperial College London}
\acrodef{ICT}{Information and Communications Technology}
\acrodef{INTEL}{Intel Ireland}
\acrodef{IPA}{International Phonetic Alphabet}
\acrodef{ISUFI}{Istituto Superiore Universitario di Formazione Interdisciplinare}
\acrodef{JDJ}{Java Developer's Journal}
\acrodef{KC}{Kolmogorov-Chaitin}
\acrodef{LAN}{local area network}
\acrodef{LSE}{London School of Economics and Political Science}
\acrodef{MAS}{Multi-Agent System}
\acrodef{MDL}{Minimum Description Length}
\acrodef{MDM2}{murine double minute 2}
\acrodef{MFT}{Mean Field Theory}
\acrodef{MoAS}{Mobile Agent System}
\acrodef{MOF}{Meta Object Facility}
\acrodef{MUH}{migration and usage history}
\acrodef{NIC}{Nature Inspired Computing}
\acrodef{NN}{Neural Network}
\acrodef{NoE}{Network of Excellence}
\acrodef{OMG}{Open Mac Grid}
\acrodef{OPAALS}{Open Philosophies for Associative Autopoietic Digital Ecosystems}
\acrodef{P2P}{peer-to-peer}
\acrodef{P53}{protein 53}
\acrodef{PDA}{Personal Digital Assistant}
\acrodef{QoS}{quality of service}
\acrodef{REST}{REpresentational State Transfer}
\acrodef{RNA}{Deoxyribose Nucleic Acid}
\acrodef{SAE}{Software Agent Ecosystem}
\acrodef{SBML}{Systems Biology Modelling Language}
\acrodef{SBVR}{Semantics of Business Vocabulary and Business Rules}
\acrodef{SDL}{Service Description Language}
\acrodef{SF}{Service Factory}
\acrodef{SIM}{Social Interaction Mechanism}
\acrodef{SM}{Service Manifest}
\acrodef{SME}{Small and Medium sized Enterprise}
\acrodef{SML}{Service Modelling Language}
\acrodef{SMO}{Sequential Minimal Optimisation}
\acrodef{SOA}{Service-Oriented Architecture}
\acrodef{SOAP}{Simple Object Access Protocol}
\acrodef{SOC}{Self-Organised Criticality}
\acrodef{SOLUTA}{SOLUTA.NET}
\acrodef{SOM}{Self-Organising Map}
\acrodef{SSL}{Semantic Service Language}
\acrodef{STU}{Salzburg Technical University}
\acrodef{SUN}{Sun Microsystems}
\acrodef{SVM}{Support Vector Machine}
\acrodef{TM}{Turing Machine}
\acrodef{UBHAM}{University of Birmingham}
\acrodef{UDDI}{Universal Description Discovery and Integration}
\acrodef{UML}{Unified Modelling Language}
\acrodef{URI}{Uniform Resource Identifier}
\acrodef{UTM}{Universal Turing Machine}
\acrodef{VLP}{variable length population}
\acrodef{VLS}{variable length sequences}
\acrodef{vls}{variable length sequence}
\acrodef{WP}{Work-Package}
\acrodef{WSDL}{Web Services Definition Language}
\acrodef{XMI}{XML Metadata Interchange}
\acrodef{XML}{eXtensible Markup Language}
\acrodef{MD5}{Message-Digest algorithm 5}
\acrodef{GA}{genetic algorithm}
\acrodef{GP}{genetic programming}
\acrodef{MASON}{Multi-Agent Simulator Of Neighbourhoods}
\acrodef{Repast}{Recursive Porous Agent Simulation Toolkit}
\acrodef{JCLEC}{Java Computing Library for Evolutionary Computing}
\acrodef{OWL-S}{Web Ontology Language - Service}
\acrodef{EGT}{Evolutionary Game Theory}
\acrodef{RBF}{Radial Basis Functions}
\acrodef{SWS}{Semantic Web Services}
\acrodef{HDD}{Hard Disk Drive}
\acrodef{SSD}{Solid-State Drive}

\acrodef{bidpCap}{starts with identifying some behaviour from a biological system, which would appear to be useful. Followed by observation to understand the mechanisms or principles by which it operates, and therefore allowing for an abstract understanding of the behaviour. This can then be mimicked in a non-biological system and its performance and effectiveness evaluated \cite{bidpSpiral}.}
\acrodef{introEcoCap}{made up of one or more communities of organisms, consisting of species in their habitats, with their populations existing in their respective micro-habitats \cite{begon96}. A community is a naturally occurring group of populations from different species that live together, and interact as a self-contained unit in the same habitat. A habitat is a distinct part of the environment \cite{begon96}}
\acrodef{im1}{are different probabilities of going from island \Circ{1} to island \Circ{2}, as there is of going from island \Circ{2} to island \Circ{1}.}
\acrodef{im2}{mirrors the naturally inspired quality that although two populations have the same physical separation, it may be easier to migrate in one direction than the other, i.e. fish migration is easier downstream than upstream.}
\acrodef{DBEdescription}{A wealthy ecosystem sees a balance between co-operation and competition in a dynamic free market.}
\acrodef{serviceCap}{lightweight entity consisting primarily of a pointer to the semantic web service it represents,}
\acrodef{service2cap}{{executable component} and {semantic description}. A software service can be a software service only, e.g. for data encryption, or a software service providing a front-end to a real-world service, e.g. selling books}
\acrodef{structureCap}{The {executable component} of a semantic web service that an Agent represents is equivalent to an organism's \acs{DNA} and is the gene (functional unit) in the evolutionary process \cite{lawrence1989hsd}. So, the Agents should be aggregated as a {sequence}, like the sequencing of genes in \acs{DNA} \cite{lawrence1989hsd}.}
\acrodef{structure2}{an unordered {set}, or, based on service orchestration, into a {tree} or {workflow}}
\acrodef{habnet}{the {agent stations} from mobile agent systems \cite{agentStation} (to provide a distributed environment in which Agent migration can occur), with {evolutionary computing} \cite{eiben2003iec} for the Agent interaction (instead of traditional agent interaction mechanisms \cite{wooldridge}), and the {island-model} of \acl{DEC} \cite{lin1994cgp} for the connectivity between Habitats.}
\acrodef{picUser}{will formulate queries to the Digital Ecosystem by creating a request as a {semantic description}, like those being used and developed in \aclp{SOA} \cite{SOAsemantic}, specifying an application they desire and submitting it to their Habitat.}
\acrodef{picUserReq}{A Population is then instantiated in the user's Habitat in response to the user's request, seeded from the Agents available at their Habitat}
\acrodef{digEco}{with the Agents, the Populations, the Agent migration for \acl{DEC}, and the environmental selection pressures provided by the user base, then the union of the Habitats creates the Digital Ecosystem}
\acrodef{archComTop}{many strongly connected clusters (communities), called {sub-networks} (quasi-complete graphs), with a few connections between these clusters (communities) \cite{swn1}. Graphs with this topology have a very high clustering coefficient and small characteristic path lengths \cite{swn1}.}
\acrodef{bizEcoCap}{As the connections between Habitats are reconfigured depending on the connectivity of the user base, the Habitat clustering will therefore be parallel to the business sector communities}
\acrodef{similarCap}{requests are evaluated on separate {islands} (Populations), with their evolution accelerated by the sharing of solutions between the evolving Populations (islands), because they are working to solve similar requests (problems).}
\acrodef{similar2}{yellow lines connecting the evolving Populations indicate similarity in the requests being managed.}
\acrodef{lifeCycleCap}{with deployment to its owner's Habitat for distribution within the Habitat network.}
\acrodef{lifeCycle2}{used in evolving the optimal Agent-sequence in response to a user request. The optimal Agent-sequence is then registered at the Habitat}
\acrodef{lifeCycle3}{If an Agent-sequence solution is then executed, an attempt is made to migrate (copy) it to every other connected Habitat, success depending on the probability associated with the connection.}
\acrodef{as3}{with an abstract representation consisting of a set of}
\acrodef{agentSemantic2}{tuple representing an {attribute} of the {semantic description}, one integer for the {attribute identifier} and one for the {attribute value}, with both ranging between one and a hundred.}
\acrodef{as4}{Each simulated Agent had a semantic description}
\acrodef{semanticRequest}{A simulated user request consisted of an abstract {semantic description}, as a list of sets of numeric tuples to represent the properties of a desired business application}
\acrodef{bmlcap1}{the {numerical semantic descriptions}, of the simulated services (Agents) and user requests, in a {human readable form}.}
\acrodef{capbml2}{The {semantic filter} translates {numerical semantic descriptions} for one community within the user base, showing it in the context of the travel industry}
\acrodef{capbml3}{The simulation still operated on the numerical representation for operational efficiency, but the {semantic filter} essentially assigns meaning to the numbers.}
\acrodef{evoGraph}{shows both the maximum and average fitness increasing over the generations of a typical Population, and as expected the {average fitness} remains below the {maximum fitness} because of variation in the Population \cite{goldberg}, showing that the evolutionary processes, which construct order in the Digital Ecosystem, are operating satisfactorily.}
\acrodef{succession}{So, it becomes increasingly more complex through this process of succession, driven by the evolution of the populations within the ecosystem \cite{connell111msn}.}
\acrodef{succession2}{The formation of a mature ecosystem}
\acrodef{succession3}{is the slow, predictable, and orderly changes in the composition and structure of an ecological community, for which there are defined stages in the increasing complexity \cite{begon96}, as shown}
\acrodef{DigEcoSuc2Cap}{the end of the simulation run, the Agent-sequences had evolved and migrated over an average of only ten user requests per Habitat, and collectively had already reached near 70\% effectiveness for the user base.}
\acrodef{DigEcoSucCap}{The formation of a mature biological ecosystem, ecological succession, is a relatively slow process \cite{begon96}, and the simulated Digital Ecosystem acted similarly in reaching a mature state.}
\acrodef{speciesAbundance}{is a measure of the proportion of all organisms in a community belonging to a particular {species} \cite{Bell}. A {relative abundance distribution} provides the inequalities in population size within an ecosystem and therefore an indicator of biodiversity, with the distribution of most biological ecosystems taking a log-normal form \cite{Bell}.}
\acrodef{specAbund2}{the Digital Ecosystem did not conform to the expected log-normal}
\acrodef{speciesArea}{In ecology the {species-area} relationship measures diversity relative to the spatial scale, showing the number of species found in a defined area of a particular habitat or habitats of different areas \cite{sizling2004pls}, and is commonly found to follow a power law in biological ecosystems}
\acrodef{ecoCapClass}{then the {Digital Ecosystem} and {biological ecosystem} classes would both inherit from the abstract {ecosystem} class, but implement its attributes differently}
\acrodef{ecoCap2Class}{So, we would argue that the apparent compromises in mimicking biological ecosystems are actually features unique to Digital Ecosystems.}
\acrodef{visOrgCap}{number of Agents, in total and of each colour, is the same in both populations. However, the Agent Population on the left intuitively shows organisation through the uniformity of the colours across the Agent-sequences, whereas the population to the right shows little or no organisation.}
\acrodef{DNAcap}{DNA sequences are made up from four nucleotides, Adenosine (A), Thymine (T), Cytosine (C) and Guanine (G). The nucleotides always pair as follows, Adenosine with Thymine, and Cytosine with Guanine. So, DNA sequences can be reduced to a {genome sequence} showing half of the paired information \cite{lawrence1989hsd}}
\acrodef{genCap2}{equalling the maximum length would be incorrect}
\acrodef{genCap}{there is an insufficient sample size for the estimated probabilities to provide an accurate calculation.}
\acrodef{orgCPcap}{are consistent with the intuitive understanding one would have for the self-organised {complexity} of the sample populations}
\acrodef{FLsingleCap}{shows the combination space (power set) of the alphabet $D$ against the fitness values from the {selection pressure} (user request)}
\acrodef{FLsingle2Cap}{The Agent-sequences of an evolving Population will evolve, moving across the {fitness landscape} and clustering around the optimal genome at the peak of the global optimum}
\acrodef{FLflatCap}{uniformly random, as any position (sequence) has the same fitness as any other. So the entropy (randomness) tends to maximum, resulting in the complexity $C_{V}$ tending to zero, and therefore the Efficiency $E$ also tending to zero}
\acrodef{FLmul2Cap}{Efficiency $E$ will tend to a maximum below one, because the population of sequences consists of more than one cluster, with each having an Efficiency tending to a maximum of one.}
\acrodef{FLmulCap}{The simplest scenario of clusters is {pure clusters}}
\acrodef{popClusShowCap}{The clusters of the population have Efficiency values tending to a maximum of one, compared to the Efficiency of the population as a whole, which is tending to a maximum significantly below one.}
\acrodef{atomCap}{of a set of Agents, such that no single Agent can functionally replace any Agent-sequence, i.e. their functionality is {mutually exclusive} to one another. It is important because non-atomicity can adversely affect}
\acrodef{atom2Cap}{the Physical Complexity measure}
\acrodef{popAtomCap}{Efficiency $E$ of the population is a half, whereas the Efficiency $E_{c}$ for populations with clusters is one, because it supports clustering and therefore non-atomicity}
\acrodef{phyComGraphCap}{Physical Complexity increases over the generations, suffering short-term decreases from the arrival of {fitter} mutants, which spread through the population over several generations and causes the uniformity of the sites to decrease temporarily, while the maximum fitness of the population increases over the generations until the global optimum is reached}
\acrodef{graph2cap}{The Physical Complexity for \aclp{vls} increases over the generations, showing short-term decreases as expected \cite{adami20002}. It increases over the generations because of the increasing information being stored, with the sharp increases occurring when the effective length $\ell_{V}$ of the Population increases.}
\acrodef{largeVisCap}{The visualisation shows that our Efficiency $E$ accurately measures the self-organised complexity of the two Populations.}
\acrodef{graph3cap}{the Population from Figure \ref{phycom}}
\acrodef{graph32cap}{The Efficiency tends to a maximum of one, indicating that the Population consists of one cluster, which is confirmed by the visualisation of the Population in Figure \ref{newphycomvis} (left). The significant decreases that occurred in the Efficiency, reducing in magnitude and frequency over the generations, came from mirroring the fluctuations that occurred in the complexity $C_V$}
\acrodef{graph4cap}{around the included best fit curve, quite significantly at the start, and then decreasing as the generations progressed.}
\acrodef{visClustersCap}{Agent-sequences were grouped to show the two clusters}
\acrodef{visClusters2Cap}{expected from (\ref{defineCluster}) each cluster had a much higher Physical Complexity and Efficiency compared to the Population as a whole. However, the Efficiency $E_{c}$}
\acrodef{visClusters3Cap}{calculated the}
\acrodef{statesCap}{possible evolutionary path through the state-space $I$}
\acrodef{capStates3}{the {selection pressure} of the evolutionary process}
\acrodef{capStates}{driving it towards the {maximal state} of the {maximum macro-state} $M_{max}$, which consists entirely of copies of the {optimal solution}, and is the equilibrium state that the system $S$ is forever {falling towards} without ever quite reaching, because of the noise (mutation) within the system.}
\acrodef{graphCap}{in the {maximum macro-state} $M_{max}$ only after generation 178 and always after generation 482. It was also observed being in the {sub-optimal macro-state} $M_{half}$ only between generations 37 and 113, with a maximum probability of 0.053 (3 d.p.) at generation 61}
\acrodef{aScap}{With the mutation rate under or equal to 60\%, the evolving Agent Population showed no instability, with $d_{ins}$ values equal to zero as the system $S$ was always in the same macro-state $M$ at infinite time, independent of the crossover rate. With the mutation rate above 60\% the instability increased significantly}
\acrodef{urlCapUnifrom}{The {observed} frequencies of the Agent-sequence length mostly matched the {expected} frequencies, which was {confirmed} by a $\chi^2$ test; with a {null hypothesis} of {no significant difference} and {sixteen degrees of freedom}, the $\chi^2$ value was 2.588 (3 d.p.), below the critical 0.95 $\chi^2$ value of 7.962.}
\acrodef{urlCapGaussian}{The {observed} frequencies of the Agent-sequence length matched the {expected} frequencies with only very minor variations, which was confirmed by a $\chi^2$ test; with a {null hypothesis} of {no significant difference} and {sixteen degrees of freedom}, the $\chi^2$ value was 2.102}
\acrodef{urlCap2Gaussian}{below the critical 0.95 $\chi^2$ value of 7.962.}
\acrodef{urlpower}{The {observed} frequencies of the Agent-sequence length matched the {expected} frequencies with some variation, which was confirmed by a $\chi^2$ test; with a {null hypothesis} of {no significant difference} and {sixteen degrees of freedom}, the $\chi^2$ value was 5.048 (3 d.p.), below the critical 0.95 $\chi^2$ value of 7.962.}
\acrodef{urvunifromCap}{The {observed} frequencies for the number of Agent attributes mostly matched the {expected} frequencies, which was confirmed by a $\chi^2$ test; with a {null hypothesis} of {no significant difference} and {ten degrees of freedom}, the $\chi^2$ value was 1.049 (3 d.p.), below the critical 0.95 $\chi^2$ value of 3.940.}
\acrodef{urvgaussianCap}{The {observed} frequencies for the number of Agent attributes {appeared} to follow the {expected} frequencies, but there was significant variation which led to a failed $\chi^2$ test; with a {null hypothesis} of {no significant difference} and {ten degrees of freedom}}
\acrodef{urvpowerCap}{The {observed} frequencies for the number of Agent attributes {appeared} to follow the {expected} frequencies, but there was significant variation which led to a failed $\chi^2$ test; with a {null hypothesis} of {no significant difference} and {ten degrees of freedom}}
\acrodef{CC1cap}{would encourage intra-cluster crossover, reducing the number of generations required for the clusters to reach their respective optimal genomes (applications), therefore directly accelerating the evolutionary self-organisation in determining applications (Agent-sequences) to user requests. So, accelerating the responsiveness of the Digital Ecosystem to the user base, and the process of {ecological succession} \cite{begon96}.}
\acrodef{com1Cap}{for the creation of potentially useful applications (Agent-sequences) or partial applications inside the Agent-pools, increasing and optimising the recombination that occurs globally within the Digital Ecosystem. So, it would help to optimise the Agent-sequences at the Agent-pools of the Habitats, which would in turn optimise the evolving Agent Populations, as they make use of the Agent-pools}
\acrodef{ch1ifAp}{effect of the different augmentations on the {evolutionary dynamics} (the evolving Agent Populations) and the {ecological dynamics} (the Habitats)}
\acrodef{ch2ifAp}{This separation of concerns is an artificial construct, but useful in summarising the potential of the different augmentations, before we decide upon which to pursue.}
\acrodef{nlccap}{The Agent life-cycle, defined in section \ref{agentLifeCycle}, will change to support the {targeted migration}, as shown}
\acrodef{nlc2cap}{by the blue circle. Specifically, there will be more opportunities for Agent migration, but more importantly these opportunities will be for {targeted migration}, which will help to optimise the set of Agents found at the Habitats}
\acrodef{svmCap}{A \acf{RBF} is recommended for text categorisation \cite{joachims1997tcs}, with the most common form of the \ac{RBF} being Gaussian \cite{gunn1998svm}.}
\acrodef{sbcf}{The evolving Agent Populations alone averaged 296 (3 s.f.)}
\acrodef{sbc2f}{while the evolving Agent Populations with the {crossover control} showed a 9\% reduction, averaging 267 (3 s.f.)}
\acrodef{sbc3f}{The evolving Agent Populations with the {hierarchical clustering based clustering catalyst} failed to provide any further optimisation, averaging 281 (3 s.f.)}
\acrodef{sbch}{The evolving Agent Populations with the {Physical Complexity based clustering catalyst} averaged 274 (3 s.f.)}
\acrodef{sb2ch}{better than the evolving Agent Populations with the {hierarchical clustering based clustering catalyst} which averaged 281 (3 s.f.) generations}
\acrodef{sb3ch}{but still worse than the evolving Agent Populations with the {crossover control} which averaged 267 (3 s.f.) generations}
\acrodef{sbrc}{As the {clustering catalyst} was unsuccessful,}
\acrodef{sbr2c}{we graphed a typical run of each scenario to observe its behaviour and so better understand why it failed. However, there was no unexpected behaviour, confirming that the evolving Agent Populations with the {clustering catalyst} were simply less efficient than the evolving Agent Populations with the {crossover control}.}
\acrodef{tmcf}{The Digital Ecosystem alone averaged a 68.0\% (3 s.f.)}
\acrodef{tmc2f}{while the Digital Ecosystem with the {migration control} showed a significant degradation to 49.6\% (3 s.f.)}
\acrodef{tmc3f}{and the Digital Ecosystem with the {pattern recognition control} showed only a small increase to 70.5\% (3 s.f.)}
\acrodef{boohoo}{The Digital Ecosystem alone performed as expected, adapting and improving over time to reach a mature state}
\acrodef{nnbtmCap}{The Digital Ecosystem alone averaged a 68.0\% (3 s.f.) response rate with a standard deviation of 2.61 (2 d.p.), while the Digital Ecosystem with the \ac{NN}-based {targeted migration} showed a significant improvement to a 92.1\% (3 s.f.) response rate with a standard deviation of 2.22 (2 d.p.).}
\acrodef{tmsf}{The Digital Ecosystem with the \ac{SVM}-based {targeted migration} averaged a 92.8\% (3 s.f.) response rate}
\acrodef{tms2f}{slightly better than the \ac{NN}-based {targeted migration} at 92.1\% (3 s.f.)}
\acrodef{tms3f}{and so significantly better than the Digital Ecosystem alone at 68.0\% (3 s.f.)}
\acrodef{gSVM2cap}{Digital Ecosystem with the \ac{SVM}-based {targeted migration}, the Digital Ecosystem with the \ac{NN}-based {targeted migration}, and the Digital Ecosystem alone.}
\acrodef{gNN2cap}{The Digital Ecosystem alone performed as expected, adapting and improving over time to reach a mature state through the process of {ecological succession} \cite{begon96}}
\acrodef{gNN3cap}{Digital Ecosystem with the {targeted migration}, \ac{NN} or \ac{SVM}-based, showed a significant improvement}
\acrodef{histogram2Cap}{the greater effectiveness of the SVM-based {targeted migration}, compared to the NN-based {targeted migration}}
\acrodef{histogramCap}{frequency of poor matches ($<$50\%) every one hundred time steps, for the Digital Ecosystem with the SVM-based {targeted migration}, compared to the Digital Ecosystem with the NN-based {targeted migration}, and the Digital Ecosystem alone}


\title{\LARGE {\bf Digital Ecosystems}\\
\vspace*{6mm}{}}
\author{Gerard Briscoe}
\narrowlinespacing
\maketitle

\preface

\chapter*{Abstract}
\addcontentsline{toc}{chapter}{Abstract}

We view Digital Ecosystems to be the \emph{digital counterparts of biological ecosystems}, which are considered to be robust, self-organising and scalable architectures that can automatically solve complex, dynamic problems. So, this work is concerned with the creation, investigation, and optimisation of Digital Ecosystems, exploiting the self-organising properties of biological ecosystems. First, we created the Digital Ecosystem, a novel optimisation technique inspired by biological ecosystems, where the optimisation works at two levels: a first optimisation, migration of agents which are distributed in a decentralised peer-to-peer network, operating continuously in time; this process feeds a second optimisation based on evolutionary computing that operates locally on single peers and is aimed at finding solutions to satisfy locally relevant constraints. We then investigated its self-organising aspects, starting with an extension to the definition of Physical Complexity to include the evolving agent populations of our Digital Ecosystem. Next, we established stability of evolving agent populations over time, by extending the Chli-DeWilde definition of agent stability to include evolutionary dynamics. Further, we evaluated the diversity of the software agents within evolving agent populations, relative to the environment provided by the user base. To conclude, we considered alternative augmentations to optimise and accelerate our Digital Ecosystem, by studying the accelerating effect of a \emph{clustering catalyst} on the evolutionary dynamics of our Digital Ecosystem, through the direct acceleration of the evolutionary processes. We also studied the optimising effect of \emph{targeted migration} on the ecological dynamics of our Digital Ecosystem, through the indirect and emergent optimisation of the agent migration patterns. Overall, we have advanced the understanding of creating Digital Ecosystems, the self-organisation that occurs within them, and the optimisation of their \acl{EOA}.

\pagebreak
\thispagestyle{plain}
\white{.}
\pagebreak

\pagestyle{uheadings}
\pdfbookmark[0]{Contents}{contents}
\begin{spacing}{1.5}
\tableofcontents
\end{spacing}
\vfill

\pagebreak
\thispagestyle{plain}

\pagestyle{uheadings}
\chapter*{Acknowledgements}
\addcontentsline{toc}{chapter}{Acknowledgements}
I would like to express my most sincere thanks to:

\begin{itemize}
\item My supervisor, Philippe De Wilde, to whom I am deeply indebted for the guidance and support he has provided throughout.
\item My examiners, Peter McBurney and Philip Treleaven, for their time and effort in the reading of this thesis and the viva.
\item My mentor, Paolo Dini, for his constant support, and originally encouraging to me pursue this research and PhD.
\item My friend and colleague, Evangelia Berdou, for her support and guidance in the writing of this document.
\item My colleagues, Sotiris Moschoyiannis and Alastair Munro, for kindly agreeing to review this document.
\item Several other researchers and friends, most notably Suzanne Sadedin, Maria Chli and Frauke Zeller from whom I have learnt much about conducting research.
\item My mother, Constance Briscoe, without whom I could never have finished this work.
\item The European Union Framework VI project, \aclp{DBE}, Contract No 507953, and the Network of Excellence, \acl{OPAALS}, Contract No 034824, for supporting this work.
\item The Open Mac Grid of \href{http://www.MacResearch.org}{\uline{MacResearch.org}}, and the Digital Ecosystems Lab of the Department of Media and Communications at the London School of Economics and Political Science, for access to their grids for the running of simulations.
\end{itemize}

\pagebreak
\thispagestyle{plain}
\white{.}
\pagebreak
\thispagestyle{plain}

\addcontentsline{toc}{chapter}{List of Figures}
\execute{./fixlof.sh}
\begin{spacing}{1.45}
\listoffigures
\end{spacing}
\vspace{10cm}

\pagebreak
\thispagestyle{plain}
\pagestyle{uheadings}
\doublespacing

\chapter{Introduction}

\section{Motivation and Objectives}

Is mimicking ecosystems the future of information systems ? 

A key challenge in modern computing is to develop systems that address complex, dynamic problems in a scalable and efficient way, because the increasing complexity of software makes designing and maintaining efficient and flexible systems a growing challenge \cite{newsArticle1, slashdot, newsArticle3}. What with the ever expanding number of services being offered online from \acp{API} being made public, there is an ever growing number of computational units available to be combined in the creation of applications. However, this is currently a task done manually by programmers, and it has been argued \cite{lyytinen2001nwn} that current software development techniques have hit a \emph{complexity wall}, which can only be overcome by automating the search for new algorithms. There are several existing efforts aimed at achieving this automated service composition \cite{reef3, reef5, reef1, reef4}, the most prevalent of which is \aclp{SOA} and its associated standards and technologies \cite{curbera2002uws, SOAstandards}. 

Alternatively, nature has been in the research business for 3.8 billion years and in that time has accumulated close to 30 million \emph{well-adjusted} solutions to a plethora of design challenges that humankind struggles to address with mixed results \cite{biomimicry}. Biomimicry is a discipline that seeks solutions by emulating nature's designs and processes, and there is considerable opportunity to learn elegant solutions for human-made problems \cite{biomimicry}. Biological ecosystems are thought to be robust, scalable architectures that can automatically solve complex, dynamic problems, possessing several properties that may be useful in automated systems. These properties include self-organisation, self-management, scalability, the ability to provide complex solutions, and the automated composition of these complex solutions \cite{Levin}.

Therefore, an approach to the aforementioned challenge would be to develop Digital Ecosystems, artificial systems that aim to harness the dynamics that underlie the complex and diverse adaptations of living organisms in \emph{biological ecosystems}. While evolution may be well understood in computer science under the auspices of \emph{evolutionary computing} \cite{eiben2003iec}, ecological models are not. The possible connections between Digital Ecosystems and their biological counterparts are yet to be closely examined, so potential exists to create an \acl{EOA} with the essential elements of \emph{biological ecosystems}, where the word \emph{ecosystem} is more than just a \emph{metaphor}. We propose that an ecosystem inspired approach, would be more effective at greater scales than traditionally inspired approaches, because it would be built upon the scalable and self-organising properties of \emph{biological ecosystems} \cite{Levin}. However, \emph{ecological succession}, the formation of a mature ecosystem from the predictable and orderly changes in the composition and structure of an ecological community \cite{begon96}, is a slow process. So, for our Digital Ecosystems it will be desirable to accelerate and optimise the equivalent process, which may be possible through the application of augmentations that interact with the ecosystem dynamics. Therefore, the primary objectives are as follows: 

\vspace{-5mm}
\begin{itemize}
\item Determine the structure of an \acl{EOA} and so create Digital Ecosystems, which are the digital counterpart of \emph{biological ecosystems}, and so have analogous properties of self-organisation, scalability and sustainability.

\item Develop an understanding of the self-organising behaviour within a Digital Ecosystem, learning where and how it occurs, what forms it can take, and how it can be quantified.

\item Investigate if we can accelerate or optimise the evolutionary and ecological self-organising dynamics of Digital Ecosystems, exploring how alternative augmentations interact with the ecosystem dynamics.
\end{itemize}

\section{Contributions}
Substantial parts of our efforts are original contributions in the area of \acl{BIC} \cite{forbes2004ilb} and the emerging field of Digital Ecosystems, with our major research contributions being as follows: 

\begin{itemize}
\item We have determined the fundamentals for a new class of system, Digital Ecosystems, created through combining understanding from theoretical ecology, evolutionary theory, \aclp{MAS}, distributed evolutionary computing, and \aclp{SOA}.

\item We have investigated where and how self-organisation occurs in Digital Ecosystems, what forms it can take and how it can be quantified, including the self-organised complexity, stability, and diversity of the evolving agent populations within. 

\item We have extended the statistical physics based definition of Physical Complexity, to include evolving agent populations. This required extending definitions for populations of variable length sequences, creating a measure for the efficiency of information storage, and an understanding of clustering within populations to support the non-atomicity of agents.

\item We have extended the Chli-DeWilde definition of agent stability to include the evolutionary dynamics of evolving agent populations. We then built upon this to construct an entropy-based definition for the degree of instability, which was used to study the stability of evolving agent populations under varying conditions.

\item We have developed an understanding and definition for the self-organised \emph{diversity}, finding no existing definition suitable because of the unique hybrid nature of Digital Ecosystems. We therefore considered the global distribution of the agents in the populations relative to the varying requirements of the user base.

\item We have investigated alternative augmentations to optimise and accelerate our Digital Ecosystems, studying the accelerating effect of a \emph{clustering catalyst} on the evolutionary dynamics, and the optimising effect of \emph{targeted migration} on the ecological dynamics.

\end{itemize}

\pagebreak
\bShell
./pubfix.sh
\eShell
\section{Publications}
\nocitepub{agentStability, phycom, de08th, dbebkpub, de07oz, bionetics, javaOne, briscoeWCAT}
The following publications support the material presented in this thesis: 
\color{white}
\bibliographystylepub{publications}
\bibliographypub{references}
\bShell
bibtex pub2
cp pub2.bbl pub.bbl
./references.sh pub.aux &
\eShell
\normalcolor

\section{Dissertation Outline}

In \bold{Chapter \ref{ch:creation}}, we explain the hybrid model created to provide the digital counterpart of a biological ecosystem. We start with the relevant theory from the domain of theoretical biology, including the fields of evolutionary and ecological theory, and from the domain of computer science, including the fields of Multi-Agent Systems, evolutionary computing and Service-Oriented Architectures. The Digital Ecosystem is then measured experimentally through simulations, with measures originating from theoretical ecology, to evaluate its likeness to a biological ecosystem. This included its responsiveness to requests for applications from the user base, as a measure of the \emph{ecological succession} (ecosystem maturity).

\bold{Chapter \ref{ch:investigation}} investigates the self-organising aspects of Digital Ecosystems. We start with the complexity of the evolving agent populations within, by extending the statistical physics based definition of Physical Complexity to support variable length populations of software agents. Next, we investigate the stability of the evolving agent populations, by extending the Chli-DeWilde definition of agent stability to include the evolutionary dynamics of Digital Ecosystems. Finally, we study the diversity of the agents within the evolving agent populations of the Digital Ecosystem, for optimality relative to the environment provided by the user base. 

In \bold{Chapter \ref{ch:optimisation}}, we start by considering alternative augmentations to optimise and accelerate Digital Ecosystems. We then further investigate the most promising, the \emph{clustering catalyst} and \emph{targeted migration}: the accelerating effect of a \emph{clustering catalyst} on the evolutionary dynamics of our Digital Ecosystem, through the direct acceleration of the evolutionary processes; and the optimising effect of \emph{targeted migration} on the ecological dynamics of our Digital Ecosystem, through the indirect and emergent optimisation of the agent migration patterns.

\bold{Chapter \ref{ch:conclusions}} provides a summary of the conclusions, and suggests possible future research into Digital Ecosystems. We also report on the status of the reference implementation for Digital Ecosystems, and the dedicated simulation framework created for its future study. After this the \bold{Bibliography} follows.

\pagebreak
\thispagestyle{plain}

\chapter{Creation of Digital Ecosystems}
\label{ch:creation}

In this chapter we create Digital Ecosystems, starting with a discussion of the relevant literature, including \acl{NIC} as a framework in which to understand this work, and the process of biomimicry to be used in mimicking the necessary biological processes to create Digital Ecosystems. We then consider the relevant theoretical ecology in creating the digital counterpart of a biological ecosystem, including the topological structure of ecosystems, and evolutionary processes within distributed environments. This leads to a discussion of the relevant fields from computer science for the creation of Digital Ecosystems, including evolutionary computing, \aclp{MAS}, and \aclp{SOA}. We then define \aclp{EOA} for the creation of Digital Ecosystems, imbibed with the properties of self-organisation, scalability and sustainability from biological ecosystems, including a novel form of \acl{DEC}. This will include a discussion of the compromises resulting from the hybrid model created, such as the network topology. We then performed simulations to compare the likeness of our Digital Ecosystem with biological ecosystems, starting with \emph{ecological succession} (development), measured by its responsiveness to requests for applications from the user base, and followed by the measures of \emph{species abundance} and the \emph{species-area relationship}, which are commonly applied to biological ecosystems. Finally, we conclude with a summary and discussion of the achievements, including the experimental results.

\section{Background Theory}
\label{background}
In this section we discuss the relevant background theory, and because of the interdisciplinary nature of our research it will cover several fields across different domains. We start with an introduction to \acl{NIC}, followed by the relevant theoretical biology and computer science. With the theoretical biology, we will consider how properties of biological ecosystems influence functions that are relevant to developing Digital Ecosystems to solve practical problems. This leads us to suggest ways in which concepts from ecology can be used in biologically inspired techniques to create Digital Ecosystems.

\subsection{Existing Digital Ecosystems}
\nocite{marrow2001adi}
Our focus is in creating the digital counterpart of biological ecosystems. However, the term \emph{digital ecosystem} has described a variety of concepts, which we shall now review. Sometimes referring to the existing networking infrastructure of the Internet \cite{debook2, ballmer, fiorina, XIMBIOTIX}, while several companies offer a \emph{digital ecosystem} service, which involves enabling customers to use existing e-business solutions \cite{accenture, syntel, xewow}. The term is also being increasingly linked to the future developments of \ac{ICT} adoption for e-business, to support \emph{business ecosystems} \cite{moore1996}. However, perhaps the most frequent references to \emph{digital ecosystems} arise in Artificial Life research, where they are created primarily to investigate aspects of biological and other complex systems \cite{sorakugun1995eas, grand1998ces, deAI1}.

The extent to which these disparate systems resemble biological ecosystems varies, and frequently the word \emph{ecosystem} is merely used for branding purposes without any inherent ecological properties. We consider Digital Ecosystems to be software systems that exploit the properties of biological ecosystems, and suggest that several key features of biological ecosystems have not been fully explored in existing \emph{digital ecosystems}. So, we will now discuss how mimicking these features can create Digital Ecosystems, which are robust, scalable, and self-organising.

\subsection{Nature-Inspired Computing}

Biomimicry (bios, meaning life, and mimesis, meaning to imitate) is the science that studies nature, its models, systems, processes, and elements, and then imitates or takes creative inspiration from them for the study and design of engineering systems and modern technology \cite{biomimicry}. This concept is far from new, with humans having long been inspired by the animals and plants of the natural world; Leonardo Da Vinci himself once said, \emph{Those who are inspired by a model other than Nature, a mistress above all masters, are labouring in vain} \cite{bramly1994laa}. Albeit overstating the point, it reminds us that the transfer of technology between life-forms and synthetic constructs is desirable because evolutionary pressures typically force living organisms to become highly optimised and efficient. A classical example is the development of dirt and water repellent paint from the observation that the surface of the lotus flower plant is practically non-sticky for anything, commonly known as the \emph{lotus effect} \cite{barthlott1997psl}. However, biomimicry, when done well, is not slavish imitation; it is inspiration using the principles which nature has demonstrated to be successful design strategies. For example, in the early days of mechanised flight the best designs were not the ornithopters, which most completely imitated birds, but the fixed-wing craft that used the principle of aerofoil cross-section in their wings \cite{andersonFlight}. 

Biomimicry in computer science is called \ac{NIC} or Natural Computation, and the benefits of natural computation technologies often mimic those found in real natural systems, and include flexibility, adaptability, robustness, and decentralised control \cite{NICbook}. The increasing demands upon current computer systems, along with technological changes, create a need for more flexible and adaptable systems. The desire to achieve this has led many computing researchers to look to natural systems for inspiration in the design of computer software and hardware, as natural systems provide many examples of the type of versatile system required \cite{NICbook}. Their sources of inspiration come from many aspects of natural systems; evolution, ecology, development, cell and molecular phenomena, behaviour, cognition, and other areas \cite{ecpaper}. The use of nature inspired techniques often results in the design of novel computing systems with applicability in many different areas \cite{ecpaper}. \ac{NIC} itself can be divided into three main branches \cite{NICbook}:

\begin{itemize}
\item \ac{BIC}: This makes use of nature as inspiration for the development of problem solving techniques. The main idea of this branch is to develop computational tools (algorithms) by taking inspiration from nature for the solution of complex problems.

\item The simulation and emulation of nature by computational means: This is basically a synthetic process aimed at creating patterns, forms, behaviours, and organisms that resemble \emph{life-as-we-know-it}. Its products can be used to mimic various natural phenomena, thus increasing our understanding of nature and insights about computer models.

\item Computing with natural materials: This corresponds to the use of natural materials to perform computation, to substitute or supplement the current silicon-based computers.
\end{itemize}

All branches share the common characteristic of human-designed computing inspired by nature, the metaphorical use of concepts, principles, and mechanisms underlying natural systems. Thus, evolutionary algorithms use the concepts of mutation, recombination, and natural selection from biology; neural networks are inspired by the highly interconnected neural structures in the brain and the nervous system; molecular computing is based on paradigms from molecular biology; and quantum computing based on quantum physics exploits quantum parallelism \cite{NICbook}. There are however, important methodological differences between various sub-areas of natural computing. For example, evolutionary algorithms and algorithms based on neural networks are presently implemented on conventional computers. However, molecular computing also aims at alternatives for silicon hardware by implementing algorithms in biological hardware, using DNA molecules and enzymes. Also, quantum computing aims at non-traditional hardware that can make use of quantum effects \cite{NICbook}.

We are concerned with \ac{BIC}, which relies heavily on the fields of biology, computer science, and mathematics. Briefly put, it is the study of nature to improve the usage of computers \cite{forbes2004ilb}, and should not to be confused with \emph{computational biology} \cite{waterman1995icb}, which is an interdisciplinary field that applies the techniques of computer science, applied mathematics, and statistics to address problems inspired by biology. \ac{BIC} has produced \aclp{NN}, swarm intelligence and evolutionary computing \cite{forbes2004ilb}. Introducing \ac{BIC}, one comes quickly to its applications, partly because this is the essence of the approach, and partly because biomimicry as a process tends to be un-formalised and ad hoc \cite{bidpSpiral}. It generally involves an engineer or scientist observing or being aware of an area of biological study, which seems applicable to a technology or research problem they are currently tackling, or which inspires the creation of a new technology \cite{NICbook}. However, there are some common steps in this process, which \setCap{starts with identifying some behaviour from a biological system, which would appear to be useful. Followed by observation to understand the mechanisms or principles by which it operates, and therefore allowing for an abstract understanding of the behaviour. This can then be mimicked in a non-biological system and its performance and effectiveness evaluated \cite{bidpSpiral}.}{bidpCap} This process is summarised Figure \ref{bidp}.

\tfigure{scale=1.0}{bidp}{graffle}{Biomimicry Design Spiral}{(modified from \cite{bidpSpiral}): The process of biomimicry \getCap{bidpCap}}{0mm}{}{}{}

\subsection{Biology of Digital Ecosystems}

\label{bioOfDE}

Natural science is the study of the universe via the rules or laws of natural order, and the term is also used to differentiate those fields using scientific method in the study of nature, in contrast with the social sciences which apply the scientific method to culture and human behaviour: economics, psychology, political economy, anthropology, etc \cite{hollis1994pss}. The fields of natural science are diverse, ranging from particle physics to astronomy \cite{salmon1999ips}, and while not all these fields of study will provide paradigms for Digital Ecosystems, the further one wishes to take the analogy of the word \emph{ecosystem}, the more one has to consider the relevance of the fields of natural science, particularly the biological sciences.

A primary motivation for our research in Digital Ecosystems is the desire to exploit the self-organising properties of biological ecosystems. Ecosystems are thought to be robust, scalable architectures that can automatically solve complex, dynamic problems \cite{Levin}. However, the biological processes that contribute to these properties have not been made explicit in Digital Ecosystems research. Here, we discuss how biological properties contribute to the self-organising features of biological ecosystems, including population dynamics, evolution, a complex dynamic environment, and spatial distributions for generating local interactions \cite{tilman1997ser}. The potential for exploiting these properties in artificial systems is then considered. We suggest that several key features of biological ecosystems have not been fully explored in existing digital ecosystems, and discuss how mimicking these features may assist in developing robust, scalable self-organising architectures.

Evolutionary computing uses natural selection to evolve solutions \cite{goldberg}; it starts with a set of possible solutions chosen arbitrarily, then selection, replication, recombination, and mutation are applied iteratively. Selection is based on conforming to a fitness function which is determined by a specific problem of interest, and so over time better solutions to the problem can thus evolve \cite{goldberg}. As Digital Ecosystems will likely solve problems by evolving solutions, they will probably incorporate some form of evolutionary computing. However, we suggest that Digital Ecosystems should also incorporate additional features, providing it with a closer resemblance to biological ecosystems. Including features such as complex dynamic fitness functions, a distributed or network environment, and self-organisation arising from interactions among organisms and their environment, such as those that we will now discuss.

Arguably the most fundamental differences between biological and digital ecosystems lie in the motivation and approach of their respective researchers. Biological ecosystems are ubiquitous natural phenomena whose maintenance is crucial to our survival \cite{balmford2002erc}, developing through the process of \emph{ecological succession} \cite{begon96}. In contrast, Digital Ecosystems will be defined here as a technology engineered to serve specific human purposes, developing to solve dynamic problems in parallel with high efficiency.

\subsubsection{Biological Ecosystems}

\tfigure{scale=1.0}{abstractEcosystem}{graffle}{Ecosystem Structure}{(redrawn from \cite{longman}): A stable, self-perpetuating system \getCap{introEcoCap}.}{-5mm}{!h}{1mm}{}

An ecosystem is a natural unit made up of living (biotic) and non-living (abiotic) components, from whose interactions emerge a stable, self-perpetuating system. It is \setCap{made up of one or more communities of organisms, consisting of species in their habitats, with their populations existing in their respective micro-habitats \cite{begon96}. A community is a naturally occurring group of populations from different species that live together, and interact as a self-contained unit in the same habitat. A habitat is a distinct part of the environment \cite{begon96}}{introEcoCap}, for example, a stream. Individual organisms migrate through the ecosystem into different habitats competing with other organisms for limited resources, with a population being the aggregate number of the individuals, of a particular species, inhabiting a specific habitat or micro-habitat \cite{begon96}. A micro-habitat is a subdivision of a habitat that possesses its own unique properties, such as a micro-climate \cite{lawrence1989hsd}. Evolution occurs to all living components of an ecosystem, with the evolutionary pressures varying from one population to the next depending on the environment that is the population's habitat. A population, in its micro-habitat, comes to occupy a niche, which is the functional relationship of a population to the environment that it occupies. A niche results in the highly specialised adaptation of a population to its micro-habitat \cite{lawrence1989hsd}.

\subsubsection{Fitness Landscapes and Agents}

\label{agents}
As described above, an ecosystem comprises both an environment and a set of interacting, reproducing entities (or agents) in that environment; with the environment acting as a set of physical and chemical constraints on reproduction and survival \cite{begon96}. These constraints can be considered in abstract using the metaphor of the \emph{fitness landscape}, in which individuals are represented as solutions to the problem of survival and reproduction \cite{wright1932}. All possible solutions are distributed in a space whose dimensions are the possible properties of individuals. An additional dimension, height, indicates the relative fitness (in terms of survival and reproduction) of each solution. The fitness landscape is envisaged as a rugged, multidimensional landscape of hills, mountains, and valleys, because individuals with certain sets of properties are \emph{fitter} than others \cite{wright1932}, as visualised in Figure \ref{figure1new}. 

\tfigure{width=170mm}{figure1new}{graffle}{Fitness Landscape}{(modified from \cite{fitland}): We can represent software development as a walk through the landscape, towards the peaks which correspond to the optimal applications. Each point represents a unique combination of software services, and the roughness of the landscape indicates how difficult it is to reach an optimal software design \cite{fitland}. In this example, there is a global optimum, and several lower local optima.}{-10mm}{}{}{}

In biological ecosystems, fitness landscapes are virtually impossible to identify. This is both because there are large numbers of possible traits that can influence individual fitness, and because the environment changes over time and space \cite{begon96}. In contrast, within a digital environment, it is normally possible to specify explicitly the constraints that act on individuals in order to evolve solutions that perform better within these constraints. Within genetic algorithms, exact specification of a fitness landscape or function is common practice \cite{goldberg}. However, within a Digital Ecosystem the ideal constraints are those that allow solution populations to evolve to meet user needs with maximum efficiency, with the user needs changing from place to place and time to time. In this sense the fitness landscape of a Digital Ecosystem is complex and dynamic, and more like that of a biological ecosystem than like that of a traditional genetic algorithm \cite{morrison2004dea, goldberg}. The designer of a Digital Ecosystem therefore faces a double challenge: firstly, to specify rules that govern the shape of the fitness function/landscape in a way that meaningfully maps landscape dynamics to user requests, and secondly, to evolve within this space, solution populations that are diverse enough to solve disparate problems, complex enough to meet user needs, and efficient enough to be preferable to those generated by other means.

The agents within a Digital Ecosystem will need to be like biological individuals in the sense that they reproduce, vary, interact, move, and die \cite{begon96}. Each of these properties contributes to the dynamics of the ecosystem. However, the way in which these individual properties are encoded may vary substantially depending on the intended purpose of the system \cite{chambers2001phg}.

\subsubsection{Networks and Spatial Dynamics}

\tfigure{width=170mm}{figure2new}{graffle}{Abstract View of An Ecosystem}{Showing different populations (by the different colours) in different spatial areas, and their connection to one another by the lines. Included are communities of populations that have become geographically separated and so are not connected to the main network of the ecosystem, and which could potentially give rise to allopatric (geographic) speciation \cite{lawrence1989hsd}.}{-12mm}{}{}{}

A key factor in the maintenance of diversity in biological ecosystems is spatial interactions, and several modelling systems have been used to represent these spatial interactions, including metapopulations\footnote{A metapopulation is a collection of relatively isolated, spatially distributed, local populations bound together by occasional dispersal between populations \cite{levins1969sda, hanski1999me, hanski2003mtf}.}, diffusion models, cellular automata and agent-based models (termed individual-based models in ecology) \cite{Greenetal2006}. The broad predictions of these diverse models are in good agreement. At local scales, spatial interactions favour relatively abundant species disproportionately. However, at a wider scale, this effect can preserve diversity, because different species will be locally abundant in different places. The result is that even in homogeneous environments, population distributions tend to form discrete, long-lasting patches that can resist an invasion by superior competitors \cite{Greenetal2006}. Population distributions can also be influenced by environmental variations such as barriers, gradients, and patches. The possible behaviour of spatially distributed ecosystems is so diverse that scenario-specific modelling is necessary to understand any real system \cite{suzie}. Nonetheless, certain robust patterns are observed. These include the relative abundance of species, which consistently follows a roughly log-normal relationship \cite{Bell}, and the relationship between geographic area and the number of species present, which follows a power law \cite{sizling2004pls}. The reasons for these patterns are disputed, because they can be generated by both spatial extensions of simple Lotka-Volterra competition models \cite{Hubbell}, and more complex ecosystem models \cite{Sole}. 

Landscape connectivity plays an important part in ecosystems. When the density of habitats within an environment falls below a critical threshold, widespread species may fragment into isolated populations. Fragmentation can have several consequences. Within populations, these effects include loss of genetic diversity and detrimental inbreeding \cite{GreenKirley}. At a broader scale, isolated populations may diverge genetically, leading to speciation, as shown in Figure \ref{figure2new}. 

From an information theory perspective, this phase change in landscape connectivity can mediate global and local search strategies \cite{Greenetal2000}. In a well-connected landscape, selection favours the globally superior, and pursuit of different evolutionary paths is discouraged, potentially leading to premature convergence. When the landscape is fragmented, populations may diverge, solving the same problems in different ways. Recently, it has been suggested that the evolution of complexity in nature involves repeated landscape phase changes, allowing selection to alternate between local and global search \cite{Greenetalinpress}. 

In a digital context, we can have spatial interactions by using a distributed system that consists of a set of interconnected locations, with agents that can migrate between these connected locations. In such systems the spatial dynamics are relatively simple compared with those seen in real ecosystems, which incorporate barriers, gradients, and patchy environments at multiple scales in continuous space \cite{begon96}. Nevertheless, depending on how the connections between locations are organised, such Digital Ecosystems might have dynamics closely parallel to spatially explicit models, diffusion models, or metapopulations \cite{suzie}. We will discuss later the use of a dynamic non-geometric spatial network, and the reasons for using this approach.

\subsubsection{Selection and Self-Organisation}

The major hypothetical advantage of Digital Ecosystems over other complex organisational models is their potential for dynamic adaptive self-organisation. However, for the solutions evolving in Digital Ecosystems to be useful, they must not only be efficient in a computational sense, but they must also solve purposeful problems. That is, the fitness of agents must translate in some sense to real-world usefulness as demanded by the users \cite{ducheyne2003fiu}.

\tfigure{width=170mm}{figure3new}{graffle}{Evolving Population of Digital Organisms}{A virtual petri dish at three successive time-steps, showing the self-organisation of the population undergoing selection. The colour shows the genetic variability of the digital organisms. Over time the fitter (purple) organisms come to dominate the population, reproducing more and essentially replacing the weaker organisms of the population \cite{petri}.}{-8mm}{}{}{}

Constructing a useful Digital Ecosystem therefore requires a balance between freedom of the system to self-organise, and constraint of the system to generate useful solutions. These factors must be balanced because the more the system's behaviour is dictated by its internal dynamics, the less it may respond to fitness criteria imposed by the users. At one extreme, when system dynamics are mainly internal, agents may evolve that are good at survival and reproduction within the digital environment, but useless in the real world \cite{ducheyne2003fiu}. At the other extreme, where the users' fitness criteria overwhelmingly dictates function, we suggest that dynamic exploration, of the solution space and complexity, is likely to be limited. The reasoning behind this argument is as follows. Consider a multidimensional solution space which maps to a rugged fitness landscape \cite{wright1932}. In this landscape, competing solution lineages will gradually become extinct through chance processes. So, the solution space explored becomes smaller over time as the population adapts and the diversity of solutions decreases. Ultimately, all solutions may be confined to a small region of the solution space. In a static fitness landscape, this situation is desirable because the surviving solution lineages will usually be clustered around an optimum \cite{goldberg}. However, if the fitness landscape is dynamic, the location of optima varies over time, and should lineages become confined to a small area of the solution space, then subsequent selection will locate only optima that are near this area \cite{morrison2004dea}. This is undesirable if new, higher optima arise that are far from pre-existing ones. A related issue is that complex solutions are less likely to be found by chance than simple ones. Complex solutions can be visualised as sharp, isolated peaks on the fitness landscape. Especially for dynamic landscapes, these peaks are most likely to be found when the system explores the solution space widely \cite{morrison2004dea}. Therefore, a self-organising mechanism other than the fitness criteria of users is required to maintain diversity among competing solutions in a Digital Ecosystem.

\newcommand{\vspacesdc}{\vspace{3mm}}

\vspacesdc

\subsubsection{Stability and Diversity in Complex Adaptive Systems}

\vspacesdc

Ecosystems are often described as \ac{CAS}, because like them, they are systems made from diverse, locally interacting components that are subject to selection. Other \ac{CAS} include brains, individuals, economies, and the biosphere. All are characterised by hierarchical organisation, continual adaptation and novelty, and non-equilibrium dynamics. These properties lead to behaviour that is non-linear, historically contingent, subject to thresholds, and contains multiple basins of attraction \cite{Levin}. 

\vspacesdc

In the previous subsections, we have advocated Digital Ecosystems that include agent populations evolving by natural selection in distributed environments. Like real ecosystems, digital systems designed in this way fit the definition of \ac{CAS}. The features of these systems, especially non-linearity and non-equilibrium dynamics, offer both advantages and hazards for adaptive problem-solving. The major hazard is that the dynamics of \ac{CAS} are intrinsically hard to predict because of the non-linear emergent self-organisation \cite{levin1999fdc}. This observation implies that designing a useful Digital Ecosystem will be partly a matter of trial and error. The occurrence of multiple basins of attraction in \ac{CAS} suggests that even a system that functions well for a long period may suddenly at some point transition to a less desirable state \cite{folke}. For example, in some types of system self-organising mass extinctions might result from interactions among populations, leading to temporary unavailability of diverse solutions \cite{newman1997mme}. This concern may be addressed by incorporating negative feedback or other mechanisms at the global scale. The challenges in designing an effective Digital Ecosystem are mirrored by the system's potential strengths. Non-linear behaviour provides the opportunity for scalable organisation and the evolution of complex hierarchical solutions, while rapid state transitions potentially allow the system to adapt to sudden environmental changes with minimal loss of functionality \cite{Levin}. 

\tfigure{width=170mm}{figure4final}{graffle}{Ecosystems as \acl{CAS}}{(modified from \cite{CASright}): (LEFT) An abstract view of an ecosystem showing the diversity of different populations by the different colours and spacing. (RIGHT) An abstract view of diversity within a population, with the space between points showing genetic diversity and the clustering prevalent.}{-8mm}{!h}{3mm}{}

A key question for designers of Digital Ecosystems is how the stability and diversity properties of biological ecosystems map to performance measures in digital systems. For a Digital Ecosystem the ultimate performance measure is user satisfaction, a system-specific property. However, assuming the motivation for engineering a Digital Ecosystem is the development of scalable, adaptive solutions to complex dynamic problems, certain generalisations can be made. Sustained diversity \cite{folke}, is a key requirement for dynamic adaptation. In Digital Ecosystems, diversity must be balanced against adaptive efficiency because maintaining large numbers of poorly-adapted solutions is costly. The exact form of this trade-off will be guided by the specific requirements of the system in question. Stability \cite{Levin}, is likewise, a trade-off: we want the system to respond to environmental change with rapid adaptation, but not to be so responsive that mass extinctions deplete diversity or sudden state changes prevent control.

\subsection{Computing of Digital Ecosystems}
\label{compOfDE}

Based on the understanding of biological ecosystems, from the theoretical biology of the previous subsection, we will now introduce fields from the domain of computer science relevant in the creation of Digital Ecosystems. As we are interested in the digital counterparts for the behaviour and constructs of biological ecosystems, instead of simulating or emulating such behaviour or constructs, we will consider what parallels can be drawn.

The value of creating parallels between biological and computer systems varies substantially depending on the behaviours or constructs being compared, and sometimes cannot be done so convincingly. For example, both have mechanisms to ensure data integrity. In computer systems, that integrity is absolute, data replication which introduces even the most minor change is considered to have failed, and is supported by mechanisms such as the \acl{MD5} \cite{rivest1992rmm}. While in biological systems, the genetic code is transcribed with a remarkable degree of fidelity; there is, approximately, only one unforced error per one hundred bases copied \cite{Kunkel2004}. There are also elaborate proof-reading and correction systems, which in evolutionary terms are highly conserved \cite{Kunkel2004}. In this example establishing a parallel is infeasible, despite the relative similarity in function, because the operational control mechanisms in biological and computing systems are radically different, as are the aims and purposes. This is a reminder that considerable finesse is required when determining parallels, or when using existing ones.

We will start by considering \aclp{MAS} to explore the references to \emph{agents} and \emph{migration}; followed by evolutionary computing and \aclp{SOA} for the references to \emph{evolution} and \emph{self-organisation}.

\subsubsection{Multi-Agent Systems}

A \emph{software agent} is a piece of software that acts, for a user in a relationship of \emph{agency}, autonomously in an environment to meet its designed objectives \cite{wooldridge}. A \ac{MAS} is a system composed of several \emph{software agents}, collectively capable of reaching goals that are difficult to achieve by an individual agent or monolithic system \cite{wooldridge}. Conceptually, there is a strong parallel between the software agents of a \ac{MAS} and the agent-based models of a biological ecosystem \cite{Greenetal2006}, despite the lack of evolution and migration in a \ac{MAS}. There is an even stronger parallel to a variant of \acp{MAS}, called \emph{mobile agent systems}, in which the mobility also mirrors the migration in biological ecosystems \cite{moaspaper}.

\tfigure{width=170mm}{mobileAgents}{graffle}{Mobile Agent System}{Visualisation that shows mobile agents as programmes that can migrate from one host to another in a network of heterogeneous computer systems and perform a task specified by its owner. On each host they visit, mobile agents need special software called an agent station, which is responsible for executing the agents and providing a safe execution environment \cite{agentStation}.}{-10mm}{}{}{}

The term \emph{mobile agent} contains two separate and distinct concepts: mobility and agency \cite{rothermel1998ma}. Hence, mobile agents are software agents capable of movement within a network \cite{moaspaper}. The mobile agent paradigm proposes to treat a network as multiple agent-\emph{friendly} environments and the agents as programmatic entities that move from location to location, performing tasks for users. So, on each host they visit mobile agents need software which is responsible for their execution, providing a safe execution environment \cite{moaspaper}.

Generally, there are three types of design for mobile agent systems \cite{moaspaper}: (1) using a specialised operating system, (2) as operating system services or extensions, or (3) as application software. The first approach has the operating system providing the requirements of mobile agent systems directly \cite{svahnberg}. The second approach implements the mobile agent system requirements as operating system extensions, taking advantage of existing features of the operating system \cite{johansen1995oss}. Lastly, the third approach builds mobile agent systems as specialised application software that runs on top of an operating system, to provide for the mobile agent functionality, with such software being called an \emph{agent station} \cite{agentStation}. In this last approach, each agent station hides the vendor-specific aspects of its host platform, and offers standardised services to visiting agents. Services include access to local resources and applications; for example, web servers or \emph{web services}, the local exchange of information between agents via message passing, basic security services, and the creation of new agents \cite{agentStation}. Also, the third approach is the most platform-agnostic, and is visualised in Figure \ref{mobileAgents}.

\subsubsection{Evolutionary Computing}

For evolving software in Digital Ecosystems evolutionary computing is the logical field from which to start. In \acl{BIC}, one of the primary sources of inspiration from nature has been evolution \cite{ecpaper}. Evolution has been clearly identified as the source of many diverse and creative solutions to problems in nature \cite{ec15, ec16}. However, it can also be useful as a problem-solving tool in artificial systems. Computer scientists and other theoreticians realised that the selection and mutation mechanisms that appear so effective in biological evolution could be abstracted to be implemented in a computational algorithm \cite{ecpaper}. Evolutionary computing is now recognised as a sub-field of artificial intelligence (more particularly computational intelligence) that involves combinatorial optimisation problems \cite{ec17}.

Evolutionary algorithms are based upon several fundamental principles from biological evolution, including reproduction, mutation, recombination (crossover), natural selection, and survival of the fittest. As in biological systems, evolution occurs by the repeated application of the above operators \cite{back1996eat}. An evolutionary algorithm operates on a set of individuals, called a population. An \emph{individual}, in the natural world, is an organism with an associated fitness \cite{lawrence1989hsd}. Candidate solutions to an optimisation problem play the role of individuals in a population, and a cost function determines the environment within which the solutions \emph{live}, analogous to the way the environment selects for the fittest individuals. Candidate solutions to an optimisation problem play the role of individuals in a population, and a cost function determines the environment by selecting for the fittest individuals. The number of individuals varies between different implementations and may also vary through time during the use of the algorithm. Each individual possesses some characteristics that are defined through its genotype, its genetic composition. These characteristics may be passed on to descendants of that individual \cite{back1996eat}. Processes of mutation (small random changes) and crossover (generation of a new genotype by the combination of components from two individuals) may occur, resulting in new individuals with genotypes different from the ancestors they will come to replace. These processes iterate, modifying the characteristics of the population \cite{back1996eat}. Which members of the population are kept, or are used as parents for offspring, will often depend upon some external characteristic, called the fitness (cost) function of the population. It is this that enables improvement to occur \cite{back1996eat}, and corresponds to the fitness of an organism in the natural world \cite{lawrence1989hsd}. Recombination and mutation create the necessary diversity and thereby facilitate novelty, while selection acts as a force increasing quality. Changed pieces of information resulting from recombination and mutation are randomly chosen. However, selection operators can be either deterministic, or stochastic. In the latter case, individuals with a higher fitness have a higher chance to be selected than individuals with a lower fitness \cite{back1996eat}.

There are different strands of what has become called evolutionary computing \cite{back1996eat}. The first is genetic algorithms. A second strand, evolution strategies, focuses strongly on engineering applications. A third strand, evolutionary programming, originally developed from machine intelligence motivations, and is related to the other two. These areas developed separately for about fifteen years, but from the early nineties they are seen as different representatives (\emph{dialects}) of one technology, called evolutionary computing \cite{eiben2003iec}. In the early nineties, another fourth stream following the general ideas had emerged, called genetic programming \cite{eiben2003iec}.

Genetic algorithms \cite{goldberg} implement a population of individuals, each of which possesses a genotype that encodes a candidate solution to a problem. Typically genotypes are encoded as bit-strings, but other encodings have been used in more recent developments of genetic algorithms. Mutation and crossover, along with selection, are then used to choose a solution to a problem. They have proven to be widely applicable, and have resulted in many applications in differing domains \cite{ec22}. Evolutionary strategies arose out of an attempt by several civil engineers to understand a problem in hydrodynamics \cite{ec24}. Evolutionary strategies \cite{ec19} differ from genetic algorithms in operating on real-valued parameters, and historically they have tended not to use crossover as a variational operator, only mutation. However, mutation rates have themselves been allowed to adapt in evolutionary strategies, which is not often the case with genetic algorithms. Evolutionary strategies have also been used for many \linebreak applications \cite{ebeling1990aes}. 

Evolutionary programming arose distinctly from the first two strands of evolutionary computation, out of an attempt to understand machine intelligence through the evolution of finite state machines \cite{fogel1966ait}. Evolutionary programming \cite{ec20} emphasises the evolution of the phenotype (instance of a solution) instead of the genotype (genetic material) of individuals, and the relation between the phenotype of parents and offspring, although crossover is not used. Thus, evolutionary programming has some differences in approach from the other major strands of evolutionary computation research. However, there have been many overlaps between the different fields and it too has been applied in many areas \cite{ec20}.

Genetic programming \cite{ec25} can be considered as a variant of genetic algorithms where individual genotypes are represented by executable programmes. Specifically, solutions are represented as trees of expressions in an appropriate programming language, with the aim of evolving the most effective programme for solving a particular problem. Genetic programming, although the newest form of evolutionary computing, has still proved to be widely applicable \cite{banzhaf1998gpi}.

Many important questions remain to be answered in understanding the performance of evolutionary algorithms. For example, current evolutionary algorithms for evolving programmes (genetic programming) suffer from some weaknesses. First, while being moderately successful at evolving simple programmes, it is very difficult to scale them to evolve high-level software components \cite{mantere2005ese}. Second, the \emph{estimated} fitness of a programme is normally given by a measure of how accurately it computes a given function, as represented by a set of input and output pairs, and therefore there is only a limited guarantee that the evolved programme actually does the intended computation \cite{mantere2005ese}. These issues are particularly important when evolving high-level, complex, structured software.

To evolve high-level software components in Digital Ecosystems, we propose taking advantage of the \emph{native} method of software advancement, human developers, and the use of \emph{evolutionary computing} for \emph{combinatorial optimisation} \cite{papadimitriou1998coa} of the available software services. This involves treating developer-produced software services as the functional building blocks, as the base unit in a genetic-algorithms-based process. Such an approach would require a modular reusable paradigm to software development, such as \aclp{SOA}, which are discussed in the following subsection.

\subsubsection{Service-Oriented Architectures}

Our approach to evolving high-level software applications requires a modular reusable paradigm to software development. \acp{SOA} are the current state-of-the-art approach, being the current iteration of interface/component-based design from the 1990s, which was itself an iteration of event-oriented design from the 1980s, and before then modular programming from the 1970s \cite{histOfProg, krafzig2004ess}. Service-oriented computing promotes assembling application components into a loosely coupled network of services, to create flexible, dynamic business processes and agile applications that span organisations and computing platforms \cite{papazoglou2003soc}. This is achieved through a \ac{SOA}, an architectural style that guides all aspects of creating and using business processes throughout their life-cycle, packaged as services. This includes defining and provisioning infrastructure that allows different applications to exchange data and participate in business processes, loosely coupled from the operating systems and programming languages underlying the applications \cite{soa1w}. Hence, a \ac{SOA} represents a model in which functionality is decomposed into distinct units (services), which can be distributed over a network, and can be combined and reused to create business applications \cite{papazoglou2003soc}.

A \acs{SOA} depends upon service-orientation as its fundamental design principle. In a \acs{SOA} environment, independent services can be accessed without knowledge of their underlying platform implementation \cite{soa1w}. Services reflect a \emph{service-oriented} approach to programming that is based on composing applications by discovering and invoking network-available services to accomplish some task. This approach is independent of specific programming languages or operating systems, because the services communicate with each other by passing data from one service to another, or by co-ordinating an activity between two or more services \cite{papazoglou2003soc}. So, the concepts of \acsp{SOA} are often seen as built upon, and the development of, the concepts of modular programming and distributed computing \cite{krafzig2004ess}.

\acsp{SOA} allow for an information system architecture that enables the creation of applications that are built by combining loosely coupled and interoperable services \cite{soa1w}. They typically implement functionality most people would recognise as a service, such as filling out an online application for an account, or viewing an online bank statement \cite{krafzig2004ess}. Services are intrinsically unassociated units of functionality, without calls to each other embedded in them. Instead of services embedding calls to each other in their source code, protocols are defined which describe how services can talk to each other, in a process known as orchestration, to meet new or existing business system requirements \cite{singh2005soc}. This is allowing an increasing number of third-party software companies to offer software services, such that \acs{SOA} systems will come to consist of such third-party services combined with others created in-house, which has the potential to spread costs over many users and uses, and promote standardisation both in and across industries \cite{chhatpar2008}. For example, the travel industry now has a well-defined, and documented, set of both services and data, sufficient to allow any competent software engineer to create travel agency software using entirely off-the-shelf software services \cite{kotok2001eng, cardoso2005isw}. Other industries, such as the finance industry, are also making significant progress in this direction \cite{zimmermann2004sgw}. 

The vision of \acsp{SOA} assembling application components from a loosely coupled network of services, that can create dynamic business processes and agile applications that span organisations and computing platforms, is visualised in Figure \ref{SOAvecFinal}. It will be made possible by creating compound solutions that use internal organisational software assets, including enterprise information and legacy systems, and combining these solutions with external components residing in remote networks \cite{SOApaper0}. The great promise of \acsp{SOA} is that the \emph{marginal cost} of creating the n-th application is virtually zero, as all the software required already exists to satisfy the requirements of other applications. Only their \emph{combination} and \emph{orchestration} are required to produce a new application \cite{tang2004ews, modi2008}. The \emph{key} is that the interactions between the \emph{chunks} are not specified within the \emph{chunks} themselves. Instead, the interaction of services (all of whom are hosted by unassociated peers) is specified by users in an ad-hoc way, with the intent driven by newly emergent business requirements \cite{leymann2002wsa}. 

\tfigure{width=170mm}{SOAvecFinal}{graffle}{Service-Oriented Architectures}{Abstract visualisations, with the first image showing the loosely joined services as cuboids, and the service orchestration as a polyhedron; and the second image showing their high interoperability and re-usability in forming applications, from the use of standardised interfaces and external service orchestration.}{-8mm}{}{}{}

The pinnacle of \acs{SOA} interoperability, is the exposing of services on the internet as \emph{web services} \cite{soa1w}. A web service is a specific type of service that is identified by a \ac{URI}, whose service description and transport utilise open Internet standards. Interactions between web services typically occur as \ac{SOAP} calls carrying \ac{XML} data content. The interface descriptions of web services are expressed using the \ac{WSDL} \cite{SOApaper2}. The \ac{UDDI} standard defines a protocol for directory services that contain web service descriptions. \ac{UDDI} enables web service clients to locate candidate services and discover their details. Service clients and service providers utilise these standards to perform the basic operations of \acsp{SOA} \cite{SOApaper2}. Service aggregators can then use the \ac{BPEL} to create new web services by defining corresponding compositions of the interfaces and internal processes of existing services \cite{SOApaper2}.

\acs{SOA} services inter-operate based on a formal definition (or contract, e.g. \ac{WSDL}) that is independent of the underlying platform and programming language. Service descriptions are used to advertise the service capabilities, interface, behaviour, and quality \cite{SOApaper2}. The publication of such information about available services provides the necessary means for discovery, selection, binding, and composition of services \cite{SOApaper2}. The (expected) behaviour of a service during its execution is described by its behavioural description (for example, as a workflow process). Also, included is a \ac{QoS} description, which publishes important functional and non-functional service quality attributes, such as service metering and cost, performance metrics (response time, for instance), security attributes, integrity (transactional), reliability, scalability, and availability \cite{SOApaper2}. Service clients (end-user organisations that use some service) and service aggregators (organisations that consolidate multiple services into a new, single service offering) utilise \emph{service descriptions} to achieve their objectives \cite{SOApaper2}. One of the most important and continuing developments in \acsp{SOA} is \acf{SWS}, which make use of \emph{semantic descriptions} for service discovery, so that a client can discover the services semantically \cite{SOAsemantic, cabral2004asw}. 

There are multiple standards available and still being developed for \acsp{SOA} \cite{SOAstandards}, most notably of recent being \ac{REST} \cite{singh2005soc}. The software industry now widely implements a thin SOAP/WSDL/UDDI veneer atop existing applications or components that implement the web services paradigm \cite{SOApaper0}, but the choice of technologies will change with time. Therefore, the fundamentals of \acsp{SOA} and its services are best defined generically, because \acsp{SOA} are technology agnostic and need not be tied to a specific technology \cite{papazoglou2003soc}. Within the current and future scope of the fundamentals of \acsp{SOA}, there is clearly potential to \emph{evolve} complex high-level software applications from the modular services of \acsp{SOA}, instead of the instruction level evolution currently prevalent in genetic programming \cite{overviewGP}.

\subsubsection{Distributed Evolutionary Computing}

Having previously introduced evolutionary computing, and the possibility of it occurring within a distributed environment, not unlike those found in mobile agent systems, leads us to consider a specialised form known as \ac{DEC}. The motivation for using parallel or distributed evolutionary algorithms is twofold: first, improving the speed of evolutionary processes by conducting concurrent evaluations of individuals in a population; second, improving the problem-solving process by overcoming difficulties that face traditional evolutionary algorithms, such as maintaining diversity to avoid premature convergence \cite{muhlenbein1991eta, stender1993pga}. The fact that evolutionary computing manipulates a population of independent solutions actually makes it well suited for parallel and distributed computation architectures \cite{cantupaz1998spg}. There are several variants of distributed evolutionary computing, leading some to propose a taxonomy for their classification \cite{nowostawski1999pga}, with there being two main forms \cite{cantupaz1998spg, stender1993pga}:

\vspace{-5mm}
\narrowlinespacing
\begin{itemize}
\item multiple-population/coarse-grained migration/island models
\item single-population/fine-grained diffusion/neighbourhood models
\end{itemize}
\normallinespacing

In the coarse-grained \emph{island} models \cite{lin1994cgp, cantupaz1998spg}, evolution occurs in multiple parallel sub-populations (islands), each running a local evolutionary algorithm, evolving independently with occasional \emph{migrations} of highly fit individuals among sub-populations. The core parameters for the evolutionary algorithm of the island-models are as follows \cite{lin1994cgp}:

\vspace{-5mm}
\narrowlinespacing
\begin{itemize}
\item number of the sub-populations: 2, 3, 4, more
\item sub-population homogeneity
\begin{itemize}
\item size, crossover rate, mutation rate, migration interval
\end{itemize}
\item topology of connectivity: ring, star, fully-connected, random
\item static or dynamic connectivity
\item migration mechanisms: 
\begin{itemize}
\item isolated/synchronous/asynchronous
\item how often migrations occur 
\item which individuals migrate 
\end{itemize}
\end{itemize}
\normallinespacing

Fine-grained \emph{diffusion} models \cite{manderick1989fgp, stender1993pga} assign one individual per processor. A local neighbourhood topology is assumed, and individuals are allowed to mate only within their neighbourhood, called a \emph{deme}\footnote{In biology a deme is a term for a local population of organisms of one species that actively interbreed with one another and share a distinct gene pool \cite{devisser2007ese}.}. The demes overlap by an amount that depends on their shape and size, and in this way create an implicit migration mechanism. Each processor runs an identical evolutionary algorithm which selects parents from the local neighbourhood, produces an offspring, and decides whether to replace the current individual with an offspring. However, even with the advent of multi-processor computers, and more recently multi-core processors, which provide the ability to execute multiple threads simultaneously \cite{newsArticle3}, this approach would still prove impractical in supporting the number of agents necessary to create a Digital Ecosystem. Therefore, we shall further consider the \emph{island} models.

An example island-model \cite{lin1994cgp, cantupaz1998spg} is visualised in Figure \ref{islandModel}, in which there \setCap{are different probabilities of going from island \Circ{1} to island \Circ{2}, as there is of going from island \Circ{2} to island \Circ{1}.}{im1} This allows maximum flexibility for the migration process, and \setCap{mirrors the naturally inspired quality that although two populations have the same physical separation, it may be easier to migrate in one direction than the other, i.e. fish migration is easier downstream than upstream.}{im2} The migration of the \emph{island} models is like the notion of migration in nature, being similar to the metapopulation models of theoretical ecology \cite{levins1969sda}. This model has also been used successfully in the determination of investment strategies in the commercial sector, in a product known as the Galapagos toolkit \cite{galapagos1, galapagos2}. However, all the \emph{islands} in this approach work on exactly the same problem, which makes it less analogous to biological ecosystems in which different locations can be environmentally different \cite{begon96}. We will take advantage of this property later when defining the \acl{EOA} of Digital Ecosystems.

\tfigure{scale=1.0}{islandModel}{graffle}{Island-Model of Distributed Evolutionary Computing}{\cite{lin1994cgp, cantupaz1998spg}: There \getCap{im1} This \getCap{im2}}{-6mm}{}{}{}

\subsection{Digital Business Ecosystems}

The questions we have raised are wide-ranging, and are motivating several interdisciplinary research teams, including those involved in an EU Framework VI project called \acp{DBE}. The \ac{DBE} is a proposed methodology for economic and technological innovation. Specifically, the \ac{DBE} is a software infrastructure for supporting large numbers of interacting business users and services \cite{dbebkintro}. The \ac{DBE} aims to be a next generation \acl{ICT} that will extend the \acl{SOA} concept with the automatic combining of available and applicable services in a scalable architecture, to meet business user requests for applications that facilitate business processes. In essence, the \ac{DBE} will be an internet-based environment in which businesses will be able to interact with each other in very effective and efficient ways \cite{dbebook}. 

The synthesis of the concept of Digital Business Ecosystems emerged by adding \cite{nachira} \emph{digital} in front of \emph{business ecosystem} \cite{moore1996}. The term Digital Business Ecosystem was used earlier, but with a focus exclusively on developing countries \cite{moore2003}. The generalisation of the term to refer to a new interpretation of what \emph{socio-economic development catalysed by \ac{ICT}} means was new, emphasising the co-evolution between the \emph{business ecosystem} and its partial digital representation: the digital ecosystem. The term Digital Business Ecosystem came to represent the combination of the two ecosystems \cite{dbebkintro}.

\tfigure{width=170mm}{DBEprojectFinal}{graffle}{Business Ecosystem}{\cite{dbebkintro}: Conceptual visualisation \cite{dbevis} showing a Business Ecosystem of interacting \acl{SME} users, via the services they provide and consume. Creating a network of business ecosystems distributed over different geographical regions, business domains, and industry sectors.}{-10mm}{}{}{}

The \emph{business ecosystem} is an economic community supported by a foundation of interacting organisations and individuals; i.e. \emph{the organisms of the business world}. This economic community produces goods and services of value to customers, who are themselves members of the ecosystem \cite{moore1996}. \setCap{A wealthy ecosystem sees a balance between co-operation and competition in a dynamic free market.}{DBEdescription} Regarding a particular \emph{business ecosystem}, two main different interpretations of its structure have been discussed in the literature. The \emph{keystone} model has a structure in which a \emph{business ecosystem} is dominated by a large firm that is surrounded by many small suppliers \cite{iansiti2004kan}. This model works well when the central firm is healthy, but represents a significant weakness for the economy of the region when the dominant economic actor experiences difficulties \cite{moore1996}. This model also matches the economic structure of the USA where there is a predominant number of large enterprises at the centre of large value networks of suppliers \cite{iansiti2004kan}. However, the model for a \emph{business ecosystem} developed in Europe is less structured and more dynamic; it is composed mainly of \acp{SME}, but can accommodate large firms \cite{eurostat2006}. All actors complement one another, leading to a more dynamic division of labour, organised along one-dimensional value chains and two-dimensional value networks \cite{corallo2007}. This model is particularly well-adapted for the service and \emph{knowledge industries}, where it is easier for small firms to reinvent themselves than, for instance, in the \emph{automotive industry} which is dominated by large enterprises \cite{dbebkintro}.

In the DBE, the \emph{digital ecosystem} is the technical infrastructure, based on a \ac{P2P} distributed software technology that transports, finds, and connects services and information over Internet links enabling networked transactions, and the distribution of all the digital \emph{objects} present within the infrastructure \cite{dbebkintro}. Such \emph{organisms of the digital world} encompass any useful digital representations expressed by languages (formal or natural) that can be interpreted and processed (by computer software and/or humans), e.g. software applications, services, knowledge, taxonomies, folksonomies, ontologies, descriptions of skills, reputation and trust relationships, training modules, contractual frameworks, laws \cite{dbebkintro}. So, the Digital Business Ecosystem is a biological metaphor that highlights the interdependence of all actors in the business environment, who \emph{co-evolve their capabilities and roles} \cite{moore1996}, and which has attempted to develop an isomorphic model between biological behaviour and the behaviour of the digital ecosystem, leading to an evolutionary, self-organising, and self-optimising environment built upon an underlying \acl{SOA} \cite{dbebook}. 

The \ac{DBE} aims to help local economic actors become active players in globalisation \cite{dini2008bid}, valorising their local culture and vocations, enabling them to interact and create value networks at the global level. Increasingly this approach, dubbed \emph{glocalisation}, is being considered a successful strategy of globalisation that preserves regional growth and identity \cite{robertson1994gog, swyngedouw1992mqg, khondker2004}, and has been embraced by the mayors and decision-makers of thousands of municipalities \cite{glocalforum2004}, because of the possible tension between globalisation and localisation when adopting \acp{ICT} \cite{castells2000}. 

The \ac{DBE} represents a \ac{B2B} interaction concept supported by a software platform (digital ecosystem) that is intended to have the desirable properties of biological ecosystems \cite{dbebokpaolo}, and its researchers also recognise the importance of \aclp{SOA} in creating Digital Ecosystems \cite{razavi2007cmd, dbebook}. So, we will consider using the concept of a \emph{business ecosystem} as a potential user base for Digital Ecosystems.

\section{Ecosystem-Oriented Architectures}

We will now define the architectural principles of Digital Ecosystems. We will use our understanding of theoretical biology from section \ref{bioOfDE}, mimicking the processes and structures of life, evolution, and ecology of biological ecosystems. We will achieve this by combining elements from mobile agents systems, \acl{DEC}, and \aclp{SOA} from section \ref{compOfDE}, to create a hybrid architecture which is the \emph{digital counterpart of biological ecosystems}.

We will refer to the agents of Digital Ecosystems as \emph{Agents}, populations as \emph{Populations}, and the habitats as \emph{Habitats}, to distinguish their new hybrid definitions from their original biological and computing definitions.

\subsection{Agents}

The Agents of the Digital Ecosystem are functionally analogous to the organisms of biological ecosystems, including the behaviour of migration and the ability to be evolved \cite{begon96}, and will be achieved through using a hybrid of different technologies. The ability to migrate is provided by using the paradigm of \emph{agent mobility} from mobile agent systems \cite{moaspaper}, with the Habitats of the Digital Ecosystem provided by the facilities of \emph{agent stations} from mobile agent systems \cite{agentStation}, i.e. a distributed network of locations to migrate to and from. The Habitats, and the Habitat network will be discussed later. The ability of the Agents to be evolved is in two parts: first, by using the interoperability of services from \aclp{SOA} \cite{soa1w} to aggregate Agents; and second, the use of \emph{evolutionary computing} \cite{eiben2003iec} for \emph{combinatorial optimisation} \cite{papadimitriou1998coa} at the Habitats to evolve optimal aggregations of Agents. The Agents will take advantage of the interoperability of \aclp{SOA} \cite{soa1w}, by \emph{acting in a relationship of agency} \cite{wooldridge} to the user supplied semantic web services, which will be \acl{SOA} compliant \cite{SOApaper2}. We can then evolve high-level software applications by using \emph{evolutionary computing} \cite{eiben2003iec} for \emph{combinatorial optimisation} \cite{papadimitriou1998coa} of the available Agents, or rather the services they represent, in a genetic-algorithms-based \cite{goldberg} process. This makes an Agent, of the Digital Ecosystem, a \setCap{lightweight entity consisting primarily of a pointer to the semantic web service it represents,}{serviceCap} including the service's \setCap{\emph{executable component} and \emph{semantic description}. A software service can be a software service only, e.g. for data encryption, or a software service providing a front-end to a real-world service, e.g. selling books}{service2cap}, as shown in Figure \ref{service}.

\tfigure{width=170mm}{service}{graffle}{Agent of the Digital Ecosystem}{A \getCap{serviceCap} which is \aclp{SOA} compliant and therefore includes an \getCap{service2cap}.}{-10mm}{}{}{}

An organism within Digital Ecosystems is an Agent, or an Agent aggregation created using evolutionary optimisation in response to a user request for an application. These Agents will migrate through the Habitat network of the Digital Ecosystem and adapt to find \emph{niches} where they are useful in fulfilling other user requests for applications. The Agents interact, evolve, and adapt over time to the environment, thereby serving the ever-changing requirements imposed by the user base.

The \emph{executable component}, of a semantic web service that an Agent represents, is equivalent to the \acs{DNA} of an organism, whose sequence encodes the genetic information of living organisms and has two primary functions \cite{lawrence1989hsd}: the holder of virtually all information in inheritance, and the controller of protein synthesis for the construction and operation of its organism. Equivalently, the \emph{executable component} is also the inheritable component from one generation to the next, and defines the objects and behaviour of its service's \emph{run-time instantiation}.

The \emph{genotype} of an individual \emph{describes} the genetic constitution (\acs{DNA}) of an individual, independent of its physical existence (the phenotype) \cite{lawrence1989hsd}. Equivalently, the \emph{semantic description}, of a semantic web service that an Agent represents, describes the functionality of the \emph{executable component}. The \emph{phenotype} of an individual arises from the combination of an organism's \acs{DNA} and the environment \cite{lawrence1989hsd}. Equivalently, the \emph{run-time instantiation}, of a service that an Agent represents, results from instantiating the \emph{executable component} in the \emph{run-time environment}. This differentiation between genotype and phenotype is fundamental for escaping local optima, and is often lacking in artificial evolutionary systems \cite{shackleton2000irg}, having instead a \emph{one-to-one genotype-phenotype mapping}, in which the phenotype is directly encoded in the genotype with no differentiation provided by instantiation (development) \cite{shackleton2000irg}. \emph{Neutral genotype-phenotype mappings} have this differentiation between the genotype and phenotype \cite{ec39}, which more strongly parallels biological evolution \cite{banzhaf1994gpm}. We therefore expect the use of a \emph{neutral genotype-phenotype mapping} to help Digital Ecosystems demonstrate behaviour more akin to biological ecosystems.

\subsubsection{Agent Aggregation}

\setCap{The \emph{executable component} of a semantic web service that an Agent represents is equivalent to an organism's \acs{DNA} and is the gene (functional unit) in the evolutionary process \cite{lawrence1989hsd}. So, the Agents should be aggregated as a \emph{sequence}, like the sequencing of genes in \acs{DNA} \cite{lawrence1989hsd}.}{structureCap} It could be argued that the Agents should be aggregated as \setCap{an unordered \emph{set}, or, based on service orchestration, into a \emph{tree} or \emph{workflow}}{structure2}, as shown in Figure \ref{agentStructures}. However, the aggregated structure of the Agents should not be the \emph{orchestration} structure of the collection of software services that the Agents represent, not only because the service orchestration of the \emph{run-time instantiation} is application domain-specific (e.g. \emph{trees} in supply chain management \cite{lambert2000isc}, \emph{workflows} in the travel industry \cite{benatallah2005frd}), but because it would also move it undesirably towards a \emph{one-to-one genotype-phenotype mapping} \cite{shackleton2000irg}.

\tfigure{width=170mm}{agentStructures}{graffle}{Structure of Aggregated Agents}{\getCap{structureCap} Instead of \getCap{structure2}.}{-10mm}{}{}{10mm}

\subsection{Habitats}

The Habitats are the nodes of the Digital Ecosystem, and are functionally analogous to the habitats of a biological ecosystem \cite{lawrence1989hsd}. Their functionality is provided by using \setCap{the \emph{agent stations} from mobile agent systems \cite{agentStation} (to provide a distributed environment in which Agent migration can occur), with \emph{evolutionary computing} \cite{eiben2003iec} for the Agent interaction (instead of traditional agent interaction mechanisms \cite{wooldridge}), and the \emph{island-model} of \acl{DEC} \cite{lin1994cgp} for the connectivity between Habitats.}{habnet} There will be a Habitat for each user, which the users will typically run locally, and through which they will submit requests for applications. Supporting this functionality, Habitats have the following core functions:

\vspace{-5mm}
\begin{itemize}
\item Provide a subset of the Agents and Agent-sequences available globally, relevant to the user that the Habitat represents, and stored in what we will call an Agent-pool (for reasons that will be explained later). 
\item Accelerate, via the Agent-pool, the Populations instantiated to evolve optimal Agent-sequences in response to user requests for applications.
\item Manage the inter-Habitat connections for Agent migration.
\item For service providers; manage the distribution of Agents (which represent their services) to other users of the Digital Ecosystem, via the network of interconnected Habitats.
\end{itemize}

The collection of Agents at each Habitat (peer) will change over time, as the more successful Agents spread throughout the Digital Ecosystem, and as the less successful Agents are deleted. Successive user requests over time to their dedicated Habitats makes this process possible, because the continuous and varying user requests for applications provide a dynamic evolutionary pressure on the Agents, which have to evolve to better satisfy those requests. So, the Agents will recombine and evolve over time, constantly seeking to increase their effectiveness for the user base. The Agent is the base unit of the evolutionary process in Digital Ecosystems, in the same way that the gene is the base unit for evolution in biological ecosystems \cite{begon96}. So, the collection of Agents at each Habitat provides an Agent-pool, similar to a gene-pool, which is all the genes in a population \cite{lawrence1989hsd}. Additionally, it also stores Agent-sequences evolved from the Habitat's Populations, and Agent-sequences that migrate to the Habitat from other users' Habitats, because they can potentially accelerate future Populations instantiated to respond to user requests.

\tfigure{width=170mm}{architecture2}{graffle}{Habitat Network}{Uses \getCap{habnet}}{-10mm}{}{}{}

The landscape, in energy-centric biological ecosystems, defines the connectivity between habitats \cite{begon96}. Connectivity of nodes in the digital world is generally not defined by geography or spatial proximity, but by information or semantic proximity. For example, connectivity in a peer-to-peer network is based primarily on bandwidth and information content, and not geography. The island-models of \acl{DEC} use an information-centric model for the connectivity of nodes (\emph{islands}) \cite{lin1994cgp}. However, because it is generally defined for one-time use (to evolve a solution to one problem and then stop) it usually has a fixed connectivity between the nodes, and therefore a fixed topology \cite{cantupaz1998spg}. So, supporting evolution in the Digital Ecosystem, with a dynamic multi-objective \emph{selection pressure} (fitness landscape \cite{wright1932} with many peaks), requires a re-configurable network topology, such that Habitat connectivity can be dynamically adapted based on the observed migration paths of the Agents between the users within the Habitat network. So, based on the island-models of \acl{DEC} \cite{lin1994cgp}, each connection between the Habitats is bi-directional and there is a probability associated with moving in either direction across the connection, with the connection probabilities affecting the rate of migration of the Agents. However, additionally, the connection probabilities will be updated by the success or failure of Agent migration using the concept of Hebbian learning \cite{hebb}: the Habitats which do not successfully exchange Agents will become less strongly connected, and the Habitats which do successfully exchange Agents will achieve stronger connections. This leads to a topology that adapts over time, resulting in a network that supports and resembles the connectivity of the user base. When we later consider an example user base, we will further discuss a resulting topology. 

When a new user joins the Digital Ecosystem, a Habitat needs to be created for them, and most importantly connected to the correct cluster(s) in the Habitat network. A new user's Habitat can be connected randomly to the Habitat network, as it will dynamically reconnect based upon the user's behaviour. User profiling can also be used to help connect a new user's Habitat to the optimal part of the network, by finding a similar user or asking the user to identify a similar user, and then cloning their Habitat's connections. Also, when a new Habitat is created, its Agent-pool should be created by merging the Agent-pools of the Habitats to which it is initially connected.

\subsubsection{Agent Migration}
\label{migrationHistory}

The Agents migrate through the interconnected Habitats combining with one another in Populations to meet user requests for applications. The migration path from the current Habitat is dependent on the \emph{migration probabilities} between the Habitats. The migration of an Agent within the Digital Ecosystem is initially triggered by deployment to its user's Habitat, for distribution to other users who will potentially make use of the service the Agent represents. When a user deploys a service, its representative Agent must be generated and deployed to their Habitat. It is then copied to the Agent-pool of the user's Habitat, and from there the migration of the Agent occurs, which involves migrating (copying) the agent probabilistically to all the connected Habitats. The Agent is copied rather than moved, because the Agent may also be of use to the providing user. The copying of an Agent to a connected Habitat depends on the associated migration probability. If the probability were one, then it would definitely be sent. When migration occurs, depending on the probabilities associated with the Habitat connections, an exact copy of the Agent is made at a connected Habitat. The copy of the Agent is identical until the new Agent's \emph{migration history} is updated, which differentiates it from the original. The successful use of the migrated Agent, in response to user requests for applications, will lead to further migration (distribution) and therefore availability of the Agent to other users.

The connections joining the Habitats are reinforced by successful Agent and Agent-sequence migration. The success of the migration, the \emph{migration feedback}, leads to the reinforcing and creation of migration links between the Habitats, just as the failure of migration leads to the weakening and negating of migration links between the Habitats. The success of migration is determined by the usage of Agents at the Habitats to which they migrate. When an Agent-sequence is found and used in responding to a user request, then the individual Agent \emph{migration histories} can be used to determine where they have come from and update the appropriate connection probabilities. If the Agent-sequence was fully or partly evolved elsewhere, then where the sequence or sub-sequences were created needs to be passed on to the connection probabilities, because the value in an Agent-sequence is the unique ordering and combination it provides of the individual Agents contained within. So, it is necessary to manage the feedback to the connection probabilities for migrating Agent-sequences, and not just the individual Agents contained within the sequence, including the partial use of an Agent-sequence in a newly evolved one. Specifically, the mechanism for \emph{migration feedback} needs to know the Habitats where migrating Agent-sequences were created, to create new connections or reinforce existing connections to these Habitats. The global effect of the Agent migration and \emph{migration feedback} on the Habitat network is the clustering of Habitats around the communities present within the user base, and will be discussed later in more detail.

The \emph{escape range} is the number of escape migrations available to an Agent upon the risk of death (deletion). If an Agent migrates to a Habitat and is not used after several user requests, then it will have the opportunity to migrate (move not copy) randomly to another connected Habitat. After this happens several times the Agent will be deleted (die). The \emph{escape range} will be dynamically responsive to the size of the Habitat cluster that the Agent exists within. This creates a dynamic \emph{time-to-live} \cite{comer1988iti} for the Agents, in which Agents that are used more will live longer and distribute farther than those that are used less.

\subsection{Populations}

\tfigure{width=170mm}{userRequestNew}{graffle}{User Request to the Digital Ecosystem}{(modified from \cite{kpic}): A user \getCap{picUser} \getCap{picUserReq} (Agent-pool).}{-10mm}{!h}{}{}

The Populations of the Digital Ecosystem are functionally equivalent to the evolving, self-organising populations of a biological ecosystem, and are achieved through using evolutionary computing. A population in biological ecosystems is all the members of a species that occupy a particular area at a given time \cite{lawrence1989hsd}. Our Population is also \emph{all the members of a species that occupy a particular area at a given time}, like an island from the island-models of \acl{DEC} \cite{lin1994cgp}. The use of \acl{DEC} to accelerate the Populations will be explained later.

The users \setCap{will formulate queries to the Digital Ecosystem by creating a request as a \emph{semantic description}, like those being used and developed in \aclp{SOA} \cite{SOAsemantic}, specifying an application they desire and submitting it to their Habitat.}{picUser} This description enables the definition of a metric for evaluating the \emph{fitness} of a composition of Agents, as a distance function between the \emph{semantic description} of the request and the Agents' \emph{semantic descriptions}. \setCap{A Population is then instantiated in the user's Habitat in response to the user's request, seeded from the Agents available at their Habitat}{picUserReq} (i.e. its Agent-pool). This allows the evolutionary optimisation to be accelerated in the following three ways: first, the Habitat network provides a subset of the Agents available globally, which is localised to the specific user it represents; second, making use of Agent-sequences previously evolved in response to the user's earlier requests; and third, taking advantage of relevant Agent-sequences evolved elsewhere in response to similar requests by other users. The Population then proceeds to evolve the optimal Agent-sequence(s) that fulfils the user request, and as the Agents are the base unit for evolution, it searches the available Agent-sequence combination space. For an evolved Agent-sequence that is executed (instantiated) by the user, it then migrates to other peers (Habitats) becoming hosted where it is useful, to combine with other Agents in other Populations to assist in responding to other user requests for applications.

\subsubsection{Evolution}

Evolution in biological ecosystems leads to both great \emph{diversity} and high \emph{specialisation} of its organisms \cite{begon96}. In Digital Ecosystems the \emph{diversity} of evolution will provide for the wide range of user needs and allow for quick responses to the changing of these user needs, while the \emph{specialisation} will simultaneously provide solutions which are tailored to fulfil specific user requests. We will consider the issue of \emph{diversity} in a later subsection, because it is achieved through evolution in a distributed environment, which will be discussed later. In biological ecosystems, evolutionary \emph{specialisation} is localised to a population within its micro-habitat, which allows for the creation of niches (high specialisation) \cite{lawrence1989hsd}. So, a Population is instantiated in the user's own Habitat, where the collection of Agents is chosen for the user, and the micro-Habitat is provided by the user request. There is nothing to preclude more than one Population being instantiated in a user's Habitat at any one time, provided there are computational resources sufficiently available.

A \emph{selection pressure} is the sum aggregate of the forces acting upon a population, resulting in genetic change through natural selection \cite{lawrence1989hsd}. Those organisms best \emph{fit} to survive the selection pressures operating upon them will pass on their biological \emph{fitness} to their progeny through the inheritance process \cite{lawrence1989hsd}. The \emph{fitness} of an individual Agent-sequence within a Population is determined by a \emph{selection pressure}, applied as a \emph{fitness function} \cite{eiben2003iec} instantiated from the user request, and works primarily on comparing the \emph{semantic descriptions} of the Agents with the \emph{semantic description} of the user request. The \emph{pressure} selects for those Agent-sequences that are \emph{fit} and capable of surviving the environment to reproduce, and against those that do not have sufficient \emph{fitness} and therefore die before passing on their \emph{genes}, thereby providing the direction for genetic change. In biology fitness is a measure of an organism's success in its environment \cite{lawrence1989hsd}, and its definition here will be further explained in the next subsection.

\newcommand{\vspacefit}{\vspace{2mm}} 

\vspacefit

\emph{Genes} are the functional unit in biological evolution \cite{lawrence1989hsd}; whereas here the functional unit is the Agent. Therefore, the evolutionary process of a Population provides a \emph{combinatorial optimisation} \cite{papadimitriou1998coa} of the Agents available, when responding to a user request. So, it does not change or mutate the Agents themselves. In biology a \emph{mutation} is a permanent transmissible change (over the generations) in the genetic material (DNA) of an individual, and recombination (e.g. crossover) is the formation within the offspring of alleles (gene combinations), which are not present in the parents \cite{lawrence1989hsd}. As in genetic algorithms \cite{goldberg}, \emph{mutations} will occur by switching Agents in and out of the Agent-sequence structure, and \emph{recombination} (crossover) will occur by performing a crossing of two Agent-sequences.

\vspacefit

As the Digital Ecosystem receives more and more sophisticated requests, so more and more complex applications are evolved and become available for use by the users. To achieve this evolution, specifically the Agent-sequence recombination and optimisation, is a very significant challenge, because of the range of services that must be catered for and the potentially huge number of factors that must be considered for creating an applicable \emph{fitness function}. First, to construct ever more complex software solutions, requires modularity, which is provided by the paradigm of service interoperability from \aclp{SOA} \cite{soa1w}. Second, two of the most important issues are that of defining \emph{fitness} and managing \emph{bloat}, which we will discuss next. Finally, there is a huge body of work and continuing research regarding theoretical approaches to evolutionary computing \cite{eiben2003iec}, including the extensive use of genetic algorithms for practical real-world problem solving \cite{ducheyne2003fiu}. In defining Digital Ecosystems we should make use of the current state-of-the-art, and future developments, in the areas of evolutionary computing \cite{jin2005csf} and service interoperability \cite{soa1w}. 

\vspacefit

\subsubsection{Fitness}

In biology \emph{fitness} is a measure of how successful an organism is in its environment, i.e. its phenotype \cite{lawrence1989hsd}. The \emph{fitness} of an Agent-sequence within a Population would also, ideally, be based upon its \emph{phenotype}, the \emph{run-time instantiation}, and nothing else. However, such an approach would be impractical, because it is currently infeasible to execute all the Agent-sequences of a Population at every generation, and not least because of the computational resources that would be required. The other concern is one of practicality, by which we mean that it may not even be possible to perform a live execution for the \emph{executable components} of an Agent-sequence; for example, if they are for buying an item from an online retailer. These are well known issues in evolutionary computing, which is why \emph{fitness functions} are often defined as simulated input/output pairs to test functionality \cite{mantere2005ese}. In Digital Ecosystems we can use historical \emph{usage information}, but this would be insufficient initially, because such information would not be available at the time of an Agent's deployment. However, because each Agent also carries a \emph{semantic description}, a specification of what it does, the \emph{fitness function} can measure a complete Agent-sequence's collective \emph{semantic descriptions} relative to the \emph{semantic description} of a user request. So, initially the \emph{fitness function} should be based primarily on comparing the \emph{semantic descriptions} of the Agent-sequences to the \emph{semantic description} of the user request, ever increasingly augmented with the growing \emph{usage information} available for the Agents. In biological terms the \emph{genotype} will be used as the \emph{phenotype}, combined with any available past fitness of the \emph{phenotype}; with the Agent's \emph{semantic description} (\emph{genotype}) therefore acting as a guarantee of its expected behaviour. So, for any newly deployed Agent a \emph{one-to-one genotype-phenotype mapping} \cite{shackleton2000irg} will initially exist, until sufficient \emph{usage information} is available. While the use of such a mapping is undesirable, it is temporary, and necessary to allow Digital Ecosystems to operate effectively.

We have already suggested that the primary driver of the evolutionary process should initially be the extent by which an Agent-sequence can verifiably satisfy the specified requirements. This could be measured probabilistically, or using theorem-proving to validate the system, though automatic theorem proving is notoriously slow \cite{slagle1974atp, schumann2001atp}. However, there will also be other pressures on the fitness. For example, one may seek the most parsimonious solution to a problem (one that provides exactly the specified features and no more), or the cheapest solution, or one with a good \emph{reputation}. Some aspects of fitness will be implicit in the evolutionary process (e.g. Agents which are often used will gain more fitness) while others will require explicit measures (e.g. price, or user satisfaction). One way to handle this multiplicity of fitness values (some qualitative) is to explicitly recognise the multi-objective nature of the optimisation problem. In this way, we are seeking not the single best solution, but a range of possible compromises that can be made most optimally. The set of solutions for which there are no better compromises is called the Pareto-set, and evolutionary techniques have been adapted to solve such problems with considerable success \cite{veldhuizen2000mea}. The main point is that selection has to be driven not by an absolute value of fitness, but rather by a notion of what it means for one solution to be better than another. We say one solution dominates another if it is better in at least one respect, and no worse in any of the others \cite{fonseca1995oea}.

\subsubsection{Bloat}

\label{bloat}

If the repetition of Agents is allowed within evolving Agent-sequences, then the search space can become countably infinite, because the nature of the problem to be solved may not allow us to determine what the length of a solution is beforehand. Therefore, a variable length approach must be adopted, which is common in genetic programming \cite{ec25}. When variable length representations of solutions are used, a well-known phenomenon arises, called \emph{bloat}, in which the individuals of an evolving population tend to grow in size without gaining any additional advantage \cite{langdon1997fcb}. The \emph{bloat} phenomenon can cause early termination of an evolutionary process due to the exhaustion of the available memory, and can also significantly reduce performance, because typically longer sequences have higher fitness computation costs \cite{ratle2001abp}. Bloat is not specific to genetic programming, and is inherent in search techniques with discrete variable length representations \cite{langdon1998esv}. It is a fundamental area of research within search-based approaches such as \aclp{GA}, \acl{GP} and other approaches not based on populations such as simulated annealing \cite{langdon1998esv}. However, considerable work on bloat has been done in connection with \acl{GP} \cite{langdon1998gpa, banzhaf1998gpi}, and we believe that the \aclp{GA} community generally, and the genetic-algorithms-based approach of our Digital Ecosystems specifically, can benefit directly from this research. While bloat is a phenomenon which was first observed in practice \cite{ec25}, theoretical analyses have been attempted \cite{banzhaf2002scr}. One should take care with these approaches as implementations will always deal with finite populations, while theoretical approaches often deal with infinite populations \cite{ec25}, and this difference can be important. Yet, both theoretical and empirical approaches are required to understand bloat. There are many factors contributing to bloat, and while the phenomenon may appear simple, the reasons are not. There are several theories to explain why this occurs, and, as we shall discuss, some measures that can be taken for its prevention.

There are several different qualitative theories which attempt to explain bloat, and they can be considered in two groups. First, protection against crossover and bias removal (which can be considered jointly) and second, the nature of programme search spaces \cite{banzhaf2002scr}. First, near the end of a run a Population consists of mostly fit individuals, and any crossover is likely to be detrimental to the fitness of the offspring. In any sequence of Agents there may be Agents that do not contribute semantically to the complete functionality of the sequence if, for example, their functionality was not requested by the user or if it is duplicated in the sequence; analogously to \acl{GP} \cite{banzhaf2002scr}, we can call these redundant Agents \emph{bloat}. The genotype can then be grown further without affecting the phenotype if Agents with similar functionality are added; but, as the genotype grows larger, crossover is more likely to transfer redundant Agents to the new off-springs (assuming uniform crossover). Second, above a certain threshold size, the distribution of functionality does not vary with the size of the search space \cite{banzhaf2002scr}. Thus, if we randomly sample long and short Agent-sequences above a length threshold, they will likely have the same functionality and fitness. So, as a search process progresses we are more likely to sample longer Agent-sequences, as mutation results in more of them (all other things being equal) and this will give rise to the bloating phenomenon.

Each of the stages of construction of a \acl{GA} (i.e. choice of fitness function, selection method and genetic operator) can affect bloat. It has been shown that even small differences in the fitness function can cause a difference: a single programme glitch in an otherwise \emph{flat fitness landscape} (from the neutral theory of molecular evolution \cite{kimura:ntm}) is sufficient to drive the average programme size of an infinite population \cite{mcphee2001EuroGP}. If a fitness-proportional selection method is used, individuals with zero fitness will be discontinued as they have zero probability of being selected as parents \cite{blickle1996css}. However, if tournament selection method is used, then there is a finite chance that individuals with zero fitness will be selected to be parents \cite{blickle1996css}. Finally, the choice of genetic operator affects the size of the programmes which are sampled; \emph{standard crossover on a flat landscape heavily oversamples the shorter programmes} \cite{poli01exact}. There are other factors that may affect bloat, for example, how the population is initialised, or the choice of representation used, such as a neutral genotype-phenotype mapping, which can actually alleviate bloat \cite{miller2001wbcgpbp}.

Bloat is a fact, whatever the reasons, happening in this type of optimisation and needs to be controlled if the space is to be searched effectively. One solution is to apply a hard limit to the size of the sets that can be sampled \cite{langdon1997fcb}: this enables the search algorithm to keep running without having out-of-memory run-time errors, but poses questions on how to set this hard limit. An alternative but similar method is to apply a \emph{parsimony pressure}, where a term is added to the fitness function which chastises big sets in preference for smaller sets \cite{soule1998ecg}. In this approach, individuals larger than the average size are evaluated with a reduced probability, biasing the search to smaller sets, while providing a dynamic limit which adapts to the average size of individuals in a changing population \cite{soule1998ecg}.

\subsection{The Digital Ecosystem}

The Digital Ecosystem supports the automatic combining of numerous Agents (which represent services), by their interaction in evolving Populations to meet user requests for applications, in a scalable architecture of distributed interconnected Habitats. The sharing of Agents between Habitats ensures the system is scalable, while maintaining a high evolutionary specialisation for each user. The network of interconnected Habitats is equivalent to the \emph{abiotic} environment of biological ecosystems \cite{begon96}; combined \setCap{with the Agents, the Populations, the Agent migration for \acl{DEC}, and the environmental selection pressures provided by the user base, then the union of the Habitats creates the Digital Ecosystem}{digEco}, which is summarised in Figure \ref{DE}. The continuous and varying user requests for applications provide a dynamic evolutionary pressure on the Agent sequences, which have to evolve to better fulfil those user requests, and without which there would be no driving force to the evolutionary self-organisation of the Digital Ecosystem.

\tfigure{scale=1.0}{DE}{graffle}{Digital Ecosystem}{A network of interconnected Habitats, combined \getCap{digEco}. Agents travel along the peer-to-peer connections; in every node (Habitat) local optimisation is performed through an evolutionary algorithm, where the search space is determined by the Agents present at the node.}{-4mm}{}{}{}

In the Digital Ecosystem, local and global optimisations concurrently operate to determine solutions to satisfy different optimisation problems. The global optimisation here is not a decentralised super-peer based control mechanism \cite{risson2006srt}, but the completely distributed peer-to-peer network of the interconnected Habitats, which are therefore not susceptible to the failure of super-peers. It provides a novel optimisation technique inspired by biological ecosystems, working at two levels: a first optimisation, migration of Agents which are distributed in a peer-to-peer network, operating continuously in time; this process feeds a second optimisation, based on evolutionary combinatorial optimisation, operating locally on single peers and is aimed at finding solutions that satisfy locally relevant constraints. So, the local search is improved through this twofold process to yield better local optima faster, as the distributed optimisation provides prior sampling of the search space through computations already performed in other peers with similar constraints. This novel form of distributed evolutionary computing will be discussed further below, once we have discussed a topology resulting from an example user base.

If we consider an example user base for the Digital Ecosystem, the use of \aclp{SOA} in its definition means that \acf{B2B} interaction scenarios \cite{krafzig2004ess} lend themselves to being a potential user base for Digital Ecosystems. So, we can consider the \emph{business ecosystem} of \acf{SME} networks from \aclp{DBE} \cite{dbebkintro}, as a specific class of examples for \ac{B2B} interaction scenarios; and in which the \ac{SME} users are requesting and providing software services, represented as Agents in the Digital Ecosystem, to fulfil the needs of their business processes. \aclp{SOA} promise to provide potentially huge numbers of services that programmers can combine, via the standardised interfaces, to create increasingly more sophisticated and distributed applications \cite{SOApaper2}. The Digital Ecosystem extends this concept with the automatic combining of available and applicable services, represented by Agents, in a scalable architecture, to meet user requests for applications. These Agents will recombine and evolve over time, constantly seeking to improve their effectiveness for the user base. From the SME users' point of view the Digital Ecosystem provides a network infrastructure where connected enterprises can advertise and search for services (real-world or software only), putting a particular emphasis on the composability of loosely coupled services and their optimisation to local and regional, needs and conditions. To support these SME users the Digital Ecosystem is satisfying the companies' business requirements by finding the most suitable services or combination of services (applications) available in the network. A composition of services is an Agent-sequence in the Habitat network that can move from one peer (company) to another, being hosted only in those where it is most useful in satisfying the SME users' business needs.

\subsubsection{Topology}

The Digital Ecosystem allows for the connectivity in the Habitats to adapt to the connectivity within the user base, with a cluster of Habitats representing a community within the user base. If a user is a member of more than one community, the user's Habitat will be in more than one cluster. This leads to a network topology that will be discovered with time, and which reflects the connectivity within the user base. Similarities in requests by different users will reinforce behavioural patterns, and lead to clustering of the Habitats within the ecosystem, which can occur over geography, language, etc. This will form communities for more effective information sharing, the creation of niches, and will improve the responsiveness of the system. The connections between the Habitats will be self-managed, through the mechanism of Agent migration defined earlier. Essentially, successful Agent migration will reinforce Habitat connections, thereby increasing the probability of future Agent migration along these connections. If a successful multi-hop migration occurs, then a new link between the start and end Habitats can be formed. Unsuccessful migrations will lead to connections (migration probabilities) decreasing, until finally the connection is closed.

\tfigure{scale=1.0}{DBE}{graffle}{Digital Business Ecosystem}{Business ecosystem, network of \aclp{SME} \cite{dbebkintro}, using the Digital Ecosystem. \getCap{bizEcoCap}.}{-5mm}{}{}{}

If we consider the \emph{business ecosystem} - a network of \aclp{SME} from \aclp{DBE} \cite{dbebkintro} - as an example user base, such business networks are typically small-world networks \cite{white2002nst, antionella}. They have \setCap{many strongly connected clusters (communities), called \emph{sub-networks} (quasi-complete graphs), with a few connections between these clusters (communities) \cite{swn1}. Graphs with this topology have a very high clustering coefficient and small characteristic path lengths \cite{swn1}.}{archComTop} \setCap{As the connections between Habitats are reconfigured depending on the connectivity of the user base, the Habitat clustering will therefore be parallel to the business sector communities}{bizEcoCap}, as shown in Figure \ref{DBE}. The communities will cluster over language, nationality, geography, etc. -- all depending on the user base. So, the Digital Ecosystem will take on a topology similar to that of the user base.

\tfigure{width=170mm}{architecture}{graffle}{Habitat Clustering}{Topology adapted to the small-world network of a business ecosystem of \acp{SME} from \aclp{DBE} \cite{dbebkintro}, having \getCap{archComTop}}{-10mm}{!b}{}{}

Fragmentation of the Habitat network can occur, but only if dictated by the structure of the user base. The issue of greater concern is when individual Habitats become totally disconnected, which can only occur under certain conditions. One condition is that the Agents within the Agent-pool consistently fail to satisfy user requests. Another condition is when the Agents and Agent-sequences they share are undesirable to the users that are within the migration range of these Agents and Agent-sequences. These scenarios can arise because the Habitat is located within the wrong cluster, in which case the user can be asked to join another cluster within the Habitat network, assuming the user base is of sufficient size to provide a viable alternative.

\newcommand{\vspaceden}{\vspace{2mm}}

\vspaceden

\subsubsection{Distributed Evolution}

\vspaceden

The Digital Ecosystem is a hybrid of \aclp{MAS}, more specifically of \emph{mobile} agent systems, \aclp{SOA}, and \emph{distributed} evolutionary computing, which leads to a novel form of evolutionary computation. The novelty comes from the creation of multiple evolving Populations in response to \emph{similar} requests, whereas in the island-models of \acl{DEC} there are multiple evolving populations in response to only one request \cite{lin1994cgp}. So, in our Digital Ecosystem different \setCap{requests are evaluated on separate \emph{islands} (Populations), with their evolution accelerated by the sharing of solutions between the evolving Populations (islands), because they are working to solve similar requests (problems).}{similarCap} This is shown in Figure \ref{similar}, where the dashed \setCap{yellow lines connecting the evolving Populations indicate similarity in the requests being managed.}{similar2}

\vspaceden

\tfigure{width=170mm}{similar}{graffle}{Distributed Evolution in the Digital Ecosystem}{Different \getCap{similarCap} The \getCap{similar2}}{-10mm}{!h}{}{}

\vspaceden

If we again consider the \emph{business ecosystem} of \aclp{SME} from \aclp{DBE} \cite{dbebkintro} as an example user base, then in Figure \ref{similar} the four Habitats, in the left cluster, could be travel agencies, and the three with linked evolving Populations are looking for similar package holidays. So, an optimal solution found and used in one Habitat will be migrated to the other connected Habitats and integrated into any evolving Populations via the local Agent-pools. This will help to optimise the search for similar package holidays at the Habitats of the other travel agencies. This also works in a time-shifted manner, because an optimal solution is stored in the Agent-pool of the Habitats to which it is migrated, being available to optimise a similar request placed later. 

\vspaceden

The distributed architecture of Digital Ecosystems favours the use of Pareto-sets for fitness determination, because Pareto optimisation for multi-objective problems is usually most effective with spatial distribution of the populations, as partial solutions (solutions to different niches) evolve in different parts of a \emph{distributed population} \cite{detoro2002ppg} (i.e. different Populations in different Habitats). By contrast, in a single population, individuals are always interacting with each other, via crossover, which does not allow for this type of specialisation \cite{back1996eat}.

\vspaceden

This approach requires the Digital Ecosystem to have a sufficiently large user base, so that there can be communities within the user base, and therefore allow for similarity in the user requests. Assuming a user base of hundreds of users, then there would be hundreds of Habitats, in which there will be potentially three or more times the number of Populations at any one time. Then there will be thousands of Agents and Agent-sequences (applications) available to meet the requests for applications from the users. In such a scenario, there would be a sufficient number of users for the Digital Ecosystem to find similarity within their requests, and therefore apply our novel form of \acl{DEC}.

\vspaceden

\subsubsection{Agent Life-Cycle}
\label{agentLifeCycle}

\vspaceden

An Agent is created to represent a user's service in the Digital Ecosystem, and its life-cycle begins \setCap{with deployment to its owner's Habitat for distribution within the Habitat network.}{lifeCycleCap} The Agent is then migrated to any Habitats connected to the owner's Habitat, to make it available in other Habitats where it could potentially be useful. The Agent is then available to the local evolutionary optimisation, to be \setCap{used in evolving the optimal Agent-sequence in response to a user request. The optimal Agent-sequence is then registered at the Habitat}{lifeCycle2}, being stored in the Habitat's Agent-pool. \setCap{If an Agent-sequence solution is then executed, an attempt is made to migrate (copy) it to every other connected Habitat, success depending on the probability associated with the connection.}{lifeCycle3} The Agent life-cycle is shown in Figure \ref{lifeCycle}. 

An Agent can also be deleted if after several successive user requests at a Habitat it remains unused; it will have a small number of \emph{escape migrations}, in which it is not copied, but is randomly moved to another connected Habitat. If the Agent fails to find a \emph{niche} before running out of \emph{escape migrations}, then it will be deleted.

\pagebreak
\tfigure{scale=1.0}{lifeCycle}{graffle}{Agent Life-Cycle}{Begins \getCap{lifeCycleCap} It can then be \getCap{lifeCycle2}. \getCap{lifeCycle3}}{-2mm}{!h}{}{}



\section{Simulation and Results}
\label{simRes}


We simulated the Digital Ecosystem, based upon our \acl{EOA}, and recorded key variables to determine whether it displayed behaviour typical of biological ecosystems. Although agent-based modelling solutions, like Repast (Recursive Porous Agent Simulation Toolkit) \cite{collier2003ref} and MASON (Multi-Agent Simulator Of Neighbourhoods) \cite{luke2004mnm}, and evolutionary computing libraries, like ECJ (Evolutionary Computing in Java) \cite{ECJ} and the JCLEC (Java Computing Library for Evolutionary Computing) \cite{ventura2008jjf}, are available, it was evident that it would take as much effort to adapt one, or a combination, of these to simulate the Digital Ecosystem, as it would to create our own simulation of the Digital Ecosystem, because the required ecological dynamics are largely absent from these and other available technologies. So, we created our own simulation, following the \acl{EOA} from the previous section (unless otherwise specified), using the \emph{business ecosystem} of \aclp{SME} from \aclp{DBE} \cite{dbebkintro} as an example user base. 

\subsection{Agents: Semantic Descriptions}
\label{descriptions}

\mfigure{\begin{center}\bold{A = \{(1,25), (2,35), (3,55), (4,6), (5,37), (6,12)\}}\end{center}}{Agent Semantic Descriptions}{\getCap{as4} \getCap{as3} between three and six numeric tuples; each \getCap{agentSemantic2}}{BMLprocess}{1mm}{-5mm}{!h}

An Agent represents a user's service, including the \emph{semantic description} of the \emph{business process} involved, and is based on existing and emerging technologies for \emph{semantically capable} \aclp{SOA} \cite{SOAsemantic}, such as the \acs{OWL-S} semantic markup for web services \cite{martin2004bsw}. We simulated a service's \emph{semantic description} \setCap{with an abstract representation consisting of a set of}{as3} numeric tuples, to simulate the properties of a \emph{semantic description}. Each \setCap{tuple representing an \emph{attribute} of the \emph{semantic description}, one integer for the \emph{attribute identifier} and one for the \emph{attribute value}, with both ranging between one and a hundred.}{agentSemantic2} \setCap{Each simulated Agent had a semantic description}{as4}, with between three and six tuples, as shown in Figure \ref{BMLprocess}.

\subsection{User Base}

\mfigure{\begin{center}\bold{R = [\{(1,23),(2,45),(3,33),(4,6),(5,8),(6,16)\}, \{(1,84),(2,48),(3,53),(4,11),(5,16)\}]}\end{center}}{User Request}{\getCap{semanticRequest}; each \getCap{agentSemantic2}}{userRequest}{4mm}{-5mm}{!h}

Throughout the simulations we assumed a hundred users, which meant that at any time the number of users joining the network equalled those leaving. The Habitats of the users were randomly connected at the start, to simulate the users going online for the first time. The users then produced Agents (services) and requests for business applications. Initially, the users each deployed five Agents to their Habitats, for migration (distribution) to any Habitats connected to theirs (i.e. their community within the \emph{business ecosystem}). Users were simulated to deploy a new Agent after the submission of three requests for business applications, and were chosen at random to submit their requests. \setCap{A simulated user request consisted of an abstract \emph{semantic description}, as a list of sets of numeric tuples to represent the properties of a desired business application}{semanticRequest}. The use of the \emph{numeric tuples} made it comparable to the \emph{semantic descriptions} of the services represented by the Agents; while the \emph{list of sets} (two level hierarchy) and a much longer length provided sufficient complexity to support the sophistication of business applications. An example is shown in Figure \ref{userRequest}.

The user requests were handled by the Habitats instantiating evolving Populations, which used evolutionary computing to find the optimal solution(s), Agent-sequence(s). It was assumed that the users made their requests for business applications \emph{accurately}, and always used the response (Agent-sequence) provided.

\subsection{Populations: Evolution}
\label{fitnessFunction}

Populations of Agents, $[A_1, A_1, A_2, ...]$, were evolved to solve user requests, seeded with Agents and Agent-sequences from the \emph{Agent-pool} of the Habitats in which they were instantiated. A dynamic population size was used to ensure exploration of the available combinatorial search space, which increased with the average length of the Population's Agent-sequences. The optimal combination of Agents (Agent-sequence) was evolved to the user request $R$, by an artificial \emph{selection pressure} created by a \emph{fitness function} generated from the user request $R$. An individual (Agent-sequence) of the Population consisted of a set of attributes, ${a_1, a_2, ...}$, and a user request essentially consisted of a set of required attributes, ${r_1, r_2, ...}$. So, the \emph{fitness function} for evaluating an individual Agent-sequence $A$, relative to a user request $R$, was
\begin{equation}
fitness(A,R) = \frac{1}{1 + \sum_{r \in R}{|r-a|}},
\label{ff}
\end{equation}
where $a$ is the member of $A$ such that the difference to the required attribute $r$ was minimised. Equation \ref{ff} was used to assign \emph{fitness} values between 0.0 and 1.0 to each individual of the current generation of the population, directly affecting their ability to replicate into the next generation. The evolutionary computing process was encoded with a low mutation rate, a fixed selection pressure and a non-trapping fitness function (i.e. did not get trapped at local optima). The type of selection used \emph{fitness-proportional} and \emph{non-elitist}, \emph{fitness-proportional} meaning that the fitter the individual the higher its probability of surviving to the next generation \cite{blickle1996css}. \emph{Non-elitist} means that the best individual from one generation was not guaranteed to survive to the next generation; it had a high probability of surviving into the next generation, but it was not guaranteed as it might have been mutated \cite{eiben2003iec}. \emph{Crossover} (recombination) was then applied to a randomly chosen 10\% of the surviving population, a \emph{one-point crossover}, by aligning two parent individuals and picking a random point along their length, and at that point exchanging their tails to create two offspring \cite{eiben2003iec}. \emph{Mutations} were then applied to a randomly chosen 10\% of the surviving population; \emph{point mutations} were randomly located, consisting of \emph{insertions} (an Agent was inserted into an Agent-sequence), \emph{replacements} (an Agent was replaced in an Agent-sequence), and \emph{deletions} (an Agent was deleted from an Agent-sequence) \cite{lawrence1989hsd}. The issue of bloat was controlled by augmenting the \emph{fitness function} with a \emph{parsimony pressure} \cite{soule1998ecg} which biased the search to shorter Agent-sequences, evaluating longer than average length Agent-sequences with a reduced \emph{fitness}, and thereby providing a dynamic control limit which adapted to the average length of the ever-changing evolving Agent Populations.

\subsection{Semantic Filter}
\label{semanticFilter}

\mfigure{\begin{center}Agent's semantic description:\end{center}
\vspace{-9mm}
\begin{quote}\begin{center}\bold{\{\red{(1,25)}, \blue{(2,35)}, \yellow{(3,55)}, \green{(4,6)}, \purple{(5,37)}, \turquoise{(6,12)}\}}\end{center}
\end{quote}
\begin{center}\vspace{-3mm}(with semantic filter):\end{center}
\vspace{-9mm}
\begin{quote}\begin{center}
\bold{\{\red{(Business, Airline)}, \blue{(Company, British Midland)}, \yellow{(Quality, Economy)}, \green{(Cost, 60)}, \purple{(Depart, Edinburgh)}, \turquoise{(Arrive, London)}\}}\end{center}\end{quote}
\begin{center}\vspace{-2mm}user request:\end{center}
\vspace{-9mm}
\begin{quote}\bold{[\{\red{(1,23)}, \blue{(2,45)}, \yellow{(3,33)}, \green{(4,6)}, \purple{(5,8)}, \turquoise{(6,16)}\}, \{\red{(1,84)}, \blue{(2,48)}, \yellow{(3,53)}, \green{(4,11)}, \grey{(7,16)}, \brown{(8,34)}\}, \{\red{(1,23)}, \blue{(2,45)}, \yellow{(3,53)}, \green{(4,6)}, \purple{(5,16)}\turquoise{(6,53)}\}, \{\red{(1,86)}, \blue{(2,48)}, \yellow{(3,33)}, \green{(4,25)}, \grey{(7,55)}\brown{(8,23)}\}, \{\red{(1,25)}, \blue{(2,52)}, \yellow{(3,53)}, \green{(4,5)}, \purple{(5,55)}, \turquoise{(6,37)}\}, \{\red{(1,86)}, \blue{(2,48)}, \yellow{(3,43)}, \green{(4,25)}, \grey{(7,37)}, \brown{(8,40)}\}, \{\red{(1,22)}, \blue{(2,77)}, \yellow{(3,82)}, \green{(4,9)}, \purple{(5,35)}, \turquoise{(6,8)}\}]}
\end{quote}
\begin{center}\vspace{-3mm}(with semantic filter):\end{center}
\vspace{-9mm}
\begin{quote}
\bold{[\{\red{(Business, Airline)}, \blue{(Company, Air France)}, \yellow{(Quality, Economy)}, \green{(Cost, 60)}, \purple{(Depart, Edinburgh)}, \turquoise{(Arrive, Paris)}\}, \{\red{(Business, Hotel)}, \blue{(Company, Continental)}, \yellow{(Quality, 3*)}, \green{(Cost, 110)}, \grey{(Location, Paris)}, \brown{(Nights, 3)}\}, \{\red{(Business, Airline)}, \blue{(Company, Air France)}, \yellow{(Quality, Economy)},\green{(Cost,60)},\purple{(Depart, Paris)}, \turquoise{(Arrive, Monte Carlo)}\}, \{\red{(Business, Hotel)}, \blue{(Company, Continental)}, \yellow{(Quality, 2*)}, \green{(Cost, 250)}, \grey{(Location, Monte Carlo)}, \brown{(Nights, 2)}\}, \{\red{(Business, Airline)}, \blue{(Company, KLM)}, \yellow{(Quality, Economy)}, \green{(Cost, 50)}, \purple{(Depart, Monte Carlo)}, \turquoise{(Arrive, London)}\}, \{\red{(Business, Hotel)}, \blue{(Company, Continental)}, \yellow{(Quality, 3*)}, \green{(Cost, 250)}, \grey{(Location, London)}, \brown{(Nights, 4)}\}, \{\red{(Business, Airline)}, \blue{(Company, Air Espana)}, \yellow{(Quality, First)}, \green{(Cost, 90)}, \purple{(Depart, London)}, \turquoise{(Arrive, Edinburgh)}\}]}
\end{quote}}{Semantic Filter}{Shows \getCap{bmlcap1} \getCap{capbml2}. \getCap{capbml3}}{BMLreal}{-2mm}{-5mm}{!h}

The simulation of the Digital Ecosystem complies with the \acl{EOA} defined in the previous section, but there was the possibility of model error in the \emph{business ecosystems} of the user base (\aclp{SME} from \aclp{DBE} \cite{dbebkintro}), because while the abstract numerical definition for the simulated \emph{semantic descriptions}, of the services and requests the users provide, makes it widely applicable, it was unclear that it could accurately represent \emph{business services}. So we created a \emph{semantic filter} to show \setCap{the \emph{numerical semantic descriptions}, of the simulated services (Agents) and user requests, in a \emph{human readable form}.}{bmlcap1} The basic properties of any \emph{business process} are cost, quality, and time \cite{davenport1990nie}; so this was followed in the \emph{semantic filter}. \setCap{The \emph{semantic filter} translates \emph{numerical semantic descriptions} for one community within the user base, showing it in the context of the travel industry}{capbml2}, as shown in Figure \ref{BMLreal}. \setCap{The simulation still operated on the numerical representation for operational efficiency, but the \emph{semantic filter} essentially assigns meaning to the numbers.}{capbml3} The output from the \emph{semantic filter}, in Figure \ref{BMLreal}, shows that the \emph{numerical semantic descriptions} are a reasonable modelling assumption that abstracts sufficiently rich textual descriptions of \emph{business services}.

\subsection{Evolutionary Dynamics}

\tfigure{}{fitnessAvgMax}{graph}{Graph of Fitness in the Evolutionary Process}{This \getCap{evoGraph}}{-5mm}{!h}{}{}

We plotted the fitness of the evolutionary process for a typical Population, to ensure that the core process that creates order within the Digital Ecosystem was operating satisfactorily. The graph in Figure \ref{fitnessAvgMax} \setCap{shows both the maximum and average fitness increasing over the generations of a typical Population, and as expected the \emph{average fitness} remains below the \emph{maximum fitness} because of variation in the Population \cite{goldberg}, showing that the evolutionary processes, which construct order in the Digital Ecosystem, are operating satisfactorily.}{evoGraph}

\subsection{Ecological Succession}
\label{secSuccession}

We then compared some of the Digital Ecosystem's dynamics with those of biological ecosystems, to determine if it had been imbibed with the properties of biological ecosystems. A biological ecosystem develops from a simpler to a more mature state, by a process of \emph{succession}, where the genetic variation of the populations changes with time \cite{begon96}. \setCap{So, it becomes increasingly more complex through this process of succession, driven by the evolution of the populations within the ecosystem \cite{connell111msn}.}{succession} Equivalently, the Digital Ecosystem's increasing complexity comes from the Agent Populations being evolved to meet the dynamic selection pressures created by the user requests. 

\tfigure{height=94mm}{successionNew}{pdf}{Ecological Succession}{(modified from \cite{davis2008}): \getCap{succession2} \getCap{succession3}. \getCap{succession}}{-3mm}{!h}{}{}

\setCap{The formation of a mature ecosystem}{succession2}, ecological succession, \setCap{is the slow, predictable, and orderly changes in the composition and structure of an ecological community, for which there are defined stages in the increasing complexity \cite{begon96}, as shown}{succession3} in Figure \ref{successionNew}. Succession may be initiated either by the formation of a new, unoccupied habitat (e.g., a lava flow or a severe landslide) or by some form of disturbance (e.g. fire, logging) of an existing community. The former case is often called \emph{primary succession}, and the latter \emph{secondary succession} \cite{begon96}. The trajectory of ecological change can be influenced by site conditions, by the interactions of the species present, and by more stochastic factors such as availability of colonists or seeds, or weather conditions at the time of disturbance. Some of these factors contribute to predictability of successional dynamics; others add more probabilistic elements \cite{gotelli1995pe}. Trends in ecosystem and community properties of succession have been suggested, but few appear to be general. For example, species diversity almost necessarily increases during early succession upon the arrival of new species, but may decline in later succession as competition eliminates opportunistic species and leads to dominance by locally superior competitors \cite{connell111msn}. Net Primary Productivity\footnote{Net Primary Productivity is the net flux of carbon from the atmosphere into green plants per unit time \cite{lawrence1989hsd}.}, biomass, and trophic level properties all show variable patterns over succession, depending on the particular system and site \cite{gotelli1995pe}. Generally, communities in early succession will be dominated by fast-growing, well-dispersed species, but as the succession proceeds these species will tend to be replaced by more competitive species \cite{begon96}.

We then considered existing theories of complexity for ecological succession and how it would apply to Digital Ecosystems, seeking a high-level understanding that would apply equally to both biological and digital ecosystems. As succession leads communities, of an ecosystem, to states of \emph{dynamic equilibrium}\footnote{Dynamic Equilibrium is when opposing forces of a system are proceeding at the same rate, such that its state is unchanging with time \cite{begon96}.} within the environment \cite{begon96}, the complexity has to increase initially or there would be no ecosystem, and presumably this increase eventually stops, because there must be a limit to how many species can be supported. The period in between is more complicated. If we consider the neutral biodiversity theory \cite{Hubbell}, which basically states network aspects of ecosystems are negligible, we would probably get a relatively smooth progression, because although you would get occasional extinctions, they would be randomly isolated events whose frequency would eventually balance arrivals, not self-organised crashes like in systems theory. In systems theory \cite{systemsTheory}, when a new species arrives in an ecological network, it can create a positive feedback loop that destabilises part of the network and drives some species to extinction. Ecosystems are constantly being perturbed, so it is reasonable to assume that a species that persists will probably be involved in a stabilising interaction with other species. So, the whole ecological network evolves to resist invasion. That would lead to a spiky succession process, perhaps getting less spiky over time. 

So, which theory is more applicable to the Digital Ecosystem depends on the extent that a species in the ecosystem acts independently, competing entities (smooth succession) \cite{Hubbell} versus tightly co-adapted ecological partners (spiky succession) \cite{systemsTheory}. Our Digital Ecosystem despite its relative complexity is quite simple compared to biological ecosystems. It has the essential and fundamental processes, but no sophisticated social mechanisms. Therefore, the smooth succession of the neutral biodiversity theory \cite{Hubbell} is more probable.

\tfigure{}{DigEcoSuc}{graph}{Graph of Succession in the Digital Ecosystem}{\getCap{DigEcoSucCap} Still, at \getCap{DigEcoSuc2Cap}}{-7mm}{!b}{2mm}{}

\label{ecosucexp}

As the increasing complexity of the Digital Ecosystem comes from its evolving Agent Populations responding to user requests, the effectiveness of the evolved Agent-sequences (responses) can provide a measure of its complexity over time. So, in simulation we measured the effectiveness of its responses over a thousand user requests, i.e. until it had reached a mature state like a biological ecosystem \cite{begon96}, and graphed a typical run in Figure \ref{DigEcoSuc}. The range and diversity of Agents at initial deployment were such that 70\% fulfilment of user requests was possible, increasing to 100\% fulfilment as more Agents were deployed. The Digital Ecosystem performed as expected, adapting and improving over time, reaching a mature state as seen in the graph of Figure \ref{DigEcoSuc}. The succession of the Digital Ecosystem followed the smooth succession of the neutral biodiversity theory \cite{Hubbell}, shown by the \emph{tight distribution} and \emph{equal density} of the points around the best fit curve of the graph in Figure \ref{DigEcoSuc}. The variation in the percentage responsiveness, over the successive user request events, came from the differential rates of adaption at the Habitats. Still, by \setCap{the end of the simulation run, the Agent-sequences had evolved and migrated over an average of only ten user requests per Habitat, and collectively had already reached near 70\% effectiveness for the user base.}{DigEcoSuc2Cap} \setCap{The formation of a mature biological ecosystem, ecological succession, is a relatively slow process \cite{begon96}, and the simulated Digital Ecosystem acted similarly in reaching a mature state.}{DigEcoSucCap}

\subsection{Species Abundance}

In ecology, \emph{relative abundance} \setCap{is a measure of the proportion of all organisms in a community belonging to a particular \emph{species} \cite{Bell}. A \emph{relative abundance distribution} provides the inequalities in population size within an ecosystem and therefore an indicator of biodiversity, with the distribution of most biological ecosystems taking a log-normal form \cite{Bell}.}{speciesAbundance} So, for Digital Ecosystems this measures globally the abundance of different solutions relative to one another.

\tfigure{}{SpeciesHistogram}{graph}{Graph of Relative Abundance in the Digital Ecosystem}{Relative abundance \getCap{speciesAbundance} However, \getCap{specAbund2}.}{-7mm}{!t}{}{7mm}

A snapshot of the Agents (organisms) within the Digital Ecosystem, for a typical simulation run, was taken after a thousand user requests, i.e. once it had reached a mature state. In biology a species is a series of populations within which significant gene flow can and does occur, so groups of organisms showing a very similar genetic makeup \cite{lawrence1989hsd}. We therefore chose to define species within Digital Ecosystems similarly, as a grouping of genetically similar digital organisms (based on their semantic descriptions), with no more than 10\% variation within the species group. Relative abundance was calculated for each species and grouped by frequency in Figure \ref{SpeciesHistogram}. In contrast to expectations from biological ecosystems, relative abundance in \setCap{the Digital Ecosystem did not conform to the expected log-normal}{specAbund2} \cite{Bell}. We suggest that the high frequency for the lowest relative abundance was caused by the dynamically re-configurable topology of the Habitat network, which allowed species of small abundance to survive as their respective Habitats were clustered by the Digital Ecosystem. Therefore, it also most likely skewed the other frequencies of the relative abundance measure.

\subsection{Species-Area Relationship}

\setCap{In ecology the \emph{species-area} relationship measures diversity relative to the spatial scale, showing the number of species found in a defined area of a particular habitat or habitats of different areas \cite{sizling2004pls}, and is commonly found to follow a power law in biological ecosystems}{speciesArea} \cite{sizling2004pls}. For Digital Ecosystems this relationship represents how similar solutions are to one another at different Habitat scales. 

\tfigure{}{speciesArea}{graph}{Graph of Species-Area in the Digital Ecosystem}{\getCap{speciesArea}, which the Digital Ecosystem also demonstrates.}{-7mm}{}{}{}

Again, a snapshot of the Agents (organisms) within the Digital Ecosystem, for a typical simulation run, was taken once it had reached a mature state, after a thousand user requests. For this experiment, we assumed each Habitat to have an area of one unit. Then, the number of species, at $n$ randomly chosen Habitats, was measured, where $n$ ranged between one and a hundred. For each $n$, ten sets of measurements were taken at different random sets of Habitats to calculate averaged results, and the $log_{10}$ values of these results are depicted in the graph of Figure \ref{speciesArea}. The distribution of species diversity over a spatial scale in the Digital Ecosystem demonstrates behaviour similar to biological ecosystems, also following a power law \cite{sizling2004pls}. However, diversity at fine spatial scales appears to be lower than predicted by the line of best fit. This may be explained by higher specialisation at some Habitats, making them more like micro-habitats in terms of a reduced species diversity \cite{lawrence1989hsd}.

\section{Summary and Discussion}
\label{endChap2}

We started by reviewing existing \emph{digital ecosystems}, and then introduced biomimicry in computing, \acl{NIC}, to create a definition that could be called the \emph{digital counterpart of biological ecosystems}. Then, by comparing and contrasting the relevant theoretical ecology with the anticipated requirements of Digital Ecosystems, we examined how ecological features may emerge in some systems designed for adaptive problem solving. Specifically, we suggested that Digital Ecosystems, like a biological ecosystems, will consist of self-replicating agents that interact both with one another and with an external environment \cite{begon96}. Population dynamics and evolution, spatial and network interactions, and complex dynamic fitness landscapes will all influence the behaviour of these systems. Many of these properties can be understood via well-known ecological models \cite{MacArthur, Hubbell}, with a further body of theory that treats ecosystems as \acl{CAS} \cite{Levin}. These models provide a theoretical basis for the occurrence of self-organisation, in digital and biological ecosystems, resulting from the interactions among the agents and their environment, leading to complex non-linear behaviour \cite{MacArthur, Hubbell, Levin}; and it is this property that provides the underlying potential for scalable problem-solving in digital environments. Based on the theoretical ecology, we considered fields from the domain of computer science, relevant in the creation of Digital Ecosystems. As we required the digital counterparts for the behaviour and constructs of biological ecosystems, and not their simulation or emulation, we considered parallels using existing and developing technologies to provide their equivalents. This included elements from mobile agent systems \cite{moaspaper} to provide a parallel to the \emph{agents} of biological ecosystems and their \emph{migration} to different habitats, and \acl{DEC} \cite{lin1994cgp} and \aclp{SOA} \cite{soa1w} for the \emph{distribution} and \emph{evolution} of these migrating agents in evolving populations. 

Our efforts culminated in the definition of \aclp{EOA} for the creation of Digital Ecosystems, where the Digital Ecosystem supports the automatic combining of numerous Agents (which represent services), by their interaction in evolving Populations to meet user requests for applications, in a scalable architecture of distributed interconnected Habitats. Agents travel along the peer-to-peer connections; in every node (Habitat) local optimisation is performed through an evolutionary algorithm, where the search space is determined by the Agents present at the node. The sharing of Agents between Habitats ensures the system is scalable, while maintaining a high evolutionary specialisation for each user. The network of interconnected Habitats is equivalent to the physical environment of biological ecosystems \cite{begon96} and - combined with the Agents, the Populations, the Agent migration for \acl{DEC}, and the environmental selection pressures provided by the user base - the union of the Habitats creates the Ecosystem-Oriented Architecture of a Digital Ecosystem. Continuous and varying user requests for applications provide a dynamic evolutionary pressure on the Agent-sequences, which have to evolve to better fulfil those requests, and without which there would be no driving force to the evolutionary self-organisation of the Digital Ecosystem. This represents a novel, cutting-edge approach to \emph{distributed} evolutionary computing, because instead of having multiple populations sharing solutions to find the optimal solution for \emph{one problem}, there are multiple populations to find optimal solutions for \emph{multiple similar problems}. The \emph{business ecosystem} of \aclp{SME} from \aclp{DBE} \cite{dbebkintro} was considered as an example user base, because of their adoption of the ecosystems paradigm, and because our use of \aclp{SOA} in defining Digital Ecosystems predisposes them to \acl{B2B} interaction scenarios. We have also dealt with critical issues which would otherwise cripple our complex system and prevent it from providing a scalable solution, like bloat in evolutionary processes and points-of-failure in networking topologies. In essence, we are making a system greater than the sum of its parts, expected to show emergent and complex behaviour that cannot be predicted until it is created.

In simulation, we compared the Digital Ecosystem's dynamics to those of biological ecosystems. The \emph{ecological succession}, measured by the responsiveness to user requests, conformed to expectations from biological ecosystems \cite{Hubbell}: improving over time, before approaching a plateau. As the evolutionary self-organisation of an ecosystem is a slow process \cite{begon96}, even the accelerated form present in Digital Ecosystems reached only 70\% responsiveness, showing \emph{potential} for improvement. In the \emph{species abundance} experiment the Digital Ecosystem did not conform to the log-normal distribution usually found in biological ecosystems \cite{Bell}. The high frequency for the lowest relative abundance was probably caused by the dynamically re-configurable topology of the Habitat network, which allowed species of small abundance to survive as their Habitats were clustered by the Digital Ecosystem. In the \emph{species-area} experiment, which measures diversity relative to spatial scale, the Digital Ecosystem did follow the power law commonly found in biological ecosystems \cite{sizling2004pls}. The \emph{species diversity} at fine spatial scales was lower than predicted by the line of best fit, and may be explained by the high specialisation at some Habitats, making them more like micro-habitats, including a reduced \emph{species diversity} \cite{lawrence1989hsd}. The \emph{majority} of the experimental results indicate that Digital Ecosystems behave like their biological counterparts, and suggest that incorporating ideas from theoretical ecology can contribute to useful self-organising properties in Digital Ecosystems, which can assist in generating scalable solutions to complex dynamic problems. 

Creating the digital counterpart of biological ecosystems was not without apparent compromises; the temporary \emph{one-to-one genotype-phenotype mapping} for Agents, the information-centric dynamically re-configurable network topology, and the \emph{species abundance} result are inconsistent with biological ecosystems. Initially, any newly deployed Agent has a \emph{one-to-one genotype-phenotype mapping} \cite{shackleton2000irg}, until sufficient \emph{usage (phenotype) information} is amassed for use in \emph{fitness functions}. While the use of such a mapping is undesirable, it is temporary, and necessary to allow the Digital Ecosystem to operate. The Digital Ecosystem requires a re-configurable network topology, to support the constantly changing multi-objective information-centric \emph{selection pressures} of the user base. Hence, using the concept of Hebbian learning \cite{hebb}, Habitat connectivity is dynamically adapted based on the observed migration paths of the Agents within the Habitat network. The dynamically re-configurable network topology probably caused the Digital Ecosystem not to conform, in the \emph{species abundance} experiment, to the log-normal distribution expected from biological ecosystems \cite{Bell}. We would argue that these differences are not compromises, but features unique to Digital Ecosystems. As we discussed earlier, biomimicry, when done well, is not slavish imitation; it is inspiration using the principles which nature has demonstrated to be successful design strategies \cite{biomimicry}. Hypothetically, if there were an abstract definition of an ecosystem, defined as an abstract \emph{ecosystem} class, \setCap{then the \emph{Digital Ecosystem} and \emph{biological ecosystem} classes would both inherit from the abstract \emph{ecosystem} class, but implement its attributes differently}{ecoCapClass}, as shown in the \acl{UML} class diagram of Figure \ref{ecoClass}. \setCap{So, we would argue that the apparent compromises in mimicking biological ecosystems are actually features unique to Digital Ecosystems.}{ecoCap2Class}

\tfigure{scale=1.0}{ecoClass}{graffle}{Hypothetical Abstract Ecosystem Definition}{If there were an abstract ecosystem class in the \acl{UML}, \getCap{ecoCapClass}. \getCap{ecoCap2Class}}{-6mm}{}{}{}

Service-oriented architectures promise to provide potentially huge numbers of services that programmers can combine via standardised interfaces, to create increasingly sophisticated and distributed applications \cite{SOApaper2}. The Digital Ecosystem extends this concept with the automatic combining of available and applicable services in a scalable architecture to meet user requests for applications. This is made possible by a fundamental paradigm shift, from a \emph{pull}-oriented approach to a \emph{push}-oriented approach. So, instead of the \emph{pull}-oriented approach of generating applications only upon request in \aclp{SOA} \cite{singh2005soc}, the Digital Ecosystem follows a \emph{push}-oriented approach of distributing and composing applications pre-emptively, as well as upon request. Although the use of \aclp{SOA} in the definition of Digital Ecosystems provides a predisposition to business \cite{krafzig2004ess}, it does not preclude other more general uses. The \acl{EOA} definition of Digital Ecosystems is intended to be inclusive and interoperable with other technologies, in the same way that the definition of \aclp{SOA} is with \emph{grid computing and other} technologies \cite{singh2005soc}. For example, Habitats could be executed using a distributed processing arrangement, such as \emph{grid computing} \cite{singh2005soc}, which would be possible because the Habitat network topology is information-centric (instead of location-centric).

In this chapter we have determined the fundamentals for a new class of system, Digital Ecosystems, created through combining understanding from theoretical ecology, evolutionary theory, \aclp{MAS}, \acl{DEC}, and \aclp{SOA}. The word \emph{ecosystem} is more than just a metaphor since it is the digital counterpart of biological ecosystems. Therefore, Digital Ecosystems have their desirable properties, such as scalability and self-organisation, and are complex systems that show emergent behaviour, since they are more than the sum of their constituent parts. Once we have further investigated its self-organising properties in the next chapter, Chapter \ref{ch:investigation}, we will attempt its optimisation, for which the experimental results have shown there is \emph{potential}, in the following chapter, Chapter \ref{ch:optimisation}.

\vfill

\pagebreak
\thispagestyle{plain}

\chapter{Investigation of Digital Ecosystems}
\label{ch:investigation}

In this chapter we investigate the self-organising behaviour of Digital Ecosystems, because a primary motivation for our research is to exploit the self-organising properties of biological ecosystems. Over time a biological ecosystem becomes increasingly self-organised through the process of \emph{ecological succession}, driven by the evolutionary self-organisation of the populations within the ecosystem. Analogously, a Digital Ecosystem's increasing self-organisation comes from the Agent Populations within being evolved to meet the dynamic \emph{selection pressures} created by the requests from the user base. We start by discussing the relevant literature on self-organisation, including the philosophical meaning of \emph{organisation} and of \emph{self}, before focusing on its application to evolving Agent Populations. The self-organisation of biological ecosystems is often defined in terms of the \emph{complexity}, \emph{stability}, and \emph{diversity}. So, we studied further to extend a definition for the \emph{complexity}, grounded in the biological sciences, called Physical Complexity; based on statistical physics, automata theory, and information theory, providing a measure of the quantity of information in an organism's genome, by calculating the entropy in a population to determine the randomness in the genome. Next, we investigate and extend a definition for the \emph{stability}, originating from the computer sciences, called Chli-DeWilde stability, which views a \acl{MAS} as a discrete time Markov chain with potentially unknown transition probabilities. With a \acl{MAS} being considered stable when its state has converged to an equilibrium distribution. Finally, we investigate a definition for the \emph{diversity}, relative to the \emph{selection pressures} provided by the user requests, considering the \emph{collective} self-organised \emph{diversity} of the evolving Agent Populations relative to the \emph{global} user request behaviour. We conclude with a summary and discussion of the achievements, including the experimental results.

\section{Background Theory}

Self-organisation is perhaps one of the most desirable features in the systems that we design, and a primary motivation for our research in Digital Ecosystems is the desire to exploit the self-organising properties of biological ecosystems \cite{Levin}, which are thought to be robust, scalable architectures that can automatically solve complex, dynamic problems. Over time a biological ecosystem becomes increasingly self-organised through the process of \emph{ecological succession} \cite{begon96}, driven by the evolutionary self-organisation of the populations within the ecosystem. Analogously, a Digital Ecosystem's increasing self-organisation comes from the Agent Populations within being evolved to meet the dynamic \emph{selection pressures} created by the requests from the user base. The self-organisation of biological ecosystems is often defined in terms of the \emph{complexity}, \emph{stability}, and \emph{diversity} \cite{king1983cda}, which we will also apply to our Digital Ecosystems.

It is important for us to be able to understand, model, and define self-organising behaviour, determining macroscopic variables to characterise this self-organising behaviour of the order constructing processes within, the evolving Agent Populations. However, existing definitions of self-organisation may not be directly applicable, because evolving Agent Populations possess properties of both computing systems (e.g. agent systems) as well as biological systems (e.g. population dynamics), and the combination of these properties makes them unique. So, to determine definitions for the self-organising \emph{complexity}, \emph{stability}, and \emph{diversity} we will start by considering the available literature on self-organisation, for its general properties, its application to \aclp{MAS} (the dominant technology in Digital Ecosystems), and its application to our evolving Agent Populations.

\subsection{Self-Organisation}

Self-organisation has been around since the late 1940s \cite{ashby}, but has escaped general formalisation despite many attempts \cite{nicolis, kohonen1989soa}. There have instead been many notions and definitions of self-organisation, useful within their different contexts \cite{heylighen2002sso}. They have come from cybernetics \cite{ashby, beer1966dac, heylighen2001cas}, thermodynamics \cite{nicolis}, mathematics \cite{lendaris1964dso}, information theory \cite{shalizi2001cac}, synergetics \cite{haken1977sin}, and other domains \cite{lehn1990psc}. The term \emph{self-organising} is widely used, but there is no generally accepted meaning, as the abundance of definitions would suggest. Therefore, the philosophy of self-organisation is complicated, because organisation has different meanings to different people. So, we would argue that any definition of self-organisation is context dependent, in the same way that a choice of statistical measure is dependent on the data being analysed.

Proposing a definition for self-organisation faces the \emph{cybernetics} problem of defining \emph{system}, the \emph{cognitive} problem of \emph{perspective}, the \emph{philosophical} problem of defining \emph{self}, and the \emph{context} dependent problem of defining \emph{organisation} \cite{gershenson}.

The \emph{system} in this context is an \emph{evolving Agent Population}, with the replication of individuals from one generation to the next, the recombination of the individuals, and a selection pressure providing a differential fitness between the individuals, which is behaviour common to any evolving population \cite{begon96}.

\emph{Perspective} can be defined as the perception of the observer in perceiving the self-organisation of a system \cite{ashby1962pso, beer1966dac}, matching the intuitive definition of \emph{I will know it when I see it} \cite{shalizi}, which despite making formalisation difficult shows that organisation is \emph{perspective} dependent (i.e. relative to the \emph{context} in which it occurs). In the \emph{context} of an evolutionary system, the observer does not exist in the traditional sense, but is the \emph{selection pressure} imposed by the environment, which \emph{selects} individuals of the population over others based on their \emph{observable} \emph{fitness}. Therefore, consistent with the theoretical biology \cite{begon96}, in an evolutionary system the self-organisation of its population is from the \emph{perspective} of its environment. 

Whether a system is \emph{self}-organising or being organised depends on whether the process causing the organisation is an internal component of the system under consideration. This intuitively makes sense, and therefore requires one to define the boundaries of the system being considered to determine if the force causing the organisation is internal or external to the system. For an evolving population the force leading to its organisation is the \emph{selection pressure} acting upon it \cite{begon96}, which is formed by the environment of the population's existence and competition between the individuals of the population \cite{begon96}. As these are internal components of an evolving Agent Population \cite{begon96}, it is a self-organising system.

Now that we have defined, for an evolving Agent Population, the \emph{system} for which its \emph{organisation} is \emph{context} dependent, the \emph{perspective} to which it is relative, and the \emph{self} by which it is caused, a definition for its \emph{self-organisation} can be considered. The \emph{context}, an evolving Agent Population in its environment, lacks a 2D or 3D metric space, so it is necessary to consider a visualisation in a more abstract form. We will let a single square, \square{-5mm}{White}{-3mm}, represent an Agent, with colours to represent different Agents. Agent-sequences will therefore be represented by a sequence of coloured squares, \square{-3.5mm}{RedGreenBlue}{-2.5mm}, with a Population consisting of multiple Agent-sequences, as shown in Figure \ref{sampleAgentPopulation}.

\tfigure{scale=1.0}{sampleAgentPopulation}{graffle}{Visualisation of Self-Organisation in Evolving Agent Populations}{The \getCap{visOrgCap}}{-2mm}{}{}{}

In Figure \ref{sampleAgentPopulation} the \setCap{number of Agents, in total and of each colour, is the same in both populations. However, the Agent Population on the left intuitively shows organisation through the uniformity of the colours across the Agent-sequences, whereas the population to the right shows little or no organisation.}{visOrgCap} Following biological ecosystems, which defines self-organisation in terms of the \emph{complexity}, \emph{stability}, and \emph{diversity} relative to the \emph{perspective} of the \emph{selection pressure} \cite{king1983cda}: the self-organised \emph{complexity} of the system is the creation of coherent patterns and structures from the Agents, the self-organised \emph{stability} of the system is the resulting stability or instability that emerges over time in these coherent patterns and structures, and the self-organised \emph{diversity} of the system is the optimal variability within these coherent patterns and structures.

\subsection{Definitions of Self-Organisation}

Many alternative definitions have been proposed for self-organisation within populations and agent systems, with each defining what property or properties demonstrate self-organisation. So, we will now consider the most applicable alternatives for their suitability in defining the self-organised \emph{complexity}, \emph{stability}, and \emph{diversity} of an evolving Agent Population. 

One possibility would be the $\in$-machine definition of evolving populations, which models the emergence of organisation in pre-biotic evolutionary systems \cite{crutchfield2006}. An $\in$-machine consists of a set of causal states and transitions between them, with symbols of an alphabet labelling the transitions and consisting of two parts: an input symbol that determines which transition to take from a state, and an output symbol which is emitted on taking that transition \cite{crutchfield2006}. $\in$-machines have several key properties \cite{crutchfieldYoung1989}: all their recurrent states form a single, strongly connected component, their transitions are deterministic in the specific sense that a causal state with the edge symbol-pair determines the successor state, and an $\in$-machine is the smallest causal representation of the transformation it implements. The $\in$-machine definition of self-organisation also identifies the forms of \emph{complexity}, \emph{stability}, and \emph{diversity} \cite{crutchfield2006}, but with definitions focused on pre-biotic evolutionary systems, i.e. the \emph{primordial soup} of \emph{chemical replicators} from the \emph{origin of life} \cite{rasmussen2004}. \emph{Complexity} is defined as a form of structural-complexity, measuring the state-machine-based information content of the $\in$-machine individuals of a population \cite{crutchfield2006}. \emph{Stability} is defined as a meta-machine, a set (composition) of $\in$-machines, that can be regarded as an autonomous and self-replicating entity \cite{crutchfield2006}. \emph{Diversity} is defined, using an interaction network, as the variability of interaction in a population \cite{crutchfield2006}. So, while these definitions of self-organisation are compatible at the higher more abstract level, i.e. in the forms of self-organisation present, the deeper definitions of these forms are not applicable because they are context dependent. As we explained in the previous subsection, definitions of self-organisation are context dependent, and so the context of pre-biotic evolutionary systems, to which the $\in$-machine self-organisation applies, is very different to the context of an evolving Agent Population from our Digital Ecosystem. Evolving Agent Populations are defined from \aclp{EOA}, which have evolutionarily surpassed the context of pre-biotic evolutionary systems, shown by the necessity of our consideration of the later evolutionary stage of \emph{ecological succession} \cite{begon96} in section \ref{secSuccession}.

The \emph{\acl{MDL}} principle \cite{barron} could be applied to the \emph{executable components} or \emph{semantic descriptions} of the Agent-sequences of a Population, with the best model, among a collection of tentatively suggested ones, being the one that provides the smallest \emph{stochastic complexity}. However, the \acl{MDL} principle does not define how to select the family of model classes to be applied for determining the stochastic complexity \cite{hansen2001msa}. This problem of model selection is well known and cannot be adequately formalised, and so in practise selection is based on human judgement and prior knowledge of the kinds of models previously chosen \cite{hansen2001msa}. Therefore, while models could be chosen to represent the self-organised \emph{complexity}, and possibly even the \emph{diversity}, there is no procedural method for determining these models, because \emph{subjective} human intervention is required for model selection on a case-by-case basis.

The \emph{Pr{\"u}gel-Bennett Shapiro formalism} models the evolutionary dynamics of a population of sequences, using techniques from statistical mechanics and focuses on replica symmetry \cite{prugel}. The individual sequences are not considered directly, but in terms of the statistical properties of the population, using a macroscopic level of description with specific statistical properties to characterise the population, that are called \emph{macroscopics}. A macroscopic formulation of an evolving population reduces the huge number of degrees of freedom to the dynamics of a few quantities, because a non-linear system of a few degrees of freedom can be readily solved or numerically iterated \cite{prugel}. However, since a macroscopic description disregards a significant amount of information, subjective human insight is essential so that the appropriate macroscopics are chosen \cite{shapiro2001smt}. So, while macroscopics could be chosen to represent the self-organised \emph{complexity}, \emph{stability}, and \emph{diversity}, there is no procedural method for determining these macroscopics, because \emph{subjective} human insight is required for macroscopic selection on a case-by-case basis.

\emph{\acl{KC}} complexity defines the complexity of binary sequences by the smallest possible \acl{UTM}, algorithm (programme and input) that produces the sequence \cite{li1997ikc}. A sequence is said to be \emph{regular} if the algorithm necessary to produce it on a \acl{UTM} is shorter than the sequence itself \cite{li1997ikc}. A \emph{regular} sequence is said to be \emph{compressible}, whereas its compression, into the most succinct \acl{UTM} possible, is said to be \emph{incompressible} as it cannot be reduced any further in length \cite{li1997ikc}. A \emph{random} sequence is said to be \emph{incompressible}, because the \acl{UTM} to represent it cannot be shorter than the random sequence itself \cite{li1997ikc}. This intuitively makes sense for algorithmic complexity, because algorithmically regular sequences require a shorter programme to produce them. So, when measuring a population of sequences, the \acl{KC} complexity would be the shortest \acl{UTM} to produce the entire population of sequences. However, Chaitin himself has considered the application of \acl{KC} complexity to evolutionary systems, and realised that although \acl{KC} complexity represents a satisfactory definition of randomness in algorithmic information theory, it is not so useful in biology \cite{chaitin}. For evolving Agent Populations the problem manifests itself most significantly when the Agents are randomly distributed within the Agent-sequences of the Population, having maximum \acl{KC} complexity, instead of the complexity it ought to have of zero. This property makes \acl{KC} complexity unsuitable as a definition for the self-organised \emph{complexity} of an evolving Agent Population.

A definition called \emph{Physical Complexity} can be estimated for a population of sequences, calculated from the difference between the maximal entropy of the population, and the actual entropy of the population when in its environment \cite{adami20002}. This Physical Complexity, based on Shannon's entropy of information, measures the information in the population about its environment, and therefore is conditional on its environment. It can be estimated by counting the number of loci that are fixed for the sequences of a population \cite{adami1998ial}. Physical Complexity would therefore be suitable as a definition of the self-organised \emph{complexity}. However, a possible limitation is that Physical Complexity is currently only formulated for populations of sequences with the same length. 

\emph{\acl{SOC}} in evolution is defined as a punctuated equilibrium in which the population's \emph{critical state} occurs when the fitness of the individuals is uniform, and for which an \emph{avalanche}, caused by the appearance and spread of advantageous mutations within the population, temporarily disrupts the uniformity of individual fitness across the population \cite{bak1988soc}. Whether an evolutionary process displays \acl{SOC} remains unclear. There are those who claim that \acl{SOC} is demonstrated by the available fossil data \cite{sneppen}, with a power law distribution on the lifetimes of genera drawn from fossil records, and by artificial life simulations \cite{adami1995}, again with a power law distribution on the lifetimes of competing species. However, there are those who feel that the fossil data is inconclusive, and that the artificial life simulations do not show \acl{SOC}, because the key power law behaviour in both can be generated by models without \acl{SOC} \cite{newman1996soc}. Also, the \acl{SOC} does not define the resulting self-organised \emph{stability} of the population, only the organisation of the events (avalanches) that occur in the population over time.

\emph{\acl{EGT}} \cite{weibull1995egt} is the application of models inspired from \emph{population genetics} to the area of \emph{game theory}, which differs from \emph{classical game theory} \cite{fudenberg1991gt} by focusing on the dynamics of strategy change more than the properties of individual strategies. In \acl{EGT}, agents of a population play a game, but instead of optimising over strategic alternatives, they inherit a fixed strategy and then replicate depending on the strategy's payoff (fitness) \cite{weibull1995egt}. The self-organisation found in \acl{EGT} is the presence of stable steady states, in which the genotype frequencies of the population cease to change over the generations. This equilibrium is reached when all the strategies have the same expected payoff, and is called a stable steady state, because a slight perturbing will not cause a move far from the state. An \emph{\acl{ESS}} leads to a stronger asymptotically stable state, as a slight perturbing causes only a temporary move away from the state before returning \cite{weibull1995egt}. So, \acl{EGT} is focused on genetic \emph{stability} between competing between individuals, rather than the stability of the population as a whole, which therefore limits its suitability for the self-organised \emph{stability} of an evolving Agent Population.

\aclp{MAS} are the dominant computational technology in the evolving Agent Populations, and while there are several definitions of self-organisation \cite{parunak, mamei2003som, tianfield2005tso, dimarzoserugendo2006som} and stability \cite{moreau2005sms, weiss1999msm, olfatisaber2007cac} defined for \aclp{MAS}, they are not applicable primarily because of the evolutionary dynamics inherent in the context of evolving Agent Populations. Whereas Chli-DeWilde stability of \aclp{MAS} \cite{chli2} may be suitable, because it models \aclp{MAS} as Markov chains, which are an established modelling approach in evolutionary computing \cite{rudolph1998fmc}. A \acl{MAS} is viewed as a discrete time Markov chain with potentially unknown transition probabilities, in which the agents are modelled as Markov processes, and is considered to be \emph{stable} when its state has converged to an equilibrium distribution \cite{chli2}. Chli-DeWilde stability provides a strong notion of self-organised \emph{stability} over time, but a possible limitation is that its current formulation does not support the necessary evolutionary dynamics.

The main concept in \emph{\acl{MFT}} is that for any single particle the most important contribution to its interactions comes from its neighbouring particles \cite{parisi1998sft}. Therefore, a particle's behaviour can be approximated by relying upon the \emph{mean field} created by its neighbouring particles \cite{parisi1998sft}, and so \acl{MFT} could be suitable as a definition for the self-organised diversity of an evolving Agent Population. Naturally, it requires a \emph{neighbourhood model} to define interaction between neighbours \cite{parisi1998sft}, and is therefore easily applied to domains such as Cellular Automata \cite{gutowitz}. While a \emph{neighbourhood model} is feasible for biological populations \cite{flyvbjerg}, evolving Agent Populations lack such \emph{neighbourhood models} based on a 2D or 3D metric space, with the only available \emph{neighbourhood model} being a distance measure on a parameter space measuring dissimilarity. However, this type of \emph{neighbourhood model} cannot represent the information-based interactions between the individuals of an evolving Agent Population, making \acl{MFT} unsuitable as a definition for the self-organised \emph{diversity} of an evolving Agent Population.

\section{Complexity}

A definition for the self-organised \emph{complexity} of an evolving Agent Population should define the creation of coherent patterns and structures from the Agents within, with no initial constraints from modelling approaches for the inclusion of pre-defined specific behaviour, but capable of representing the appearance of such behaviour should it occur.

None of the proposed definitions are directly applicable for the self-organised \emph{complexity} of an evolving Agent Population. The $\in$-machine modelling \cite{crutchfield2006} is not applicable, because it is only defined within the context of pre-biotic populations. Neither is the \acl{MDL} principle \cite{barron} or the Pr{\"u}gel-Bennett Shapiro formalism \cite{prugel}, because they require the involvement of \emph{subjective} human judgement at the critical stage of model and quantifier selection \cite{hansen2001msa, shapiro2001smt}. \acl{KC} complexity \cite{chaitin} is also not applicable as randomness is given maximum complexity.

Physical Complexity \cite{adami20002} fulfils abstractly the required definition for the self-organised \emph{complexity} of an evolving Agent Population, estimating complexity based upon the individuals of a population within the context of their environment. However, its current formulation is problematic, primarily because it is only defined for populations of fixed length, but as this is not a fundamental property of its definition \cite{adami20002} it should be feasible to redefine and extend it as needed. So, the use of Physical Complexity as a definition for the self-organised \emph{complexity} of evolving Agent Populations will be investigated further to determine its suitability.

\subsection{Physical Complexity}
\label{measureSelfOrg}

Physical Complexity was born \cite{adami1998ial} from the need to determine the proportion of information in sequences of DNA, because it has long been established \cite{thomasjr1971goc} that the information contained is not directly proportional to the length, known as the C-value enigma/paradox \cite{gregory2001cco}. Understanding DNA requires knowing the environment (context) in which it exists, which may initially appear obvious as DNA is considered to be the \emph{language of life} \cite{searls2002lg} and the purpose of life is to procreate or replicate \cite{dawkins2006sg}. Virtually all activities of biological life-forms are towards this aim \cite{dawkins2006sg}, with a few exceptions (e.g. suicide before procreation), and to achieve replication requires resources, energy and matter to be harvested \cite{marais1999seo}. So, for any individual the environment represents the problem of extracting energy for replication, and so their DNA sequence represents a solution to this problem. Furthermore, an individual DNA \emph{solution} is not necessarily a simple inverse of the \emph{problem} that the environment represents, with forms of life having evolved specialised, specific and effective ways (niches) to acquire the necessary energy and matter for replication \cite{lawrence1989hsd}. Even with this understanding it would seem we still need to define the environment to be able to distinguish the information from the redundancy in a solution (DNA sequence). However, because Physical Complexity analyses a group of solutions to the same problem, the consistency between the different solutions shows the information, and the differences the redundancy \cite{adami2003}. Entropy, a measure of disorder \cite{vonbertalanffy1973gst}, is used to determine the redundancy from the information in a population of solutions. Physical Complexity therefore provides a context-relative definition for the self-organised \emph{complexity} of a population without needing to define the context (environment) explicitly \cite{adami2000}.

\label{defPhyCom}
Physical Complexity was derived \cite{adami2000} from the notion of \emph{conditional complexity} defined by Kolmogorov, which is different from traditional Kolmogorov complexity and states that the determination of complexity of a sequence is conditional on the environment in which the sequence is interpreted \cite{li1997ikc}. In contrast, traditional \ac{KC} complexity is only conditional on the implicit rules of mathematics necessary to interpret a programme on the tape of a \ac{TM}, and nothing else \cite{li1997ikc}. So, if we consider a \ac{TM} that takes a tape $e$ as input (which represents its physical environment), including the particular rules of mathematics of this \emph{world}; without such a tape, this \ac{TM} is incapable of computing anything, except for writing to the output what it reads in the input. Thus, without tape $e$ all sequences $s$ have maximal \ac{KC}-complexity, because there is nothing by which to determine regularity \cite{adami2000}. However, \emph{conditional complexity} can be stated as the length of the smallest programme that computes sequence $s$ from an environment $e$,
\begin{equation}
K(s|e) = \min \left\{ {|p|:s = C_T(p,e)} \right\},
\label{komCom}
\end{equation}
where $C_T(p,e)$ denotes the result of running programme $p$ on \acl{TM} $T$ with the input sequence $e$ \cite{adami2000}. This is not yet Physical Complexity, but rather, it is the smallest programme that computes the sequence $s$ from an environment $e$, in the limit of sequences of infinite length, containing only the bits that are entirely unrelated to $e$, since, if they were not, they could be obtained from $e$ with a programme of a size tending to zero \cite{adami2000}. The Physical Complexity $K(s:e)$ can now be defined as the number of bits that are meaningful in sequence $s$ (that can be obtained from $e$ with a programme of vanishing size), and is given by the \emph{mutual complexity} \cite{kolmogorov},
\begin{equation}
K(s:e) = K(s|\emptyset ) - K(s|e), 
\label{komNot}
\end{equation}
where $K(s|\emptyset)$ is the unconditional complexity with an empty input tape, $e \equiv \emptyset$ \cite{adami2000}. This is different from the Kolmogorov complexity, because in Kolmogorov's construction the rules of mathematics were given to the \ac{TM} \cite{li1997ikc}. As argued above, every sequence $s$ is random if no environment $e$ is specified, as non-randomness can only exist for a specific world or environment. Thus, $K(s|\emptyset )$ is always maximal,
\begin{equation}
K(s|\emptyset ) = |s|,
\label{komEmpty}
\end{equation}
and is given by the length of $s$ \cite{adami2000}. So (\ref{komNot}) represents the length of the sequence $s$, minus those bits that cannot be obtained from $e$. So, conversely (\ref{komNot}) represents the number of bits that can be obtained in a sequence $s$, by a computation with vanishing programme size, from $e$. Thus, $K(s: e)$ represents the Physical Complexity of $s$ \cite{adami2000}. The determination of the Physical Complexity, $K(s: e)$, of a sequence $s$ with a description of the environment $e$ is not practical. Meaning that it cannot generally be determined by inspection, because its impossible to determine which, and how many, of the bits of sequence $s$ correspond to information about the environment $e$. The reason is that we are generally unaware of the coding used to code information about $e$ in $s$, and therefore coding and non-coding bits look entirely alike \cite{adami2000}. However, it is possible to distinguish coding from non-coding bits if we are given multiple copies of sequences that have adapted to the environment, or more generally, if a statistical ensemble (population) of sequences is available to us. Then, coding bits are revealed by non-uniform probability distributions across the population (\emph{conserved sites}), whereas random bits have uniform distributions (\emph{volatile sites}) \cite{adami2000}. The determination of complexity then becomes an exercise in information theory, because the average complexity $\langle K \rangle$, in the limit of infinitely long strings, tends to the entropy of the ensemble of strings $S$\footnote{This holds for near-optimal codings. For strings $s$ that do not code perfectly we have $\langle K \rangle \ge H$ \cite{zurek1990aic}.} \cite{adami2000-14},
\begin{equation}
 \left\langle {K(s)} \right\rangle _S = \sum\limits_{s \in S} {p(s)K(s)} \approx H(S),
\label{avgCompPart1}
\end{equation}
where $H$ is defined from Shannon's (information) entropy \cite{mackay}, and is given by
\begin{equation}
H(S)=log_{n}(S),
\end{equation}
where $n$ is the number of symbols available for encoding. If each symbol is equally probable, we can rewrite the above function as
\begin{eqnarray}
H(S)&=&-log_{n}(1/S) \nonumber \\
&=&-log_{n}(p),
\end{eqnarray}
where $p$ is the probability of occurrence of any one of the symbols. For a source that outputs an infinite sequence of bits, to communicate a finite set of symbols $S$, Shannon generalised the above function to express an average symbol length \cite{mackay}. This derivation is easier to see for a large, but finite, number of symbols $N$,
\begin{eqnarray}
 H(S) &=& \frac{{\sum\limits_{i = 1}^S {N_i \left[ { - \log_{N} (1/S_i )} \right]} }}{{\sum\limits_{i = 1}^S {N_i } }} = \frac{{\sum\limits_{i = 1}^S {N_i \left[ { - \log_{N} (1/S_i )} \right]} }}{N} \nonumber \\
\white{.} & \white{.} \nonumber \\
&=& - \sum\limits_{i = 1}^S {\frac{{N_i }}{N}\left[ {\log_{N} (1/S_i )} \right]} = - \sum\limits_{i = 1}^S {p_i \log_{N} (p_i )},
\label{Hderivation}
\end{eqnarray} 
where $N_i$ is the number of occurrences of the symbol $S_i$. So, given (\ref{avgCompPart1}) and (\ref{Hderivation}), the \emph{average complexity} of the sequences $s$ of a population $S$, $\left\langle {K(s)} \right\rangle _S$, tends to the entropy of the sequences $s$ in the ensemble $S$ \cite{adami2000},
\begin{equation}
 \left\langle {K(s)} \right\rangle _S = -\sum\limits_{s \in S} {p(s)\log p(s)}.
\label{avgComp}
\end{equation}

(\ref{avgComp}) remains consistent with (\ref{komEmpty}) as the determination of $K(s|\emptyset )$, sequence $s$ without an environment $e$, must equal the sequence's length $|s|$, because Shannon's formula for entropy is an average logarithmic measure of the symbol sets \cite{mackay}, and so the maximum entropy of a population is equivalent to the length of the sequences in the population, $H_{max}(S) = |s|$. Indeed, if nothing is known about the environment to which a sequence $s$ pertains, then according to the \emph{principle of indifference}\footnote{The \emph{principle of indifference} states that if there are $n>1$ mutually exclusive and collectively exhaustive possibilities, which are indistinguishable except for their names then each possibility should be assigned an equal probability $\frac{1}{n}$ \cite{jaynes2003ptl}.}, the probability distribution $p(s)$ must be uniformly random. However, if an environment $e$ is given we have some information about the system, and the probability distribution will be nonuniform. Indeed, it can be shown that for every probability distribution $p(s|e)$, to find sequence $s$ given environment $e$, we have
\begin{equation}
H(S|e) \le H(S|\emptyset ) = |s|,
\label{shannonEntropy}
\end{equation}
because of the concavity of Shannon entropy \cite{adami2000}. So, the difference between the maximal entropy $H(S|\emptyset ) = |s|$ and $H(S|e)$, according to the construction outlined above, represents the average number of bits in sequence $s$ taken from the population $S$ that can be obtained by zero-length universal programmes from the environment $e$. Therefore, the average mutual complexity of sequences $s$ in a population $S$, given an environment $e$, is	
\begin{eqnarray}
\left\langle {K(s:e)} \right\rangle _S &=& \sum\limits_{s \in S} {p(s)K(s:e)} \nonumber \\
&\approx& H(S|\emptyset ) - H(S|e) \nonumber \\
&\equiv& I(S|e),
\label{avgInfo}
\end{eqnarray}
where $I(S|e)$ is the information about the environment $e$ stored in the population $S$, which we identify with the Physical Complexity \cite{adami2000}. To estimate $I(S|e)$ it is necessary to estimate the entropy $H(S|e)$ using a representative population of sequences $S$ for a given environment $e$, by summing, over the sequences $s$ of the population $S$, the probability $p(s|e)$ multiplied by the logarithm of the probability $p(s|e)$,
\begin{equation}
H(S|e) = - \sum\limits_{s \in S} {p(s|e)\log p(s|e)}.
\label{popEntropy}
\end{equation}
The entropy $H(S|e)$ can be estimated by summing the per-site $H(i)$ entropies of the sequence,
\begin{equation}
H(S|e) \approx \sum\limits_{i = 1}^{|s|} {H(i)},
\label{estEntropySumSite}
\end{equation}
where $i$ is a site in the sequence $s$ \cite{adami2000}. Random sites are identified by a nearly uniform probability distribution, and contribute positively to the entropy, whereas non-random sites (which have strongly peaked distributions) contribute very little \cite{adami2000}. So, the Physical Complexity, the average mutual complexity of sequences $s$ in a population $S$ for an environment $e$, $\left\langle {K(s:e)} \right\rangle _S$, abbreviated as $C$, is the maximal entropy $H(S|\emptyset )$ minus the sum of the per-site entropies,
\begin{equation}
C = H(S|\emptyset ) - \sum\limits_{i = 1}^{|s|} {H(i)}.
\label{avgCom}
\end{equation}
If the sequences $s$ are constructed from an alphabet, a set $D$, then the per site entropy $H(i)$ for the sequences is
\begin{equation}
H(i) = - \sum\limits_{d \in D} {p_d (i)\log _{|D|} p_d (i)}, \\
\label{persite}
\end{equation}
where $i$ is a site in the sequences ranging between one and the length of the sequences $\ell$, $D$ is the alphabet of characters found in the sequences, and $p_d(i)$ is the probability that site $i$ (in the sequences) takes on character $d$ from the alphabet $D$, with the sum of the $p_d(i)$ probabilities for each site $i$ equalling one, $\sum\limits_{d \in D} {p_d (i) = 1}$ \cite{adami2000}. Taking the log to the base $|D|$ conveniently normalises $H(i)$ to range between zero and one,
\begin{equation}
0 \le H(i) \le 1.
\label{persiteMinMax}
\end{equation}
If the site $i$ is identical across the population it will have no entropy,
\begin{equation}
H_{\min } (i) = 0.
\label{persiteMin}
\end{equation}
If the content of site $i$ is uniformly random, i.e. the $p_d (i)$ probabilities all equal to $\frac{1}{{|D|}}$, it will have maximum entropy,
\begin{equation}
H_{\max } (i) = 1.
\label{persiteMax}
\end{equation}
When the entropy of $H(i)$ is at its minimum of zero, then the site $i$ holds information, as every sample shows the same character of the alphabet. When the entropy of $H(i)$ is at its maximum of one, the character found in the site $i$ is uniformly random and therefore holds no information. So, the amount of information is the maximal entropy of the site (\ref{persiteMax}) minus the actual per-site entropy (\ref{persite}) \cite{adami2000},
\begin{eqnarray}
 I(i) &=& H_{\max } (i) - H(i) \nonumber \\
 &=& 1 - H(i) .
\label{persiteInfo}
\end{eqnarray}

\tfigure{width=170mm}{sampleDNAsequencesOrig}{numbers}{DNA Samples from a Population}{\getCap{DNAcap}.}{-7mm}{!b}{}{}

DNA, whose sequence encodes the genetic information of living organisms \cite{lawrence1989hsd}, was the original driver for the creation of Physical Complexity \cite{adami1998ial}, and so is a good example upon which to demonstrate the definition. \setCap{DNA sequences are made up from four nucleotides, Adenosine (A), Thymine (T), Cytosine (C) and Guanine (G). The nucleotides always pair as follows, Adenosine with Thymine, and Cytosine with Guanine. So, DNA sequences can be reduced to a \emph{genome sequence} showing half of the paired information \cite{lawrence1989hsd}}{DNAcap}, and with a sufficiently sized sample population, the $p_d (i)$ probabilities can be estimated by the frequencies of the nucleotides at the sites. Considering the genome samples, in Figure \ref{sampleDNAsequencesOrig}, the per-site entropy for site 11 will have maximum entropy, as the nucleotides (characters of the alphabet) all have equal probability,

\vspace{-7mm}

\begin{eqnarray*}
H(11) &=& - \sum\limits_{d \in D}^{A,T,C,G} {p_d (11)\log _{|D|} p_d (11)} \\
&=& - \left( {\frac{1}{4}\log _4 \frac{1}{4} + \frac{1}{4}\log _4 \frac{1}{4} + \frac{1}{4}\log _4 \frac{1}{4} + \frac{1}{4}\log _4 \frac{1}{4}} \right) = 1,
\end{eqnarray*}

given that the alphabet $D$ equals $\{ A, T, C, G\}$, and that the probabilities all equal a quarter, $p_A (11) = p_T (11) = p_C (11) = p_G (11) = \frac{1}{4}$. As the per-site entropy (randomness) is maximum, the information content is its minimum of zero,

\vspace{-10mm}

\begin{equation*}
I(11) = 1 - H(11) = 0.
\end{equation*}

This intuitively makes sense, as it states that if the site content is random across the population, then it contains no information. At the other extreme, if we calculate the per-site entropy for site 16 in Figure \ref{sampleDNAsequencesOrig}, it will have no entropy,

\vspace{-7mm}

\begin{eqnarray*}
H(16) &=& - \sum\limits_{d \in D}^{A,T,C,G} {p_d (16)\log _{|D|} p_d (16)} \\ 
&=& - \left( {0\log _4 0 + 1\log _4 1 + 0\log _4 0 + 0\log _4 0} \right) = 0,
\end{eqnarray*}

as the nucleotide Thymine has a probability of one, $p_T (16)=1$, while the other three nucleotides have a probability of zero, $p_A (16) = p_C (16) = p_T (16) = 0$. As the per-site entropy is minimum, the information content is its maximum of one,

\vspace{-5mm}

\begin{equation*}
 I(16) = 1 - H(16) = 1.
\end{equation*}

This also intuitively makes sense, as it states that if the site is identical across the entire population (no randomness), then the site holds definitive information. Finally, the per-site entropy for site 19 is at neither extreme, but is entropically in the middle,
\begin{eqnarray*}
H(19) &=& - \sum\limits_{d \in D}^{A,T,C,G} {p_d (19)\log _{|D|} p_d (19)} \\ 
&=& - \left( {0\log _4 0 + 0\log _4 0 + \frac{1}{2}\log _4 \frac{1}{2} + \frac{1}{2}\log _4 \frac{1}{2}} \right) = \frac{1}{2},
\end{eqnarray*}
as the probabilities $p_A (19)=p_T (19)=0$, and $p_C (19)=p_G (19)=\frac{1}{2}$. Intuitively, this states that if there is some entropy (randomness) in the samples of the site, then there is only partial information, 
\begin{equation*}
I(19) = 1 - H(19) = \frac{1}{2}.
\end{equation*}

For clarity the length of the sequences $|s|$ will be abbreviated to $\ell$ \cite{adami2000},
\begin{equation}
|s| \equiv \ell .
\label{clarity}
\end{equation}
So, the complexity of a population $S$, of sequences $s$, is the maximal entropy of the population (equivalent to the length of the sequences) $\ell$, minus the sum, over the length $\ell$, of the per-site entropies $H(i)$,
\begin{equation}
C = \ell - \sum\limits_{i = 1}^\ell {H(i)}, 
\label{complexity}
\end{equation}
given (\ref{avgCom}), (\ref{shannonEntropy}) and (\ref{clarity}) \cite{adami2000}. The equivalence of the maximum complexity to the length matches the intuitive understanding that if a population of sequences of length $\ell$ has no redundancy, then their complexity is their length $\ell$.

If $G$ represents the set of all possible genotypes constructed from an alphabet $D$ that are of length $\ell$, then the size (cardinality) of $|G|$ is equal to the size of the alphabet $|D|$ raised to the length $\ell$,
\begin{equation}
|G| = |D|^\ell.
\label{recPopSize}
\end{equation}
For the complexity measure to be accurate, a	sample size of $|D|^\ell$ is suggested to minimise the error \cite{adami2000, basharin}, but such a large quantity can be computationally infeasible. The definition's creator, for practical applications, chooses a population size of roughly $1.29|D|\ell $ \cite{adami20002}. We suggest that a population size of $|D|\ell $ is sufficient to show any trends present, but that the population size will fluctuate when simulated, and so a population size slightly larger than $|D|\ell $ is chosen for simulations to ensure that the necessary minimum of $|D|\ell $ is maintained. So, for a population of sequences $S$ we choose, with the definition's creator, a computationally feasible population size of $|D|$ times $\ell $,
\begin{equation}
|S|\ \ge \ |D|\ell.
\label{popSize}
\end{equation}
The size of the alphabet, $|D|$, depends on the domain to which Physical Complexity is applied. For RNA the alphabet is the four nucleotides, $D = \{A, C, G, U\}$, and therefore $|D|=4$ \cite{adami2000}. When Physical Complexity was applied to the Avida simulation software, there was an alphabet size of twenty-eight, $|D|=28$, as that was the size of the instruction set for the self-replicating programmes \cite{adami20002}.

\subsection{Extending to Agent Populations}

Reformulating Physical Complexity for an evolving Agent Population requires consideration of the following issues: the mapping of the sequence sites to the Agent-sequences, the managing of Populations of variable length sequences, and the non-atomicity of Agents leading to clustering within populations.

\subsubsection{Mapping Sequence Sites}
The first concern is mapping the Population's Agent-sequences to the sequence \emph{sites} of Physical Complexity, with the intuitive approach being to map the \emph{sites} to the Agents, because they are the functional unit of processing, the base unit for evolution in the evolving Agent Populations. Physical Complexity has been applied to RNA sequences \cite{adami2000}, and populations of self-replicating programmes in the artificial life simulator Avida \cite{adami20032}; for the RNA the \emph{sites} were mapped to the nucleotides from which it is constructed, and for the artificial life simulator the \emph{sites} were mapped to the programme instructions which made up the self-replicating programmes. So, the only alternative, of mapping the \emph{sites} to the Agents, would be mapping to the programme instructions of the \emph{executable components} of the services that the Agents represent, similarly to the populations of self-replicating programmes in the artificial life simulator Avida \cite{adami20032}. However, mapping to the \emph{executable components} in the evolving Agent Populations would be like mapping to the binary representation of the instruction set in the Avida simulator, or to the molecules that make up the nucleotides in RNA, which in all cases would be unsuitable as they are the components that make up the functional units, and not the functional units themselves. Therefore, mapping the sequence \emph{sites} of Physical Complexity to the Agents is the most suitable approach for evolving Agent Populations.

\subsubsection{Variable Length Sequences}

Physical Complexity is currently formulated for a population of sequences of the same length \cite{adami2000}, and so we will now investigate an extension to include populations of \aclp{vls}, which will include Populations of variable length Agent-sequences. This will require changing and re-justifying the fundamental assumptions, specifically the conditions and limits upon which Physical Complexity operates. In (\ref{complexity}) the Physical Complexity, $C$, is defined for a population of sequences of length $\ell$ \cite{adami2000}. The most important question is what does the length $\ell$ equal if the population of sequences is of variable length? The issue is what $\ell$ represents, which is the maximum possible complexity for the population \cite{adami2000}, which we will call the \emph{complexity potential} $C_P$. The maximum complexity in (\ref{complexity}) occurs when the per-site entropies sum to zero, $\sum\limits_{i = 1}^\ell {H(i)} \to 0$, as there is no randomness in the sites (all contain information), i.e. $C \to \ell$ \cite{adami2000}. So, the \emph{complexity potential} equals the length,
\begin{equation}
C_P = \ell,
\label{comPot2} 
\end{equation}
provided the population $S$ is of sufficient size for accurate calculations, as found in (\ref{popSize}), i.e. $|S|$ is equal or greater than $|D|\ell$. For a population of \aclp{vls}, $S_{V}$, the complexity potential, $C_{V_P}$, cannot be equivalent to the length $\ell$, because it does not exist. However, given the concept of minimum sample size from (\ref{popSize}), there is a length for a population of \aclp{vls}, $\ell_V$, between the minimum and maximum length, such that the number of per-site samples up to and including $\ell_V$ is sufficient for the per-site entropies to be calculated. So the \emph{complexity potential} for a population of \aclp{vls}, $C_{V_P}$, will be equivalent to its \emph{calculable} length, 
\begin{equation}
\label{potential}
C_{V_P} = \ell_V.
\end{equation}

\tfigure{width=170mm}{sampleDNAsequences}{numbers}{Alternative DNA Samples of the Population}{from Figure \ref{sampleDNAsequencesOrig}: Calculating the entropy for site 19 provides a value of zero, but as evident \getCap{genCap} So, having the length of a population of \aclp{vls} \getCap{genCap2}.}{-7mm}{}{}{}

If $\ell_V$ where to be equal to the length of the longest individual(s) $\ell _{max}$ in a population of \aclp{vls} $S_{V}$, then the operational problem is that for some of the later sites, between one and $ \ell _{max}$, the sample size will be less than the population size $|S_{V}|$. So, having the length $\ell_v$ \setCap{equalling the maximum length would be incorrect}{genCap2}, as there would be an insufficient number of samples at the later sites, and therefore $\ell _V \not\equiv \ell _{max}$. Consider the alternative samples of DNA sequences shown in Figure \ref{sampleDNAsequences}; if the entropy is calculated again for site 19, $H(19) = 0$, but \setCap{there is an insufficient sample size for the estimated probabilities to provide an accurate calculation.}{genCap} Therefore, the length for a population of \aclp{vls}, $\ell _V $, is the highest value within the range of the minimum (one) and maximum length, $1 \le \ell _V \le \ell _{\max } $, for which there are sufficient samples to calculate the entropy. A function which provides the sample size at a given site is required to specify the value of $\ell_V$ precisely,
\begin{equation}
sampleSize(i\ :site)\ :int,
\end{equation}
where the output varies between $1$ and the population size $|S_{V}|$ (inclusive). Therefore, the length of a population of \aclp{vls}, $\ell_{V}$, is the highest value within the range of one and the maximum length for which the sample size is greater than or equal to the alphabet size multiplied by the length $\ell _{V}$,
\begin{equation}
sampleSize(\ell _V ) \ge |D|\ell _V \wedge sampleSize(\ell _V + 1) < |D|\ell _V,
\label{lengthVLP}
\end{equation}
where $\ell _V$ is the length for a population of \aclp{vls}, and $\ell _{max} $ is the maximum length in a population of \aclp{vls}, $\ell _V$ varies between $ 1 \le \ell _V \le \ell _{max }$, $D$ is the alphabet and $|D| > 0$. This definition intrinsically includes a minimum size for populations of \aclp{vls}, $|D|\ell _V$, and therefore is the counterpart of (\ref{popSize}), which is the minimum population size for populations of fixed length.

The length $\ell$ used in the limits of (\ref{persite}) no longer exists, and therefore (\ref{persite}) must be updated; so, the per-site entropy calculation for \aclp{vls} will be denoted by $H_{V}(i)$, and is, 
\begin{equation}
H_V (i) = - \sum\limits_{d \in D} {p_d (i)\log _{|D|} p_d (i)},
\label{perSiteVLP}
\end{equation}
where $D$ is still the alphabet, $\ell _V$ is the length for a population of \aclp{vls}, with the site $i$ now ranging between $ 1 \le i \le \ell _V $, while the $p_d (i)$ probabilities still range between $ 0 \le p_d (i) \le 1$, and still sum to one. It remains algebraically almost identical to (\ref{persite}), but the conditions and constraints of its use will change, specifically $\ell$ is replaced by $\ell_{V}$. Naturally, $H_{V}(i)$ ranges between zero and one, as did $H(i)$ in (\ref{persiteMinMax}),
\begin{equation}
0 \le H_V (i) \le 1,
\label{perSiteVLPrange}
\end{equation}
where $i$ ranges between $1$ and $\ell _V$, and so the condition on the site $i$ changes at the upper limit from $\ell$ to $\ell_{V}$. As before in (\ref{persiteMin}), if a site $i$ is identical across the population, it will have no entropy,
\begin{equation}
H_{V_{min}} (i) = 0,
\label{perSiteVLPrangeMin}
\end{equation}
where, again, $i$ ranges between $1$ and $\ell _V$. Analogously to (\ref{persiteMax}), if the $p_{d}(i)$ probabilities in (\ref{perSiteVLP}) are equal, then the site $i$ has maximum entropy. In effect, the content of the site is uniformly random and therefore
\begin{equation}
H_{V_{max}} (i) = 1
\label{perSiteVLPrangeMax}
\end{equation}
is true for all $i$, where the $p_d (i)$ probabilities are $\frac{1}{{|D|}}$, and where $i$ continues to range between $1$ and $\ell _V$. Analogously to (\ref{persiteInfo}), when the entropy is its minimum of zero then the site $i$ holds information, as every sample shows the same character of the alphabet. However, when the entropy is maximum the character found in the site $i$ is uniformly random, and therefore holds no information. So, the amount of information is the maximal entropy of the site (\ref{perSiteVLPrangeMax}) minus the actual per-site entropy (\ref{perSiteVLP}),
\begin{eqnarray}
 I_V (i) &=& H_{V_{\max } } (i) - H_V (i) \nonumber \\
&=& 1 - H_V (i),
\end{eqnarray}
where $i$ again now ranges between $1$ and $\ell _V$. Therefore, the complexity for a population of \aclp{vls}, $C_{V}$, is the \emph{complexity potential} of the population of \aclp{vls} minus the sum, over the length of the population of \aclp{vls}, of the per-site entropies (\ref{perSiteVLP}),

\vspace{-20mm}

\begin{equation}
\label{newComplexity}
C_{V} = \ell _V - \sum\limits_{i = 1}^{\ell _V } {H_V (i)},
\end{equation}

\vspace{-10mm}

where $\ell_V$ is the length for the population of \aclp{vls}, and $H_V(i)$ is the entropy for a site $i$ in the population of \aclp{vls}. 

\begin{figure}
	\vspace{3mm}
	\centering	
	\execute{cd images; ./pdfcrop.sh abstractAgentPopulation}
			\ifthenelse{\boolean{final}}
				{\includegraphics[scale=1.0]{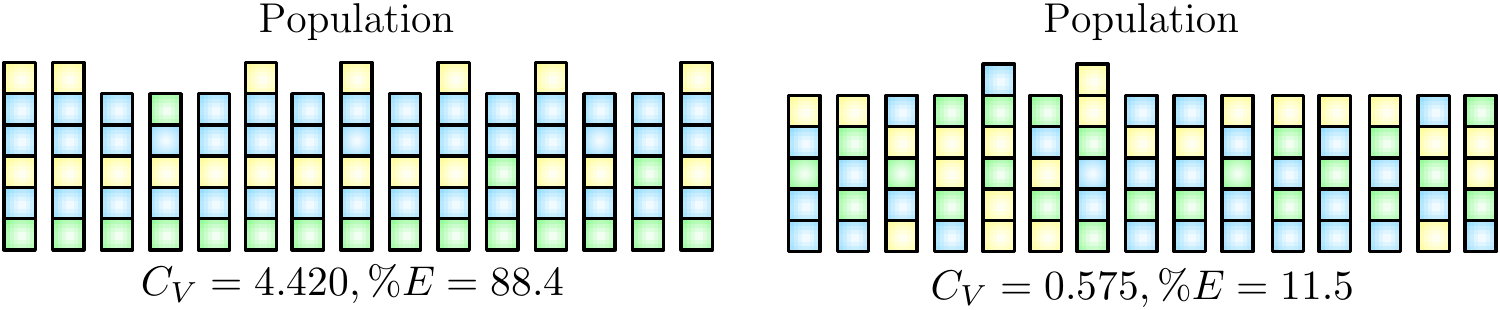}}
				{\href{file://localhost/Users/g/Desktop/PhDthesis/images/abstractAgentPopulation.graffle}{\includegraphics[scale=1.0]{images/abstractAgentPopulation-crop.pdf}}}
	\vspace{-2mm}
	\caption[Abstract Visualisation for Populations of Variable Length Sequences]
		{
		\label{orgCompPop}
\italics{
Abstract Visualisation for Populations of Variable Length Sequences: The alphabet $D$ is the set \{\squaresub{-2mm}{Yellow}{-2mm}{-2pt} , \squaresub{-3mm}{Green}{-1mm}{-2pt}, \squaresub{-3mm}{Purple}{-1mm}{-2pt}\} , the maximum length $\ell _{max}$ is $6$ and the length for populations of \aclp{vls} $\ell _V$ is calculated as $5$ from (\ref{lengthVLP}). The Physical Complexity and Efficiency values \getCap{orgCPcap}.
}
		}
	\end{figure}

Physical Complexity can now be applied to populations of \aclp{vls}, so we will consider the abstract example populations in Figure \ref{orgCompPop}. We will let a single square, \square{-5mm}{White}{-3mm}\nolinebreak[4],\linebreak represent a site $i$ in the sequences, with different colours to represent the different values. Therefore, a sequence of sites will be represented by a sequence of coloured squares, \square{-6mm}{YellowGreenPurple}{-3.5mm}.\linebreak Furthermore, the alphabet $D$ is the set \{\squaresub{-3.5mm}{Yellow}{-1.5mm}{-2pt}, \squaresub{-4mm}{Green}{-1.5mm}{-2pt}, \squaresub{-4mm}{Purple}{-1.5mm}{-2pt}\}, the maximum length $\ell _{max}$ is 6 and the \linebreak length for populations of \aclp{vls} $\ell _V$ is calculated as 5 from (\ref{lengthVLP}). The Physical Complexity values in Figure \ref{orgCompPop} \setCap{are consistent with the intuitive understanding one would have for the self-organised \emph{complexity} of the sample populations}{orgCPcap}; the population with high Physical Complexity has a little randomness, while the population with low Physical Complexity is almost entirely random. 

Using our extended Physical Complexity we can construct a measure showing the use of the information space, called the Efficiency $E$, which is calculated by the Physical Complexity $C_{V}$ over the complexity potential $C_{V_P }$,
\begin{equation}
E = \frac{{C_V }}{{C_{V_P } }}.
\label{efficiencyEQ}
\end{equation}
The Efficiency $E$ will range between zero and one,
\begin{equation}
0 \le E \le 1,
\label{efficienyRange}
\end{equation}
only reaching its maximum of one when the actual complexity $C_{V}$ equals the complexity potential $C_{V_{P}}$, indicating that there is no randomness in the population. In Figure \ref{orgCompPop} the populations of sequences are shown with their respective Efficiency values as percentages, and the values are as one would expect.

The complexity $C_{V}$ (\ref{newComplexity}) is an absolute measure, whereas the Efficiency $E$ (\ref{efficiencyEQ}) is a relative measure (based on the complexity $C_{V}$). So, the Efficiency $E$ can be used to compare the self-organised complexity of populations, independent of their size, their length, and whether their lengths are variable or not (as its equally applicable to the fixed length populations of the original Physical Complexity).

\subsubsection{Clustering}
\label{cluster123}

The \emph{self-organised complexity} of an evolving Agent Population is the \emph{clustering}, amassing of same or similar sequences, around the optimum genome \cite{begon96}. This can be visualised on a \emph{fitness landscape} \cite{wright1932}, which \setCap{shows the combination space (power set) of the alphabet $D$ against the fitness values from the \emph{selection pressure} (user request)}{FLsingleCap}. \setCap{The Agent-sequences of an evolving Population will evolve, moving across the \emph{fitness landscape} and clustering around the optimal genome at the peak of the global optimum}{FLsingle2Cap}, assuming that its evolutionary process does not become trapped while clustering over local optima, and as shown in Figure \ref{fitLandSingleConverted}.

\tfigure{scale=1.0}{fitLandSingleConverted}{ai}{3D Fitness Landscape with a Global Optimum}{This \getCap{FLsingleCap}, resulting in a global optimum. \getCap{FLsingle2Cap}.}{-3mm}{!h}{5mm}{}

Clustering is indicated by the Efficiency $E$ tending to its maximum, as the population's Physical Complexity $C_{V}$ tends to the \emph{complexity potential} $C_{V_{P}}$, because an optimal sequence is becoming dominant in the population, and therefore increasing the uniformity of the sites across the population. With a global optimum, the Efficiency $E$ tends to a maximum of one, indicating that the \emph{evolving population of sequences} is tending to a \emph{set of clusters $T$ of size one},
\begin{equation} 
E = \frac{C_V}{C_{V_P}} \to 1 \ as\ |T| \to 1,
\label{clusters1}
\end{equation}
assuming its evolutionary process does not become trapped at local optima. So, the \emph{tending} of the Efficiency $E$ provides a \emph{clustering coefficient}. It \emph{tends}, never quite reaching its maximum, because of the mutation inherent in the evolutionary process.

\tfigure{scale=1.0}{fitLandFlatConverted}{graffle}{3D Fitness Landscape with No Optimum}{Theoretical extreme scenario in which the selection pressure is non-discriminating. So, the population occupancy of the fitness landscape would then be \getCap{FLflatCap}.}{-3mm}{!h}{5mm}{}

The other extreme scenario occurs when the number of clusters equals the size of the population, which would only occur with a \emph{flat fitness landscape} \cite{kimura:ntm} resulting from a non-discriminating \emph{selection pressure}, as shown in Figure \ref{fitLandFlatConverted}. The population occupancy is \setCap{uniformly random, as any position (sequence) has the same fitness as any other. So the entropy (randomness) tends to maximum, resulting in the complexity $C_{V}$ tending to zero, and therefore the Efficiency $E$ also tending to zero}{FLflatCap}, while the number of clusters $|T|$ tends to the number of sequences in the population $|S|$,
\begin{eqnarray}
E = \frac{{C_V }}{{C_{V_P } }} \to 0 \ as \ |T| \to |S|.
\label{clusters0}
\label{maxClusters}
\end{eqnarray}
So the number of clusters $|T|$ tends to the population size $|S|$, with each cluster consisting of only one unique sequence (individual).

\tfigure{scale=1.0}{fitLandMultipleConverted}{ai}{3D Fitness Landscape with Global Optima}{Clustering scenario, in which the Efficiency $E$ of the population $S$ tends to a value based on the number of clusters $|T|$, because the population of sequences is clustering around more than one global optima, with each cluster having an Efficiency $E$ tending to a maximum of one.}{-3mm}{}{}{}

If there are global optima, as there are in Figure \ref{fitLandMultipleConverted}, the \setCap{Efficiency $E$ will tend to a maximum below one, because the population of sequences consists of more than one cluster, with each having an Efficiency tending to a maximum of one.}{FLmul2Cap} \setCap{The simplest scenario of clusters is \emph{pure clusters}}{FLmulCap}; \emph{pure} meaning that each cluster uses a distinct (mutually exclusive) subset of the alphabet $D$ relative to any other cluster. In this scenario the Efficiency $E$ tends to a value based on the number of clusters $|T|$, because a \emph{number} of the $p_d (i)$ probabilities at each \emph{site} in (\ref{perSiteVLP}) are the reciprocal of the number of clusters, $\frac{1}{|T|}$. So, given that the \emph{number} of the $p_d (i)$ probabilities taking the value $\frac{1}{|T|}$ is equal to the number of clusters, while the other $p_d (i)$ probabilities take a value of zero, then the per-site entropy calculation of $H_V (i)$ from (\ref{perSiteVLP}) becomes
\begin{equation}
H_V (i) = \log _{|D|} |T|,
\label{calcNumClusters}
\end{equation}
where $i$ is the site, $|D|$ is the alphabet size, and $|T|$ is the number of clusters. Hence, given (\ref{calcNumClusters}), (\ref{newComplexity}), and (\ref{potential}), then the Efficiency $E$ from (\ref{efficiencyEQ}) becomes 
\begin{equation}
E \to 1 - (\log _{|D|} |T|),
\label{calcNumClusters2}
\end{equation}
where $|D|$ is the alphabet size and $|T|$ is the number of clusters. Therefore, the Efficiency $E$, the \emph{clustering coefficient}, tends to a value that can be used to determine the number of \emph{pure clusters} in an evolving population of sequences.

For a population $S$ with clusters, each cluster is a sub-population with an Efficiency $E$ tending to a maximum of one. To specify this relationship we require a function that provides the Efficiency $E$ (\ref{efficiencyEQ}) of a population or sub-population of sequences,
\begin{equation}
\mbox{\emph{efficiency(input :population) :int}}.
\end{equation}
So, for a population $S$ consisting of a set of clusters $T$, each member (cluster) $t$ is therefore a sub-population of the population $S$, and is defined as
\begin{equation}
t \in T \to \left(t \subseteq S \wedge \mbox{\textit{efficiency(t)}} \to 1 \wedge |t| \approx \frac{|S|}{|T|} \wedge \sum\limits_{t \in T} {|t|} = |S|\right),
\label{defineCluster}
\end{equation}
where a cluster $t$ has an Efficiency $E$ tending to a maximum of one, and the cluster size $|t|$ is approximately equal to the population size $|S|$ divided by the number of clusters $|T|$. It is only \emph{approximately equal} because of variation from mutation, and because the population size may not divide to a whole number. These conditions are true for all members $t$ of the set of clusters $T$, and therefore the summation of the cluster sizes $|t|$ equals the size of the population $|S|$.

The population of sequences from the \emph{fitness landscape} of Figure \ref{fitLandMultipleConverted} is visualised in Figure \ref{populationHiddenClusters}, but the clusters within cannot be seen. So, the population is arranged to show the clustering in Figure \ref{populationWithClusters}, in which the two clusters are clearly evident. \setCap{The clusters of the population have Efficiency values tending to a maximum of one, compared to the Efficiency of the population as a whole, which is tending to a maximum significantly below one.}{popClusShowCap} This is the expected behaviour of clusters as defined in (\ref{defineCluster}).

\tfigure{scale=1.0}{populationHiddenClusters}{graffle}{Population with Hidden Clusters}{Visualisation for the population of sequences from the fitness landscape of Figure \ref{fitLandMultipleConverted}, with clusters visually hard to identify. The clusters lead to a low complexity $C_V$ relative to the maximum $C_{V_{P}}$, and hence the Efficiency $E$ (\ref{efficiencyEQ}) tends to a maximum significantly below one.}{-2mm}{!h}{}{}

\tfigure{scale=1.0}{populationWithClusters}{graffle}{Population with Clusters Visible}{Visualisation for the population of sequences from Figure \ref{populationHiddenClusters}, which has been arranged to show the clusters present. \getCap{popClusShowCap}}{-2mm}{!h}{2mm}{}

The population size $|S|$, in Figures \ref{populationHiddenClusters} and \ref{populationWithClusters}, is double the minimum requirement specified in (\ref{lengthVLP}), so that the complexity $C_{V}$ (\ref{newComplexity}) and Efficiency $E$ (\ref{efficiencyEQ}) could be used in defining the principles of clustering without redefining the \emph{length of a population of \aclp{vls}} $\ell_{V}$ (\ref{lengthVLP}). However, when determining the variable length $\ell_{V}$ of a cluster $t$, the sample size requirement is different, specifically a cluster $t$ is a sub-population of $S$, and therefore by definition cannot have a population size equivalent to $S$ (unless the population consists of only one cluster). Therefore, to manage clusters requires a reformulation of $\ell _V$ (\ref{lengthVLP}) to
\begin{equation}
\ell _V = \left(sampleSize(\ell _V ) \approx \frac{|D|\ell _V }{|T|} \wedge sampleSize(\ell_{V} + 1) < \frac{|D|\ell _V }{|T|}\right),
\label{clustersSampleSize}
\end{equation}
where $\ell _{max} $ is the maximum length in a population of \aclp{vls}, $\ell _V$ varies between $ 1 \le \ell _V \le \ell _{max }$, $D$ is the alphabet, $|D| > 0$, and $T$ is the set of clusters in the population $S$.

A population with clusters will always have an Efficiency $E$ tending towards a maximum significantly below one. Therefore, managing populations with clusters requires a reformulation of the Efficiency (\ref{efficiencyEQ}) to
\begin{equation}
E_{c} (S) = \left\{
\begin{array}{cl}
\frac{C_V }{C_{V_P}} & \mbox{if $|T| = 1$} \\
\vspace{-3mm}\white{.} & \white{.}\\
\frac{\sum\limits_{t \in T} {E_{c} (t)}}{|T|} & \mbox{if $|T| > 1$}
\end{array}\right.,
\label{efficiencyMultiple}
\end{equation}
where $t$ is a cluster, and a member of the set of clusters $T$ of the population $S$. So, the Efficiency $E_{c}$ is equivalent to the Efficiency $E$ if the population consists of only one cluster, but if there are clusters then the Efficiency $E_{c}$ is the average of the Efficiency $E$ values of the clusters.

\tfigure{scale=1.0}{atomicity}{graffle}{Agent Atomicity}{Property \getCap{atomCap} \getCap{atom2Cap}. In this example, the alphabet is non-atomic, with the yellow Agent able to functionally replace a green blue Agent-sequence.}{-5mm}{!h}{5mm}{}

\label{atomicitySection} 
\emph{Atomicity} is the property \setCap{of a set of Agents, such that no single Agent can functionally replace any Agent-sequence, i.e. their functionality is \emph{mutually exclusive} to one another. It is important because non-atomicity can adversely affect}{atomCap} the uniformity of the calculated per-site entropies, which is the main construct of \setCap{the Physical Complexity measure}{atom2Cap}, and so non-atomicity risks introducing error when calculating the information content. Our extensions to Physical Complexity to support clustering are also necessary to manage non-atomicity, because it leads to the formation of clusters within evolving Agent Populations. The presence of clusters can be identified by the \emph{clustering coefficient}, the Efficiency $E$ tending to a value below one, with the Efficiency $E_{c}$ (\ref{efficiencyMultiple}) used to calculate the actual Efficiency as it supports clustering and therefore non-atomicity.

\tfigure{scale=1.0}{nonAtomicPopulation}{graffle}{Population Constructed from a Non-Atomic Alphabet}{The population is constructed from the alphabet shown in Figure \ref{atomicity}, with the yellow Agent able to functionally replace a green blue Agent-sequence. So, the \getCap{popAtomCap}.}{-2mm}{}{}{}

If we consider the example population shown in Figure \ref{nonAtomicPopulation}, which is constructed from the alphabet shown in Figure \ref{atomicity}, the yellow Agent \square{-4.25mm}{Yellow}{-3.25mm} can functionally replace a green blue \linebreak Agent-sequence \square{-3.5mm}{GreenBlue}{-3mm}, and so the uniformity across site two is lost. Therefore, the \setCap{Efficiency $E$ of the population is a half, whereas the Efficiency $E_{c}$ for populations with clusters is one, because it supports clustering and therefore non-atomicity}{popAtomCap}.

\subsection{Simulation and Results}
\label{specs}

We simulated an evolving Agent Population from the Digital Ecosystem, using our simulation from section \ref{simRes} (unless otherwise specified), seeded with an alphabet (Agent-pool) of 15 Agents for the evolutionary process. We also added the classes and methods necessary to calculate our extended Physical Complexity and Efficiency, which required implementing the $C_V$ of (\ref{newComplexity}), the $\ell_V$ of (\ref{clustersSampleSize}) and the $H_V$ of (\ref{perSiteVLP}) for the per-site entropies. The Efficiency $E_{c}$ (\ref{efficiencyMultiple}), for populations with clusters, was also implemented in the simulation.

\subsubsection{Physical Complexity}

Our extended Physical Complexity has the same structure and properties as the original Physical Complexity \cite{adami2000}, and so the relationship between fitness and our extended Physical Complexity should be the same as the relationship between fitness and the original Physical Complexity \cite{adami20002}. If we consider the original Physical Complexity and fitness graphs, reprinted from \cite{adami20002} in Figure \ref{origPhyComUpd}, we can define the relationship as follows; the \setCap{Physical Complexity increases over the generations, suffering short-term decreases from the arrival of \emph{fitter} mutants, which spread through the population over several generations and causes the uniformity of the sites to decrease temporarily, while the maximum fitness of the population increases over the generations until the global optimum is reached}{phyComGraphCap}, provided that there is a static selection pressure and a low mutation rate (making it unlikely that the maximum fitness will decrease) \cite{adami20002}. The original Physical Complexity starts uncharacteristically high in Figure \ref{origPhyComUpd}, because the population is seeded with a single sequence that temporarily takes over the population \cite{adami20002}.

\tfigure{width=170mm}{origPhyComUpd}{graffle}{Original Physical Complexity Graphs}{(reprinted from \cite{adami20002}): The \getCap{phyComGraphCap}.}{-8mm}{}{}{7mm}

\vspace{7mm}
Figure \ref{phycom} shows, for a typical evolving Agent Population, the Physical Complexity $C_V$ (\ref{newComplexity}) for \aclp{vls} and the \emph{maximum fitness} $F_{max}$ over the generations. \setCap{The Physical Complexity for \aclp{vls} increases over the generations, showing short-term decreases as expected \cite{adami20002}. It increases over the generations because of the increasing information being stored, with the sharp increases occurring when the effective length $\ell_{V}$ of the Population increases.}{graph2cap} The temporary decreases, such as the one beginning at generation 138, are preceded by the advent of a new \emph{fitter} mutant, as indicated by a corresponding sharp increase in the \emph{maximum fitness} in the immediately preceding generations, which temporarily disrupt the self-organised complexity of the population, until this new fitter mutant becomes dominant and leads to a new higher level of self-organised complexity. Figure \ref{phycom} shows that the fitness and our extended Physical Complexity; both increase over the generations, synchronised with one another, until generation 160 when the \emph{maximum fitness} tapers off more slowly than the Physical Complexity. At this point the optimal length for the sequences is reached within the simulation, and so the advent of new fitter sequences (of the same of similar length) creates only minor fluctuations in the Physical Complexity, while having a more significant effect on the \emph{maximum fitness}.

\tfigure{}{phycom}{graph}{Graph of Physical Complexity and Maximum Fitness}{over the Generations: \getCap{graph2cap}}{-7mm}{}{}{}

The similarity of the graph in Figure \ref{phycom} to the graphs in Figure \ref{origPhyComUpd} confirms that the Physical Complexity measure has been successfully extended to \aclp{vls}. The temporary decreases in the Physical Complexity $C_{V}$ for \aclp{vls} were not as severe as the original \cite{adami20002}, because our simulation's mutation rate was relatively low at only 10\%. Also, our Physical Complexity $C_{V}$ does not start uncharacteristically high like the original, because at the start the entire population was randomly seeded, instead of being seeded by just a single individual as in the original \cite{adami20002}.

\subsubsection{Efficiency}

Figure \ref{newphycomvis} is a visualisation of the simulation, showing two alternate Populations that were run for a thousand generations, with the one on the left from Figure \ref{phycom} run under normal conditions, while the one on the right was run with a non-discriminating selection pressure; each multi-coloured line represents an Agent-sequence, while each colour represents an Agent (site). \setCap{The visualisation shows that our Efficiency $E$ accurately measures the self-organised complexity of the two Populations.}{largeVisCap} It also shows significant variation in the Population run under normal conditions, as the evolutionary computing process creates the opportunity to find fitter (better) sequences, providing potential to avoid getting trapped at local optima.

\tfigure{width=170mm}{newphycomvis}{graffle}{Visualisation of Evolving Agent Populations}{at the 1000th Generation: Each multi-coloured line represents an Agent-sequence, while each colour represents an Agent (site). The population on the left from Figure \ref{phycom} was run under normal conditions, while the one on the right was run with a non-discriminating selection pressure. \getCap{largeVisCap}}{-8mm}{!h}{3mm}{}

\tfigure{}{efficiency}{graph}{Graph of Population Efficiency over the Generations}{for \getCap{graph3cap}: \getCap{graph32cap}.}{-7mm}{}{}{}

Figure \ref{efficiency} shows the Efficiency $E$ (\ref{efficiencyEQ}), over the generations, for \setCap{the Population from Figure \ref{phycom}}{graph3cap}. \setCap{The Efficiency tends to a maximum of one, indicating that the Population consists of one cluster, which is confirmed by the visualisation of the Population in Figure \ref{newphycomvis} (left). The significant decreases that occurred in the Efficiency, reducing in magnitude and frequency over the generations, came from mirroring the fluctuations that occurred in the complexity $C_V$}{graph32cap}, because the Efficiency $E$ (\ref{efficiencyEQ}) is the complexity $C_V$ (\ref{newComplexity}) over the complexity potential $C_{V_{P}}$(\ref{potential}). These falls are caused by the creation of fitter (better) mutants within the population, which eventually become the dominant genotype, but during the process causes the Physical Complexity and the Efficiency to fall in the short-term.

\subsubsection{Clustering}

\tfigure{}{coefficient}{graph}{Graph of the Clustering Coefficient over the Generations}{The Efficiency oscillated \getCap{graph4cap} It tended to 0.744, as expected from (\ref{calcNumClusters2}) given the alphabet size was fifteen, $|D|$=15, and the number of clusters was two, $|T|$=2, indicating more than one cluster.}{-7mm}{}{}{}

\tfigure{width=170mm}{newCluster}{graffle}{Visualisation of Clusters in an Evolving Agent Population}{at the 1000th Generation: The \getCap{visClustersCap}, and as \getCap{visClusters2Cap} \getCap{visClusters3Cap} complexity correctly.}{-7mm}{!b}{-10mm}{}

To further investigate the self-organised \emph{complexity} of evolving Agent Populations, we simulated a typical Population with a multi-objective \emph{selection pressure} that had two independent global optima (like the fitness landscape of Figure of \ref{fitLandMultipleConverted}), and so the potential to support two \emph{pure clusters} (each cluster using a unique subset of the alphabet $D$). The graph in Figure \ref{coefficient} shows the Efficiency $E$ over the generations acting as a \emph{clustering coefficient}, oscillating \setCap{around the included best fit curve, quite significantly at the start, and then decreasing as the generations progressed.}{graph4cap} The initial severe oscillations were caused by the creation and spread of fitter \emph{longer} mutants (Agent-sequences) in the Population, causing the Physical Complexity and therefore the Efficiency to fluctuate significantly. The Efficiency tended to 0.744, as expected from (\ref{calcNumClusters2}) given the alphabet size was fifteen, $|D|$=15, and the number of clusters was two, $|T|$=2. The tending itself indicated clustering, while the value it tended to indicated, as expected, the presence of two clusters in the Population. A visualisation of the Population is shown in Figure \ref{newCluster}, in which the \setCap{Agent-sequences were grouped to show the two clusters}{visClustersCap}. As \setCap{expected from (\ref{defineCluster}) each cluster had a much higher Physical Complexity and Efficiency compared to the Population as a whole. However, the Efficiency $E_{c}$}{visClusters2Cap} is immune to the clusters and therefore \setCap{calculated the}{visClusters3Cap} self-organised \emph{complexity} of the Population correctly.

\subsection{Summary}

None of the existing definitions we considered \cite{crutchfield2006, barron, prugel, chaitin, adami20002} were directly applicable as a definition for the self-organised \emph{complexity} of an evolving Agent Population, but the properties of Physical Complexity \cite{adami20002} closely matched our intuitive understanding, and so was chosen for further investigation. Based upon information theory and entropy, it provides a measure of the quantity of information in a population's genome, relative to the environment in which it evolves, by calculating the entropy in the population to determine the randomness in the genome \cite{adami20002}. Reformulating Physical Complexity for evolving Agent Populations required consideration of the following issues: the mapping of the sequence sites to the Agent-sequences, and the managing of Populations of variable length sequences. We then built upon this to construct a variant of the Physical Complexity called the \emph{Efficiency}, because it was based on the efficiency of information storage in Physical Complexity, which we then used to develop an understanding of clustering and atomicity within evolving Agent Populations.

We then investigated the self-organised \emph{complexity} of evolving Agent Populations through experimental simulations, for which our extended Physical Complexity was consistent with the original. We then investigated the \emph{Efficiency}, which performed as expected, confirmed by the numerical results and Population visualisations matching our intuitive understanding. We then applied the \emph{Efficiency} to the determination of clusters when subjecting an evolving Agent Population to a multi-objective \emph{selection pressure}. The numerical results, combined with the visualisation of the multi-cluster Population, confirmed the ability of the \emph{Efficiency} to act as a \emph{clustering coefficient}, not only indicating the occurrence of clustering, but also the number of clusters (for pure clusters). We also confirmed that the \emph{Efficiency $E_{c}$ for populations with clusters} was able to calculate correctly the self-organised \emph{complexity} of evolving Agent Populations with clusters.

Collectively, the experimental results confirm that Physical Complexity has been successfully extended to evolving Agent Populations. Most significantly, Physical Complexity has been reformulated algebraically for populations of \aclp{vls}, which we have confirmed experimentally through simulations. Our \emph{Efficiency} definition provides a macroscopic value to characterise the level of \emph{complexity}. Furthermore, the \emph{clustering coefficient} defined by the tending of the Efficiency, not only indicates clustering, but can also distinguish between a single cluster population and a population with clusters. The number of clusters can even be determined, for \emph{pure clusters}, from the value to which the \emph{clustering coefficient} tends. Combined, this allows the \emph{Efficiency} $E_{c}$ definition to provide a normalised \emph{universally applicable} macroscopic value to characterise the \emph{complexity} of a population, independent of clustering, atomicity, length (variable or same), and size.
 
We have determined an effective understanding and quantification for the self-organised \emph{complexity} of the evolving Agent Populations of our Digital Ecosystem. Furthermore, the understanding and techniques we have developed have applicability beyond evolving Agent Populations, as wide as the original Physical Complexity, which has been applied from \acs{DNA} \cite{adami2000} to simulations of self-replicating programmes \cite{adami20032}.

\section{Stability} 

A definition for the self-organised \emph{stability} of an evolving Agent Population should define the resulting stability or instability that emerges over time, with no initial constraints from modelling approaches for the inclusion of pre-defined specific behaviour, but capable of representing the appearance of such behaviour should it occur.

None of the proposed definitions are directly applicable for the self-organised \emph{stability} of an evolving Agent Population. The $\in$-machine modelling \cite{crutchfield2006} is not applicable, because it is only defined within the context of pre-biotic populations. The Pr{\"u}gel-Bennett Shapiro formalism \cite{prugel} is not suitable, because it necessitates the involvement of \emph{subjective} human judgement at the critical stage of quantifier selection. \acl{SOC} \cite{bak1988soc} is also not applicable as it only models the events of genetic change in the population over time, rather than measuring the resulting stability or instability of the population. Neither is \acl{EGT} \cite{weibull1995egt}, which only defines the \emph{genetic stability} of the genotypes, in terms of \emph{equilibrium} and \emph{non-equilibrium} dynamics, instead of the stability of the population as a whole. 

Chli-DeWilde stability of \aclp{MAS} \cite{chli2} does fulfil the required definition of the self-organised \emph{stability}, measuring convergence to an equilibrium distribution. However, its current formulation does not include \aclp{MAS} that make use of \emph{evolutionary computing} algorithms, i.e. our evolving Agent Populations, but it could be extended to include such \aclp{MAS}, because its \emph{Markov-based modelling} approach is well established in \emph{evolutionary computing} \cite{rudolph1998fmc}. While there has been past work on modelling \emph{evolutionary computing} algorithms as Markov chains \cite{rudolph, nix, goldberg2, eibenAarts}, we have found none including \aclp{MAS} despite both being mature research areas \cite{masOverviewPaper, ecpaper}, because their integration is a recent development \cite{smith1998fec}. So, the use of Chli-DeWilde stability as a definition for the self-organised \emph{stability} of evolving Agent Populations will be investigated further to determine its suitability.

\subsection{Chli-DeWilde Stability}
\label{chlidewilde}

Chli-DeWilde stability was created to provide a clear notion of stability in \aclp{MAS} \cite{chli2}, because stability is perhaps one of the most desirable features of any engineered system, given the importance of being able to predict its response to various environmental conditions prior to actual deployment; and while computer scientists often talk about stable or unstable systems \cite{mspaper5ThomasSycara1998, mspaper9Balakrishnan1997}, they did so without having a concrete or uniform definition of stability. Also, other properties had been widely investigated, such as openness \cite{mspaperAbramov2001}, scalability \cite{mspaperMarwala2001} and adaptability \cite{mspaperSimoesMarques2003}, but stability had not. So, the Chli-DeWilde definition of stability for \aclp{MAS} was created \cite{chli2}, based on the stationary distribution of a stochastic system, modelling the agents as Markov processes, and therefore viewing a \acl{MAS} as a discrete time Markov chain with a potentially unknown transition probability distribution. The \acl{MAS} is considered to be stable once its state has converged to an equilibrium distribution \cite{chli2}, because stability of a system can be understood intuitively as exhibiting bounded behaviour.

Chli-DeWilde stability was derived \cite{chlithesis} from the notion of stability defined by De Wilde \cite{mspaperDeWilde1999a, mspaperLee1998}, based on the stationary distribution of a stochastic system, making use of discrete-time Markov chains, which we will now introduce\footnote{A more comprehensive introduction to Markov chain theory and stochastic processes is available in \cite{msthesisNorris1997} and \cite{msthesisCoxMiller1972}.}. If we let $I$ be a \emph{countable set}, in which each $i \in I$ is called a \emph{state} and $I$ is called the \emph{state-space}. We can then say that $\lambda = (\lambda_i : i \in I)$ is a \emph{measure on} $I$ if $0 \le \lambda_i < \infty$ for all $i \in I$, and additionally a \emph{distribution} if $\sum_{i \in I}{\lambda_i=1}$ \cite{chlithesis}. So, if $X$ is a \emph{random variable} taking values in $I$ and we have $\lambda_i = \Pr(X = i)$, then $\lambda$ is \emph{the distribution of $X$}, and we can say that a matrix $P = (p_{ij} : i,j \in I)$ is \emph{stochastic} if every row $(p_{ij} : j \in I)$ is a \emph{distribution} \cite{chlithesis}. We can then extend familiar notions of matrix and vector multiplication to cover a general index set $I$ of potentially infinite size, by defining the multiplication of a matrix by a measure as $\lambda P$, which is given by
\begin{equation}
(\lambda P)_i = \sum\limits_{j \in I}{\lambda_{j}p_{ij}}.
\label{ms3dot1}
\end{equation}
We can now describe the rules for a Markov chain by a definition in terms of the corresponding matrix $P$ \cite{chlithesis}.\\

\begin{definition}
We say that $(X^t)_{t\ge0}$ is a Markov chain with initial distribution $\lambda = (\lambda_i : i \in I)$ and transition matrix $P = (p_{ij} : i,j \in I)$ if:
\narrowlinespacing
\begin{enumerate}
\item $\Pr(X^0 = i_0) = \lambda_{i_0}$ and
\item $\Pr(X^{t+1} = i_{t+1}\ |\ X^0 = i_0, \ldots, X^t = i_t) = p_{i_t i_{t+1}}$.
\end{enumerate}
\vspace{-3mm}
\normallinespacing
We abbreviate these two conditions by saying that $(X^t)_{t\ge0}$ is Markov$(\lambda, P)$.
\end{definition}

In this first definition the Markov process is \emph{memoryless}, resulting in only the current state of the system being required to describe its subsequent behaviour. We say that a Markov process $X^0, X^1, \ldots, X^t$ has a \emph{stationary distribution} if the probability distribution of $X^t$ becomes independent of the time $t$ \cite{chli2}. So, the following theorem is an \emph{easy consequence} of the second condition from the first definition.\\

\begin{theorem}
A discrete-time random process $(X^t)_{t\ge0}$ is Markov$(\lambda,P)$, if and only if for all $t$ and $i_0, \ldots, i_t$ we have
\narrowlinespacing
\begin{equation}
\Pr(X^0 = i_0, \ldots, X^t = i_t) = \lambda_{i_0}p_{i_0 i_1} \cdots p_{i_{t-1}i_t}.
\label{ms3dot2}
\end{equation}
\vspace{-6mm}
\normallinespacing
\end{theorem}
 
This first theorem depicts the structure of a Markov chain, illustrating the relation with the stochastic matrix $P$, and defining its time-invariance property \cite{chlithesis}.\\

\pagebreak
\begin{theorem}
Let $(X^t)_{t\ge0}$ be $Markov(\lambda,P)$, then for all $t,s\ge0$:
\narrowlinespacing
\begin{enumerate}
\item $\Pr(X^t = j) = (\lambda P^t)_j$ and
\item $\Pr(X^t = j\ |\ X^0 = i) = \Pr(X^{t+s} = j\ |\ X^s = i) = (P^t)_{ij}$.
\end{enumerate}
\vspace{-3mm}
\normallinespacing
\label{ms3dot3dot2}
For convenience $(P^t)_{ij}$ can be more conveniently denoted as $p^{(t)}_{ij}$.
\end{theorem}

Given this second theorem we can define $p^{(t)}_{ij}$ as the t-step transition probability from the state $i$ to $j$ \cite{chlithesis}, and we can now introduce the concept of an \emph{invariant distribution} \cite{chlithesis}, in which we say that $\lambda$ is invariant if
\begin{equation}
\lambda P = \lambda .
\end{equation}
The next theorem will link the existence of an \emph{invariant distribution}, which is an algebraic property of the matrix $P$, with the probabilistic concept of an \emph{equilibrium distribution}. This only applies to a restricted class of Markov chains, namely those with \emph{irreducible} and \emph{aperiodic} stochastic matrices. However, there is a multitude of analogous results for other types of Markov chains to which we can refer \cite{msthesisNorris1997, msthesisCoxMiller1972}, and the following theorem is provided as an indication of the family of theorems that apply. An \emph{irreducible} matrix $P$ is one for which, for all $i,j \in I$ there are sufficiently large $t,p^{(t)}_{ij} > 0$, and is \emph{aperiodic} if for all states $i \in I$ we have $p^{(t)}_{ii} > 0$ for all sufficiently large $t$ \cite{chlithesis}.\\

\begin{theorem}
Let $P$ be irreducible, aperiodic and have an invariant distribution, $\lambda$ can be any distribution, and suppose that $(X^t)_{t\ge0}$ is Markov$(\lambda, P)$ \cite{chlithesis}, then
\narrowlinespacing
\begin{eqnarray}
& \Pr(X^t = j) \to p_{j}^\infty \ as\ t \to \infty\ \mbox{for all}\ j \in I & \\
& and & \nonumber \\
& p^{(t)}_{ij} \to p_{j}^\infty \ as\ t \to \infty\ \mbox{for all}\ i,j \in I. &
\end{eqnarray}
\vspace{-9mm}
\normallinespacing
\end{theorem} 
 
We can now view a system $S$ as a countable set of states $I$ with implicitly defined transitions $P$ between them, and at time $t$ the state of the system is the random variable $X^t$, with the key assumption that $(X^t)_{t,0}$ is Markov$(\lambda,P)$ \cite{chlithesis}.\\

\begin{definition}
The system $S$ is said to be stable when the distribution of the its states converge to an \emph{equilibrium distribution},
\narrowlinespacing
\begin{equation}
\Pr(X^t = j) \to p_{j}^\infty \ as\ t \to \infty\ for\ all j\ \in I.
\end{equation}
\vspace{-9mm}
\normallinespacing
\end{definition}

More intuitively, the system $S$, a stochastic process $X^0$,$X^1$,$X^2$,... is \emph{stable} if the probability distribution of $X^t$ becomes independent of the time index $t$ for large $t$ \cite{chli2}. Most Markov chains with a finite state-space and positive transition probabilities are examples of stable systems, because after an initialisation period they settle down on a stationary distribution \cite{chlithesis}.

A \acl{MAS} can be viewed as a system $S$, with the system state represented by a finite vector $\bx$, having dimensions large enough to manage the agents present in the system. The state vector will consist of one or more elements for each agent, and a number of elements to define general properties of the system state. We can then model an agent as being \emph{dead}, i.e. not being present in the system, by setting the vector elements for that agent to some predefined null value \cite{chlithesis}.

\subsection{Extensions for Evolving Populations}
\label{def}

Extending Chli-DeWilde stability to the \emph{class} of \aclp{MAS} that make use of \emph{evolutionary computing} algorithms, including our evolving Agent Populations, requires consideration of the following issues: the inclusion of \emph{population dynamics}, and an understanding of population \emph{macro-states}.

\subsubsection{Population Dynamics}

First, the \acl{MAS} of an evolving Agent Population is composed of $n$ Agent-sequences, with each Agent-sequence $i$ in a state $\xi_i^t$ at time $t$, where $i=1, 2, \ldots, n$. The states of the Agent-sequences are \emph{random variables}, and so the state vector for the \acl{MAS} is a vector of random variables $\bxi^t$, with the time being discrete, $t=0, 1, \ldots$ . The interactions among the Agent-sequences are noisy, and are given by the probability distributions
\be
\Pr(X_i | \by) = \Pr(\xi_i^{t+1} = X_i | \bxi^t = \by) , \quad \ra,
\eeq{eq1}
where $X_i$ is a value for the state of Agent-sequence $i$, and $\by$ is a value for the state vector of the \acl{MAS}. The probabilities implement a Markov process \cite{suzuki}, with the noise caused by mutations. Furthermore, the Agent-sequences are individually subjected to a \emph{selection pressure} from the environment of the system, which is applied equally to all the Agent-sequences of the population. So, the probability distributions are statistically independent, and
\be
\Pr(\bx | \by) = \Pi_{i=1}^n \Pr(\xi_i^{t+1} = X_i | \bxi^t = \by).
\eeq{eq5}
If the occupation probability of state $\bx$ at time $t$ is denoted by $p_{\bx}^t$, then
\be
p_{\bx}^t = \sum_{\by} \Pr(\bx | \by) p_{\by}^{t-1}.
\eeq{eq5.1}
This is a discrete time equation used to calculate the evolution of the state occupation probabilities from $t=0$, while equation (\ref{eq5}) is the probability of moving from one state to another. The \acl{MAS} (evolving Agent Population) is self-stabilising if the limit distribution of the occupation probabilities exists and is non-uniform, i.e.
\be
p_\bx^\infty = lim_{t \rightarrow \infty} p_{\bx}^t
\eeq{eq2}
exists for all states $\bx$, and there exist states $\bx$ and $\by$ such that
\be
p_\bx^\infty \neq p_\by^\infty.
\eeq{eq3}
These equations define that some configurations of the system, after an extended time, will be more likely than others, because the likelihood of their occurrence no longer changes. Such a system is \emph{stable}, because the likelihood of states occurring no longer changes with time, and is the definition of stability developed in \cite{chli2}. While equation (\ref{eq2}) is the \emph{probabilistic equivalence} of an \emph{attractor}\footnote{An attractor is a set of states, invariant under the dynamics, towards which neighbouring states asymptotically approach during evolution \cite{weisstein2003cce}.} in a system with deterministic interactions, which we had to extend to a stochastic process because mutation is inherent in evolutionary dynamics.

Although the number of agents in the Chli-DeWilde formalism can vary, we require it to vary according to the \emph{selection pressure} acting upon the evolving Agent Population. We must therefore formally define and extend the definition of \emph{dead} agents, by introducing a new state $d$ for each Agent-sequence. If an Agent-sequence is in this state, $\xi_i^t=d$, then it is \emph{dead} and does not affect the state of other Agent-sequences in the population. If an Agent-sequence $i$ has low fitness then that Agent-sequence will likely die, because
\be
\Pr(d | \by) = \Pr(\xi_i^{t+1} = d | \bxi^t = \by)
\eeq{eq4}
will be high for all $\by$. Conversely, if an Agent-sequence has high fitness, then it will likely replicate, assuming the state of a similarly successful Agent-sequence (mutant), or crossover might occur changing the state of the successful Agent-sequence and another Agent-sequence.

\subsubsection{Population Macro-States}

As we defined earlier, the state of an evolving Agent Population is determined by the collection of Agent-sequences of which it consists at a specific time $t$, and potentially changing state as the time $t$ increases. So, we can define a macro-state $M$ as a set of states with a common property, here possessing at least one copy of the \emph{current maximum fitness individual}. Therefore, by its definition, each macro-state $M$ must also have a \emph{maximal state} composed entirely of copies of the \emph{current maximum fitness individual}. There must also be a macro-state consisting of all the states that have at least one copy of the \emph{global maximum fitness individual}, which we will call the \emph{maximum macro-state} $M_{max}$.

\tfigure{scale=1.0}{states}{graffle}{State-Space of an Evolving Agent Population}{A \getCap{statesCap} is shown, with \getCap{capStates3} \getCap{capStates}}{-2mm}{!h}{3mm}{}

We can consider the \emph{macro-states} of an evolving Agent Population visually through the representation of the state-space $I$ of the system $S$ shown in Figure \ref{states}, which includes a \setCap{possible evolutionary path through the state-space $I$}{statesCap}. Traversal through the state-space $I$ is directed by \setCap{the \emph{selection pressure} of the evolutionary process}{capStates3} acting upon the Population $S$, \setCap{driving it towards the \emph{maximal state} of the \emph{maximum macro-state} $M_{max}$, which consists entirely of copies of the \emph{optimal solution}, and is the equilibrium state that the system $S$ is forever \emph{falling towards} without ever quite reaching, because of the noise (mutation) within the system.}{capStates} So, while this \emph{maximal state} will never be reached, the \emph{maximum macro-state} $M_{max}$ itself is certain to be reached, provided the system does not get trapped at local optima, i.e. the probability of being in the \emph{maximum macro-state} $M_{max}$ at infinite time is one, $p^{\infty}_{\bone}=1$, as defined from equation (\ref{eq5.1}).

Furthermore, we can define quantitatively the probability distribution of the macro-states that the system occupies at infinite time. For a stable system, as defined by equation (\ref{eq3}), the \emph{degree of instability}, $d_{ins}$, can be defined as the entropy of its probability distribution at infinite time,
\be
d_{ins} = H(p^\infty) = -\sum\limits_{\bx}p_{\bx}^{\infty}log_{N}(p_{\bx}^{\infty}),
\eeq{eq6}
where $N$ is the number of possible states, and taking $log$ to the base $N$ normalises the \emph{degree of instability}. The \emph{degree of instability} will range between zero (inclusive) and one (exclusive), because a maximum instability of one would only occur during the theoretical extreme scenario of a \emph{non-discriminating selection pressure} \cite{kimura:ntm} (as shown in Figure \ref{fitLandFlatConverted}).\\

\subsection{Simulation and Results}

We simulated an evolving Agent Population from the Digital Ecosystem, using our simulation from section \ref{simRes} (unless otherwise specified), seeded with an Agent-pool of 20 Agents\footnote{From optimisation improvements to the code base of the simulation, we were able to increase the size of the Agent-pool from 15 to 20 without any significant degradation in performance, within the scope of running tens of thousands of simulation runs.} for the evolutionary process. We also added the classes and methods necessary to implement our extended Chli-DeWilde stability and \emph{degree of instability}, which required calculating $p_{\bx}^{(t)}$ of (\ref{eq5.1}) to estimate the stability, and $p_\bx^\infty$ of (\ref{eq2}) to prove the existence of $p_\bx^\infty \neq p_\by^\infty$ from (\ref{eq3}). The \emph{degree of instability}, $d_{ins}$ of (\ref{eq6}), was also implemented in the simulation.

\subsubsection{Stability}

Our evolving Agent Population (a \acl{MAS} with evolutionary dynamics) is stable if the distribution of the limit probabilities exists and is non-uniform, as defined by equations (\ref{eq2}) and (\ref{eq3}). The simplest case is a typical evolving Agent Population with one global optimal solution, which is stable if there are at least two macro-states with different limit occupation probabilities. We shall consider the \emph{maximum macro-state} $M_{max}$ and the \emph{sub-optimal macro-state} $M_{half}$. Where the states of the macro-state $M_{max}$ each possess at least one individual with global maximum fitness,
\begin{equation*}
p_{\bone}^\infty = lim_{t \rightarrow \infty} p_{\bone}^{(t)} = 1,
\end{equation*}
while the states of the macro-state $M_{half}$ each possess at least one individual with a fitness equal to \emph{half} of the global maximum fitness,
\begin{equation*}
p_{\bfif}^\infty = lim_{t \rightarrow \infty} p_{\bfif}^{(t)} = 0,
\end{equation*}
thereby fulfilling the requirements of equations (\ref{eq2}) and (\ref{eq3}). The \emph{sub-optimal macro-state} $M_{half}$, having a lower fitness, is predicted to be seen earlier in the evolutionary process before disappearing as higher fitness macro-states are reached. The system $S$ will take longer to reach the \emph{maximum macro-state} $M_{max}$, but once it does will likely remain, leaving only briefly depending on the strength of the mutation rate, as the \emph{selection pressure} is \emph{non-elitist}\footnote{\emph{Non-elitist} meaning that the best individual from one generation was not guaranteed to survive to the next generation; it had a high probability of surviving into the next generation, but it was not guaranteed as it might have been mutated \cite{eiben2003iec}.} (as defined in section \ref{simRes}).

\vspace{3mm}

A value of $t=1000$ was chosen to represent $t=\infty$ experimentally, because the simulation has often been observed to reach the \emph{maximum macro-state} $M_{max}$ within 500 generations. Therefore, the probability of the system $S$ being in the \emph{maximum macro-state} $M_{max}$ at the thousandth generation is expected to be one, $p^{1000}_{\bone} = 1$. Furthermore, the probability of the system being in the \emph{sub-optimal macro-state} $M_{half}$ at the thousandth generation is expected to be zero, $p^{1000}_{\bfif} = 0$.

\vspace{3mm}

Figure \ref{macrostates} shows, for a typical evolving Agent Population, a graph of the probability as defined by equation (\ref{eq5.1}) of the \emph{maximum macro-state} $M_{max}$ and the \emph{sub-optimal macro-state} $M_{half}$ at each generation, averaged from ten thousand simulation runs for statistical significance. The behaviour of the simulated system $S$ was as expected, being \setCap{in the \emph{maximum macro-state} $M_{max}$ only after generation 178 and always after generation 482. It was also observed being in the \emph{sub-optimal macro-state} $M_{half}$ only between generations 37 and 113, with a maximum probability of 0.053 (3 d.p.) at generation 61}{graphCap}, and was such because the evolutionary path (state transitions) could avoid visiting the macro-state. As expected the probability of being in the \emph{maximum macro-state} $M_{max}$ at the thousandth generation was one, $p^{1000}_{\bone} = 1$, and so the probability of being in any other macro-state, including the \emph{sub-optimal macro-state} $M_{half}$, at the thousandth generation was zero, $p^{1000}_{\bfif} = 0$. 

\tfigure{}{macrostates}{graph}{Graph of the Probabilities of the Macro-States}{$M_{max}$ and $M_{half}$ at each Generation: The system $S$, a typical evolving Agent Population, was \getCap{graphCap}.}{-7mm}{}{}{}

A visualisation for the state of a typical evolving Agent Population at the thousandth generation is shown in Figure \ref{vis62}, with each line representing an Agent-sequence and each colour representing an Agent, with the identical Agent-sequences grouped for clarity. It shows that the evolving Agent Population reached the \emph{maximum macro-state} $M_{max}$ and remained there, but as expected never reached the \emph{maximal state} of the \emph{maximum macro-state} $M_{max}$, where all the Agent-sequences are identical and have maximum fitness, which is indicated by the lack of total uniformity in Figure \ref{vis62}. This was expected, because of the mutation (noise) within the evolutionary process, which is necessary to create the opportunity to find fitter (better) sequences and potentially avoid getting trapped at any local optima that may be present.

\pagebreak

\tfigure{width=170mm}{vis62}{graffle}{Visualisation of an Evolving Agent Population}{at the 1000th Generation: The Population consists of multiple Agent-sequences, with each line representing an Agent-sequence, and therefore each colour representing an Agent. The identical Agent-sequences were grouped for clarity, and as expected the system $S$ reached the {maximum macro-state} $M_{max}$ and remained there, but never reached the {maximal state} of the {maximum macro-state} $M_{max}$.}{-8mm}{!h}{}{}

\subsubsection{Degree of Instability}

Given that our simulated evolving Agent Population is stable as defined by equations (\ref{eq2}) and (\ref{eq3}), we can determine the \emph{degree of instability} as defined by equation (\ref{eq6}). So, calculated from its limit probabilities, the \emph{degree of instability} was

\vspace{-5mm}

\begin{eqnarray}
d_{ins} = H(p^{1000}) &=& -\sum\limits_{\bx}p^{1000}_\bx log_{N}(p^{1000}_\bx) \nonumber \\ 
 &=& -1log_{N}(1) \nonumber \\
 &=& 0, \nonumber
\label{result2}
\end{eqnarray}

where $t=1000$ is an effective estimate for $t=\infty$, as explained earlier. The result was as expected because the \emph{maximum macro-state} $M_{max}$ at the thousandth generation was one, $p^{1000}_{\bone} = 1$, and so the probability of being in the other macro-states at the thousandth generation was zero. The system therefore shows no instability, as there is no entropy in the occupied macro-states at infinite time.

\subsubsection{Stability Analysis}

\tfigure{}{stabilityAnalysis}{graph}{Graph of Stability with Different Mutation and Crossover Rates}{\getCap{aScap}.}{-7mm}{!h}{}{}

We then performed a \emph{stability analysis} (similar to a \emph{sensitivity analysis} \cite{cacuci2003sau}) of a typical evolving Agent Population, varying key parameters within the simulation. We varied the mutation and crossover rates from 0\% to 100\% in 10\% increments, calculating the \emph{degree of instability}, $d_{ins}$ from (\ref{eq6}), at the thousandth generation. These \emph{degree of instability} values were averaged over ten thousand simulation runs, and graphed against the mutation and crossover rates in Figure \ref{stabilityAnalysis}. It shows that the crossover rate had little effect on the stability of our simulated evolving Agent Population, whereas the mutation rate did significantly affect the stability. \setCap{With the mutation rate under or equal to 60\%, the evolving Agent Population showed no instability, with $d_{ins}$ values equal to zero as the system $S$ was always in the same macro-state $M$ at infinite time, independent of the crossover rate. With the mutation rate above 60\% the instability increased significantly}{aScap}, with the system being in one of several different macro-states at infinite time; with a mutation rate of 70\% the system was still very stable, having low $d_{ins}$ values ranging between 0.08 (2 d.p.) and 0.16 (2 d.p.), but once the mutation rate was 80\% or greater the system became quite unstable, shown by high $d_{ins}$ values nearing 0.5. 

As one would have expected, an \emph{extremely} high mutation rate has a destabilising effect on the \emph{stability} of an evolving Agent Population. The crossover rate had only a minimal effect, because variation from \emph{crossover} was limited when the Population had \emph{matured}, consisting of Agent-sequences identical or very similar to one another. It should also be noted that the stability of the system is different to its performance, because although showing no instability with mutation rates below 60\% (inclusive), it only reached the \emph{maximum macro-state} $M_{max}$ with a mutation rate of 10\% or above, while at 0\% it was stable at a sub-optimal macro-state.

\subsection{Summary}

None of the existing definitions we considered \cite{crutchfield2006, prugel, bak1988soc, chli2} were directly applicable as a definition for the self-organised \emph{stability} of an evolving Agent Population, but the properties of Chli-DeWilde stability \cite{chli2} closely matched our intuitive understanding, and so was chosen for further investigation. It views a \acl{MAS} as a discrete time Markov chain (with potentially unknown transition probabilities) that is considered to be stable when its state, a stochastic process, has converged to an equilibrium distribution. Extending Chli-DeWilde stability to the \acl{MAS} of an evolving Agent Population required consideration of the following issues: the inclusion of \emph{population dynamics}, and an understanding of population \emph{macro-states}. We then built upon this to construct an entropy-based definition for the \emph{degree of instability} (entropy of the limit probabilities), which was later used to perform a \emph{stability analysis} of an evolving Agent Population. 

We then investigated the self-organised \emph{stability} of evolving Agent Populations through experimental simulations, and the results showed that there was a limit probability distribution, and that it was non-uniform. Furthermore, the reaching of the \emph{maximum macro-state} was confirmed by a visualisation matching the numerical results. We then applied our \emph{degree of instability} to determine that there was no instability under normal conditions, and then performed a \emph{stability analysis} (similar to a \emph{sensitivity analysis} \cite{cacuci2003sau}) showing the variation of the self-organised \emph{stability} under varying conditions. Collectively, the experimental results confirm that Chli-Dewilde stability has been successfully extended to evolving Agent Populations, while our definition for the \emph{degree of instability} provides a macroscopic value to characterise the level of \emph{stability}. 

We have determined an effective understanding and quantification for the self-organised \emph{stability} of the evolving Agent Populations of our Digital Ecosystem. Also, our extended Chli-DeWilde stability is applicable to other \aclp{MAS} with evolutionary dynamics. Furthermore, our \emph{degree of instability} provides a definition for the \emph{level of stability}, applicable to \aclp{MAS} with or without evolutionary dynamics.

\section{Diversity}

A definition for the self-organised \emph{diversity} of an evolving Agent Population should define the optimal variability, of the Agents and Agent-sequences, that emerge over time, with no initial constraints from modelling approaches for the inclusion of pre-defined specific behaviour, but capable of representing the appearance of such behaviour should it occur.

None of the proposed definitions are applicable for the self-organised \emph{diversity} of an evolving Agent Population. The $\in$-machine modelling \cite{crutchfield2006} is not applicable, because it is only defined within the context of pre-biotic populations. Neither is the \acl{MDL} \cite{barron} principle or the Pr{\"u}gel-Bennett Shapiro formalism \cite{prugel} suitable, because they necessitate the involvement of \emph{subjective} human judgement at the critical stages of model or quantifier selection. \acl{MFT} is also not applicable because of the necessity of a \emph{neighbourhood model} for defining interaction, and evolving Agent Populations lack a 2D or 3D metric space for such models. So, the only available \emph{neighbourhood model} becomes a distance measure on a parameter space that measures dissimilarity. However, this type of \emph{neighbourhood model} cannot represent the information-based interactions between the individuals of an evolving Agent Population. 

We suggest that the uniqueness of Digital Ecosystems makes the application of existing definitions inappropriate for the self-organised \emph{diversity}, because while we could extend a biology-centric definition for the self-organised \emph{complexity}, and a computing-centric definition for the self-organised \emph{stability}, we found neither of these approaches, or any other, appropriate for the self-organised \emph{diversity}. The Digital Ecosystem being the digital counterpart of a biological ecosystem gives it unique properties, as discussed earlier in section \ref{endChap2}. So, the evolving Agent Populations possess properties of both computing systems (e.g. agent systems) as well as biological systems (e.g. population dynamics), and the combination of these properties makes them unique. So, we will further consider the evolving Agent Populations to create a definition for their self-organised \emph{diversity}.

\subsection{Evolving Agent Populations}

\newcommand{\vspaceeap}{\vspace{-1mm}}

\vspaceeap

The self-organised \emph{diversity} of an evolving Agent Population comes from the Agent-sequences it \emph{evolves}, in response to the \emph{selection pressure}, seeded with Agents and Agent-sequences from the Agent-pool of the Habitat in which it is instantiated. The set of Agents and Agent-sequences available when seeding an evolving Agent Population is regulated over time by other evolving Agent Populations, instantiated in response to other user requests, leading to the death and migration of Agents and Agent-sequences, as well as the formation of new Agent-sequence combinations. The seeding of existing Agent-sequences provides a direction to accelerate the evolutionary process, and can also affect the self-organised \emph{diversity}; for example, if only a proportion of any available global optima is favoured. So, the set of Agents available when seeding an evolving Agent Population provides potential for the self-organised \emph{diversity}, while the \emph{selection pressure} of a \emph{user request} provides a constraining factor on this \emph{potential}. Therefore, the \emph{optimality} of the self-organised \emph{diversity} of an evolving Agent Population is \emph{relative} to the \emph{selection pressure} of the user request for which it was instantiated.

While we could measure the self-organised \emph{diversity} of individual evolving Agent Populations, or even take a random sampling, it will be more informative to consider their \emph{collective} self-organised \emph{diversity}. Also, given that the Digital Ecosystem is required to support a range of user behaviour, we can consider the \emph{collective} self-organised \emph{diversity} of the evolving Agent Populations relative to the \emph{global} user request behaviour. So, when varying a behavioural property of the user requests according to some distribution, we would expect the corresponding property of the evolving Agent Populations to follow the same distribution. We are not intending to prescribe the expected user behaviour of the Digital Ecosystem, but investigate whether the Digital Ecosystem can adapt to a range of user behaviour in terms of the self-organised \emph{diversity}. So, we will consider Uniform, Gaussian (Normal) and Power distributions for the parameters of the user request behaviour. The Uniform distribution will provide a control, while the Normal (Gaussian) distribution will provide a reasonable assumption for the behaviour of a large group of users, and the Power distribution will provide a relatively extreme variation in user behaviour.

\vspaceeap

\subsection{Simulation and Results}

\vspaceeap

We simulated the Digital Ecosystem, using our simulation from section \ref{simRes} (unless otherwise specified). We also added the classes and methods necessary to vary aspects of the user behaviour according to different distributions, and a way to measure the related aspects of the evolving Agent Populations. This consisted of a mechanism to vary the user request properties of \emph{length} and \emph{modularity}, according to Uniform, Gaussian (normal) and Power distributions, and a mechanism to measure the corresponding Agent(-sequence) properties of \emph{length} and \emph{number of attributes}. For statistical significance each scenario (experiment) will be averaged from ten thousand simulation runs. We expect it will be obvious whether the \emph{observed} behaviour of the Digital Ecosystem matches the \emph{expected} behaviour from the user base. Nevertheless, we will also implement a chi-squared ($\chi^2$) test to determine if the observed behaviour (distribution) of the Agent(-sequence) properties matches the expected behaviour (distribution) from the user request properties.

Given the requirement to run a minimum of sixty thousand simulation runs, ten thousand for each experiment, we adapted the code base of the simulation to take advantage of the Xgrid \cite{kramer2004ulg} distributed computing technology, and therefore make use of the \emph{grids} mentioned in the acknowledgements.

\subsubsection{User Request Length}

We started by varying the \emph{user request length} according to the available distributions, expecting the length of the Agent-sequences to be distributed according to the length of the user requests, i.e. the longer the user request, the longer the Agent-sequence needed to fulfil it.

We first applied the Uniform distribution as a control, and graphed the results in Figure \ref{urluniform}. \setCap{The \emph{observed} frequencies of the Agent-sequence length mostly matched the \emph{expected} frequencies, which was \emph{confirmed} by a $\chi^2$ test; with a \emph{null hypothesis} of \emph{no significant difference} and \emph{sixteen degrees of freedom}, the $\chi^2$ value was 2.588 (3 d.p.), below the critical 0.95 $\chi^2$ value of 7.962.}{urlCapUnifrom} 

We then applied the Gaussian distribution as a reasonable assumption for the behaviour of a large group of users, and graphed the results in Figure \ref{urlgaussian}. \setCap{The \emph{observed} frequencies of the Agent-sequence length matched the \emph{expected} frequencies with only very minor variations, which was confirmed by a $\chi^2$ test; with a \emph{null hypothesis} of \emph{no significant difference} and \emph{sixteen degrees of freedom}, the $\chi^2$ value was 2.102}{urlCapGaussian} (3 d.p.), \setCap{below the critical 0.95 $\chi^2$ value of 7.962.}{urlCap2Gaussian} 

\pagebreak

\tfigure{}{urluniform}{graph}{Graph of Uniformly Distributed Agent-Sequence Length Frequencies}{\getCap{urlCapUnifrom}}{-7mm}{}{}{}

\tfigure{}{urlgaussian}{graph}{Graph of Gaussian Distributed Agent-Sequence Length Frequencies}{\getCap{urlCapGaussian}, \getCap{urlCap2Gaussian}}{-7mm}{!b}{-10mm}{}

\tfigure{}{urlpower}{graph}{Graph of Power Distributed Agent-Sequence Length Frequencies}{\getCap{urlpower}}{-7mm}{}{}{}

Finally, we applied the Power distribution to represent a relatively extreme variation in user behaviour, and graphed the results in Figure \ref{urlpower}. \setCap{The \emph{observed} frequencies of the Agent-sequence length matched the \emph{expected} frequencies with some variation, which was confirmed by a $\chi^2$ test; with a \emph{null hypothesis} of \emph{no significant difference} and \emph{sixteen degrees of freedom}, the $\chi^2$ value was 5.048 (3 d.p.), below the critical 0.95 $\chi^2$ value of 7.962.}{urlpower} 

There were a couple of minor discrepancies, similar to all the experiments. First, there were a small number of \emph{individual} Agents at the thousandth time step, caused by the typical user behaviour of continuously creating new services (Agents). Second, while the chi-squared tests confirmed that there was no significant difference between the \emph{observed} and \emph{expected} frequencies of the Agent-sequence length, there was still a \emph{bias} to longer Agent-sequences (solutions). Evident visually in the graphs of the experiments, and evident numerically in the chi-squared test of the Power distribution experiment as it favoured shorter Agent-sequences. The cause of this \emph{bias} was most likely some aspect of \emph{bloat} (as we discussed in section \ref{bloat}) that was not fully controlled.

\newcommand{\vspaceurm}{\vspace{-1.5mm}}

\subsubsection{User Request Modularity}

\vspaceurm

Next, we varied the \emph{user request modularity} according to the available distributions, expecting the \emph{sophistication} of the Agents to be distributed according to the modularity of the user requests, i.e. the more complicated (in terms of modular non-reducible tasks) the user request, the more sophisticated (in terms of the number of attributes) the Agents needed to fulfil it.

We first applied the Uniform distribution as a control, and graphed the results in Figure \ref{urvuniform}. \setCap{The \emph{observed} frequencies for the number of Agent attributes mostly matched the \emph{expected} frequencies, which was confirmed by a $\chi^2$ test; with a \emph{null hypothesis} of \emph{no significant difference} and \emph{ten degrees of freedom}, the $\chi^2$ value was 1.049 (3 d.p.), below the critical 0.95 $\chi^2$ value of 3.940.}{urvunifromCap}

\tfigure{}{urvuniform}{graph}{Graph of Uniformly Distributed Agent Attribute Frequencies}{\getCap{urvunifromCap}}{-8mm}{!h}{}{}

\vspaceurm

We then applied the Gaussian distribution as a reasonable assumption for the behaviour of a large group of users, and graphed the results in Figure \ref{urvgaussian}. \setCap{The \emph{observed} frequencies for the number of Agent attributes \emph{appeared} to follow the \emph{expected} frequencies, but there was significant variation which led to a failed $\chi^2$ test; with a \emph{null hypothesis} of \emph{no significant difference} and \emph{ten degrees of freedom}}{urvgaussianCap}, the $\chi^2$ value was 50.623 (3 d.p.), not below the critical 0.95 $\chi^2$ value of 3.940.

\pagebreak

\tfigure{}{urvgaussian}{graph}{Graph of Gaussian Distributed Agent Attribute Frequencies}{\getCap{urvgaussianCap}.}{-7mm}{!t}{}{}

\label{divmodexp}

\tfigure{}{urvpower}{graph}{Graph of Power Distributed Agent Attribute Frequencies}{\getCap{urvpowerCap}.}{-7mm}{!b}{-10mm}{}

Finally, we applied the Power distribution to represent a relatively extreme variation in user behaviour, and graphed the results in Figure \ref{urvpower}. \setCap{The \emph{observed} frequencies for the number of Agent attributes \emph{appeared} to follow the \emph{expected} frequencies, but there was significant variation which led to a failed $\chi^2$ test; with a \emph{null hypothesis} of \emph{no significant difference} and \emph{ten degrees of freedom}}{urvpowerCap}, the $\chi^2$ value was 61.876 (3 d.p.), not below the critical 0.95 $\chi^2$ value of 3.940.

In all the experiments the \emph{observed} frequencies of the number of Agent attributes \emph{appeared} to follow the \emph{expected} frequencies, but this could only be confirmed statistically, by a $\chi^2$ test, for the Uniform distribution experiment. In the Gaussian and Power distribution experiments the $\chi^2$ tests failed by considerable margins, most likely because the evolving Agent Populations were still self-organising to match the user behaviour, shown by the \emph{observed} frequencies \emph{approaching} the \emph{expected} frequencies, but not yet sufficiently to meet $\chi^2$ tests, because by the thousandth \emph{time step} (user request event) each user had placed an average of only ten requests.

\subsection{Summary}

None of the existing definitions we considered \cite{crutchfield2006, barron, prugel, flyvbjerg} were applicable as a definition for the self-organised \emph{diversity} of the evolving Agent Populations. So, we further considered the unique properties resulting from information-centric Digital Ecosystems being the digital counterpart of energy-centric biological ecosystems, creating our own definition for the self-organised \emph{diversity} of an evolving Agent Population, relative to the \emph{selection pressure} provided by a user request. We then considered the \emph{collective} self-organised \emph{diversity} of the evolving Agent Populations relative to the \emph{global} user request behaviour. Therefore, when varying a behavioural property of the \emph{user requests} according to some \emph{distribution}, we expected the corresponding property of the \emph{evolving Agent Populations} to follow the same \emph{distribution}. We used the Uniform distribution to provide a control, the Normal (Gaussian) distribution to provide a reasonable assumption for the behaviour of a large group of users, and the Power distribution to represent a relatively extreme distribution in user behaviour. 

We then investigated the self-organised \emph{diversity} of evolving Agent Populations through experimental simulations. First, varying the \emph{user request length} according to the different distributions, and testing whether the \emph{observed} frequencies of the Agent-sequence length matched the \emph{expected} frequencies, which we confirmed with successful chi-squared tests. Second, varying the \emph{user request modularity} according to the different distributions, and testing whether the \emph{observed} frequencies for the number of Agent attributes matched the \emph{expected} frequencies, again confirming with chi-squared tests. Under the Gaussian and Power distributions the chi-squared tests failed, most likely because the evolving Agent Populations were still self-organising to match the user behaviour, because at the time the Digital Ecosystem was sampled each user had placed an average of only ten requests.

Collectively, the experimental results confirm that the self-organised \emph{diversity} of the evolving Agent Populations is relative to the \emph{selection pressures} of the user base, which was confirmed statistically for most of the experiments. So, we have determined an effective understanding and quantification for the self-organised \emph{diversity} of the evolving Agent Populations of our Digital Ecosystem. While the minor experimental failures, in which the Digital Ecosystem responded more slowly than in the other experiments, have shown that there is potential to optimise the Digital Ecosystem, because the evolutionary self-organisation of an ecosystem is a slow process \cite{begon96}, even the accelerated form present in our Digital Ecosystem.

\section{Summary and Discussion}

We have investigated the self-organising behaviour of Digital Ecosystems, because a primary motivation for our research is the desire to exploit the self-organising properties of biological ecosystems \cite{Levin}, which are thought to be robust, scalable architectures that can automatically solve complex, dynamic problems. Over time a biological ecosystem becomes increasingly self-organised through the process of \emph{ecological succession} \cite{begon96}, driven by the evolutionary self-organisation of the populations within the ecosystem. Analogously, a Digital Ecosystem's increasing self-organisation comes from the Agent Populations being evolved to meet the dynamic \emph{selection pressures} created by requests from the user base. The self-organisation of biological ecosystems is often defined in terms of the \emph{complexity}, \emph{stability}, and \emph{diversity} \cite{king1983cda}, which we also applied in defining the self-organisation of our Digital Ecosystems. We started by discussing the relevant literature, including the philosophical meaning of \emph{organisation} and of \emph{self}, learning that self-organisation is context dependent, and that a system is only \emph{self}-organising if the process or force causing the organisation is within its boundaries. So, we compared and contrasted alternative definitions \cite{crutchfield2006, barron, prugel, chaitin, adami20002, bak1988soc, chli2, flyvbjerg} for the self-organised \emph{complexity}, \emph{stability}, and \emph{diversity} of the evolving Agent Populations, examining their suitability and application, because possessing properties of computing systems (e.g. agent systems) as well as biological systems (e.g. population dynamics), the combination of these properties makes the evolving Agent Populations unique.

None of the existing definitions we considered \cite{crutchfield2006, barron, prugel, chaitin, adami20002} were directly applicable as a definition for the self-organised \emph{complexity} of evolving Agent Populations, but the properties of Physical Complexity \cite{adami20002} closely matched our intuitive understanding, and so was chosen for further investigation. Based upon information theory and entropy, it provides a measure of the quantity of information in the genome of a population, relative to the environment in which it evolves, by calculating the entropy in the population to determine the randomness in the genome \cite{adami20002}. Reformulating Physical Complexity for an evolving Agent Population required consideration of the following issues: the mapping of the sequence sites to the Agent-sequences, and the managing of populations of variable length sequences. We then built upon this to construct a variant of the Physical Complexity called the \emph{Efficiency}, because it was based on the efficiency of information storage in Physical Complexity, which we then used to develop an understanding of clustering and atomicity within evolving Agent Populations. Collectively, the experimental results confirm that Physical Complexity has been successfully extended to evolving Agent Populations. Most significantly, Physical Complexity has been reformulated algebraically for populations of \aclp{vls}, which we have confirmed experimentally through simulations. Our \emph{Efficiency} definition provides a \emph{universally applicable} macroscopic value to characterise the \emph{complexity} of a population, independent of clustering, atomicity, length (variable or same), and size. So, we have determined an effective understanding and quantification for the self-organised \emph{complexity} of the evolving Agent Populations of our Digital Ecosystem. The understanding and techniques we have developed have applicability beyond evolving Agent Populations, as wide as the original Physical Complexity, which has been applied from \acs{DNA} \cite{adami2000} to simulations of self-replicating programmes \cite{adami20032}. 

None of the existing definitions we considered \cite{crutchfield2006, prugel, bak1988soc, bak1988soc, chli2} were directly applicable as a definition for the self-organised \emph{stability} of evolving Agent Populations, but the properties of Chli-DeWilde stability \cite{chli2} closely matched our intuitive understanding, and so was chosen for further investigation. It views a \acl{MAS} as a discrete time Markov chain (with potentially unknown transition probabilities) that is considered to be stable when its state, a stochastic process, has converged to an equilibrium distribution. Extending Chli-DeWilde stability to the \acl{MAS} of an evolving Agent Population required consideration of the following issues: the inclusion of \emph{population dynamics}, and an understanding of population \emph{macro-states}. We then built upon this to construct an entropy-based definition for the \emph{degree of instability} (entropy of the limit probabilities), which was used to perform a \emph{stability analysis} (similar to a \emph{sensitivity analysis} \cite{cacuci2003sau}) of an evolving Agent Population. Collectively, the experimental results confirm that Chli-Dewilde stability has been successfully extended to evolving Agent Populations, while our definition for the \emph{degree of instability} provides a macroscopic value to characterise the level of \emph{stability}. So, we have determined an effective understanding and quantification for the self-organised \emph{stability} of the evolving Agent Populations of our Digital Ecosystem. Also, our extended Chli-DeWilde stability is applicable to other \aclp{MAS} with evolutionary dynamics. Furthermore, our \emph{degree of instability} is applicable to all \aclp{MAS}, with or without evolutionary dynamics.

None of the existing definitions we considered \cite{crutchfield2006, barron, prugel, flyvbjerg} were applicable as a definition for the self-organised \emph{diversity} of evolving Agent Populations. So, we further considered the unique properties resulting from information-centric Digital Ecosystems being the digital counterpart of energy-centric biological ecosystems, creating our own definition for the self-organised \emph{diversity} of an evolving Agent Population relative to the \emph{selection pressure} provided by a user request. We then considered the \emph{collective} self-organised \emph{diversity} of the evolving Agent Populations relative to the \emph{global} user request behaviour. Therefore, when varying a behavioural property of the \emph{user requests} according to some \emph{distribution}, we expected the corresponding property of the \emph{evolving Agent Populations} to follow the same \emph{distribution}. We used the Uniform distribution to provide a control, the Normal (Gaussian) distribution to provide a reasonable assumption for the behaviour of a large group of users, and the Power distribution to provide a relatively extreme distribution in user behaviour. Collectively, the experimental results confirm that the self-organised \emph{diversity} of the evolving Agent Populations is relative to the \emph{selection pressures} of the user base, which was confirmed statistically for most of the experiments. So, we have determined an effective understanding and quantification for the self-organised \emph{diversity} of the evolving Agent Populations of our Digital Ecosystem. While the minor experimental failures, in which the Digital Ecosystem responded more slowly than in the other experiments, have shown that there is potential to optimise the Digital Ecosystem, because the evolutionary self-organisation of an ecosystem is a slow process \cite{begon96}, even the accelerated form present in our Digital Ecosystem.

Overall an insight has been achieved into where and how self-organisation occurs in our Digital Ecosystem, and what forms this self-organisation can take and how it can be quantified. The hybrid nature of the Digital Ecosystem resulted in the most suitable definition for the self-organised \emph{complexity} coming from the biological sciences, while the most suitable definition for the self-organised \emph{stability} coming from the computer sciences. However, we were unable to use any existing definition for the self-organised \emph{diversity}, because the hybrid nature of the Digital Ecosystem makes it unique, and so we constructed our own definition based on variation relative to the user base. The (Physical) \emph{complexity} definition applies to a single point in time of the evolving Agent Populations, whereas the (Chli-DeWilde) \emph{stability} definition applies at the end of these instantiated evolutionary processes, while our \emph{diversity} definition applies to the optimality of the distribution of the Agents within the evolving Agent Populations of the Digital Ecosystem. The experimental results have generally supported the hypotheses, and have provided more detail to the behaviour of the self-organising phenomena under investigation, showing some of its properties and for the self-organised \emph{diversity} has shown that there is potential for optimising the Digital Ecosystem.

In this chapter we have investigated the emergent self-organising properties of Digital Ecosystems, and with the greater and more in-depth understanding we have developed and gained of the order constructing processes (the evolving Agent Populations), including a clearer identification of the potential areas and scopes for augmentation, we will attempt the optimisation of Digital Ecosystems in the following chapter, Chapter \ref{ch:optimisation}, for which the results here have confirmed the potential for optimisation identified in the previous chapter,\linebreak Chapter \ref{ch:creation}.

\chapter{Optimisation of Digital Ecosystems}

\label{ch:optimisation}

In this chapter we attempt the acceleration and optimisation of Digital Ecosystems, because the evolutionary self-organisation of \emph{ecological succession} (the formation of a mature ecosystem) is a slow process, even the accelerated form present in our Digital Ecosystem. First, we consider the scope for optimisation identified in the previous chapters, and the potential for augmentations from the biological sciences. Consolidating this understanding we propose, construct and explore alternative augmentations to accelerate or optimise the evolutionary and ecological self-organising dynamics of our Digital Ecosystems. The most promising, the \emph{clustering catalyst} and the \emph{targeted migration}, were completed theoretically, before being investigated experimentally to determine their improvement on the evolutionary and ecological dynamics in responding to the needs of the user base; the first aiming to optimise the evolutionary dynamics, while the second aiming to optimise the ecological dynamics. First, the \emph{clustering catalyst} operates upon an evolutionary process, encouraging intra-cluster crossover to accelerate reaching the optimal solution, directly accelerating a core operation of the Digital Ecosystem. A suitable existing clustering algorithm and a \emph{Physical Complexity} based one were both evaluated for the required clustering. Second, the \emph{targeted migration} operates on the ecological dynamics, allowing the Agents to interact for additional highly targeted migration, indirectly optimising a global operation of the Digital Ecosystem. Both \aclp{NN} and \aclp{SVM} were evaluated for the required pattern recognition and learning functionality. We conclude with a summary and discussion of the achievements, including the experimental results.

\section{Background Theory}

We proposed that an ecosystem inspired approach would be more effective at greater scales than traditionally inspired approaches, because it would be built upon the scalable and self-organising properties of biological ecosystems \cite{Levin}. So, a Digital Ecosystem, being the digital counterpart of biological ecosystems, possesses their scalable self-organising behaviour, properties and processes. However, the self-organising process of \emph{ecological succession} is a slow one, the orderly and predictable changes in the composition and structure of an ecological community in forming a mature ecosystem \cite{begon96}, even the accelerated form present in our Digital Ecosystem. Therefore, it may be possible to accelerate and optimise this equivalent process of our Digital Ecosystem.

The scope for optimisation and acceleration was identified and confirmed in the previous chapters. First, identified by the results of Chapter \ref{ch:creation}, specifically the \emph{ecological succession} experiment (section \ref{ecosucexp}) in which the Digital Ecosystem reached only 70\% responsiveness, clearly showing potential for improvement. Second, confirmed by the results of Chapter \ref{ch:investigation}, specifically two of the modularity scenarios of the self-organised diversity experiment (section \ref{divmodexp}), in which the Digital Ecosystem responded more slowly in these scenarios than others, confirming the potential for improvement. Therefore, there is scope for optimising and accelerating the equivalent process of \emph{ecological succession} in our Digital Ecosystem.

In \emph{biological ecosystems} the trajectory of ecological change can be influenced by site conditions, by the interactions of the species present, and by more stochastic factors such as the availability of colonists or seeds, or weather conditions at the time of disturbance \cite{turner1998fis}. So, ecological optimisation is generally concerned with the maintenance of diversity and stability, for the survival of populations, species, habitats, etc \cite{walters1978eoa, abrams1984fto, naeem1997bee}, and ecological acceleration is similarly concerned with the re-establishment of diversity and stability, through optimal species selection and promotion \cite{mcclanahan1993afs, lugo1997apr, wunderle1997ras}. Therefore, \emph{biological ecosystems} research has no focus on the type of optimisation or acceleration we require, which is unsurprising, because one of the fundamental differences between biological and digital ecosystems lie in the motivation and approach of their researchers; given that biological ecosystems are ubiquitous natural phenomena whose maintenance is crucial to our survival \cite{balmford2002erc}, whereas Digital Ecosystems are a technology engineered to serve specific human purposes. So, we are unlikely to find augmentations from \emph{biological ecosystems} to optimise our Digital Ecosystem.

The optimisation of Digital Ecosystems sought is not that of parameter optimisation, which is achievable through \emph{exploratory programming} \cite{sommerville2006se}, but an augmentation to the \acl{EOA} that provides a significant improvement in performance, i.e. better solutions for the users than the Digital Ecosystem alone could achieve. In the previous chapter we have investigated the emergent self-organising properties of Digital Ecosystems, and with the greater and more in-depth understanding we have developed and gained of the order constructing processes (the evolving Agent Populations), including a clearer identification of the potential areas and scopes for augmentation, we will now propose, construct and explore alternative augmentations to accelerate or optimise the evolutionary and ecological self-organising dynamics of our Digital Ecosystem. The most promising will be completed theoretically and then investigated experimentally through simulations.

\newcommand{\vspaceaas}{\vspace{2mm}}
\vspaceaas

\section{Alternative Augmentations}

\vspaceaas

Any proposed augmentation should improve the process of \emph{ecological succession} \cite{begon96} for our Digital Ecosystem. So, based on the understanding and results from the previous chapters, our general knowledge, and our intuition, we will now propose, construct, and explore possible alternative augmentations for our Digital Ecosystem that fulfils this requirement.

\vspaceaas

\addtocontents{lof}{\protect\pagebreak}

\subsection{Clustering Catalyst}
\label{ccsection}

\vspaceaas

A significant proportion of user requests will be returned multiple optimal responses (applications), by evolving Agent Populations consisting of clusters as defined in the previous chapter. So, potential exists to accelerate these \emph{evolving Agent Populations with clusters} by a \emph{clustering catalyst}, which \setCap{would encourage intra-cluster crossover, reducing the number of generations required for the clusters to reach their respective optimal genomes (applications), therefore directly accelerating the evolutionary self-organisation in determining applications (Agent-sequences) to user requests. So, accelerating the responsiveness of the Digital Ecosystem to the user base, and the process of \emph{ecological succession} \cite{begon96}.}{CC1cap}

\vspaceaas

Crossover involves the crossing of two Agent-sequences, leading to recombination in the creation of new Agent-sequences, during the replication stage of the evolutionary cycle \cite{back1996eat}. This augmentation would encourage inter-cluster crossover within the evolving Agent Populations, with the aim of directly accelerating them in to find the optimal Agent-sequence(s) in fewer generations. As each \emph{evolving Agent Population} within the Digital Ecosystem would be accelerated, the entire ecosystem would operate more efficiently.

\tfigure{scale=1.0}{CCfinal}{graffle}{Clustering Catalyst}{This \getCap{CC1cap}}{-6mm}{!h}{}{}

This augmentation has considerable potential to optimise the evolving Agent Populations of Digital Ecosystems, but to be effective the determination of clusters needs to be computationally negligible, otherwise the overall effect would be counterproductive. While evolving Agent Populations would find the optimal application (Agent-sequence) within fewer generations, more time overall would be required. 

Our work on clustering with Physical Complexity, from section \ref{cluster123}, may prove useful for this augmentation.

\subsection{Replacement Aggregator}

\emph{Evolutionary computing} \cite{eiben2003iec} was chosen exclusively for the aggregation (\emph{combinatorial optimisation} \cite{papadimitriou1998coa}) of the Agents into optimal Agent-sequences (applications), without comparison to other techniques, because are focus was creating the digital counterpart of \emph{biological ecosystems}. If we were to assume it might not be the optimal technique, we could consider a \emph{replacement aggregator} to perform the aggregation of the Agents with an alternative technique, potentially accelerating the responsiveness of the Digital Ecosystem to the user base, and the process of \emph{ecological succession} \cite{begon96}.

\tfigure{scale=1.0}{NNreplace}{graffle}{Replacement Aggregator}{This would work by treating the {evolving Agent Population}, the embodiment of {evolutionary computing} in Digital Ecosystems, as an interchangeable {module}, and considering a {replacement aggregator} to perform the aggregation of the Agents with an alternative technique, potentially {accelerating} the responsiveness of the Digital Ecosystem to the user base, and the process of {ecological succession} \cite{begon96}.}{-6mm}{!h}{-5mm}{}

This augmentation would work by treating the \emph{evolving Agent Population}, the embodiment of \emph{evolutionary computing} in Digital Ecosystems, as an interchangeable module. It also assumes a more effective aggregator can be found to perform the \emph{combinatorial optimisation} \cite{papadimitriou1998coa} that occurs in response to a user request, on the set of Agents and Agent-sequences available from the Agent-pool of a Habitat. As each Agent aggregation process within the Digital Ecosystem would be accelerated, the entire ecosystem would operate more efficiently.

This augmentation could even allow for a range of available aggregators, choosing the most effective depending on the user, or on a case-by-case basis. However, replacing the evolutionary mechanism of the Digital Ecosystem with an alternative technique would weaken its \acl{EOA}; potentially risking the loss of valuable behaviour, such as emergent self-organisation, scalability and sustainability, imbibed from creating the digital counterpart of biological ecosystems. So, while the modular nature of the \acl{EOA} of Digital Ecosystems makes this augmentation possible, it would not be prudent.

\subsection{Agent-Pool Aggregation}

\tfigure{scale=1.0}{complementNEW}{graffle}{Agent-Pool Aggregation}{This augmentation allows \getCap{com1Cap}.}{-6mm}{!b}{}{}

The appealing vision of the \emph{Agent-pool aggregation} is that of the Agents intelligently recombining with one another, joining and leaving Agent-sequences of their own accord to improve the responsiveness of the Digital Ecosystem, allowing \setCap{for the creation of potentially useful applications (Agent-sequences) or partial applications inside the Agent-pools, increasing and optimising the recombination that occurs globally within the Digital Ecosystem. So, it would help to optimise the Agent-sequences at the Agent-pools of the Habitats, which would in turn optimise the evolving Agent Populations, as they make use of the Agent-pools}{com1Cap} when determining applications (Agent-sequences) to user requests. Therefore, accelerating the process of \emph{ecological succession} \cite{begon96}, and so the responsiveness of the Digital Ecosystem to the user base.

This augmentation would work by providing the Agents with the opportunity to interact inside the Agent-pools and the ability to determine whether to recombine with one another, outside the evolutionary optimisation of the evolving Agent Populations. For the Agents to judge potential re-combinations, they will require an understanding of the context in which they would operate, most importantly the past user requests of the Habitat where the recombination would occur. So, this augmentation would optimise the set of Agents and Agent-sequences at the Agent-pools, and therefore indirectly optimise and accelerate the evolving Agent Populations within the Digital Ecosystem. As each evolving Agent Population within the Digital Ecosystem would be accelerated, the entire ecosystem would operate more efficiently.

This augmentation would strengthen the Agent concept within the \acl{EOA} of Digital Ecosystems, endowing the individual Agents with some intelligence and control over their behaviour. However, a sophisticated process would be required for the Agents to evaluate a potential recombination, considering their descriptions with other Agents' descriptions collectively, within the context of the past user requests handled by the Habitat where the recombination is to occur. Additionally, a scalable mechanism would be required to determine which re-combinations the Agents should evaluate, because generally it will be impractical to evaluate all the re-combinations possible at any one time. Interestingly, the effectiveness of this augmentation relies on the local interactions of the Agents, producing an emergent global optimising effect on the evolving Agent Populations to accelerate the \emph{ecological succession} of a Digital Ecosystem.

\subsection{Targeted Migration}

\label{tmsection}

The \emph{self-organised diversity} experiments from the previous chapter, showed that the Digital Ecosystem can be slow to optimally distribute the Agents, within the Habitat network, relative to the user request behaviour. So, potential exists to optimise the distribution of the Agents within the Habitat network, through additional \emph{targeted migration} of the Agents, which would indirectly optimise the evolving Agent Populations. The \emph{migration probabilities} between the Habitats produces the existing passive Agent migration, allowing the Agents to spread in the correct general direction within the Habitat network, based primarily upon success at their current location. This augmentation will work in a more active manner, allowing the Agents highly targeted migration to specific Habitats, in addition to the generally directed passive migration. It will help to optimise the Agents found at the Agent-pools of the Habitats, which would in turn optimise the evolving Agent Populations as they make use of the Agent-pools when determining applications (Agent-sequences) to user requests. So, accelerating the process of \emph{ecological succession} \cite{begon96}, and therefore the responsiveness of the Digital Ecosystem to the user base.

\tfigure{scale=1.0}{complementMIGRATION}{graffle}{Targeted Migration}{This augmentation would optimise the distribution of the Agents within the Habitat network, through additional {targeted migration} of the Agents, helping to {optimise} the Agents found at the Agent-pools of the Habitats, which would in turn optimise the evolving Agent Populations. So, {accelerating} the process of {ecological succession} \cite{begon96}, and therefore the responsiveness of the Digital Ecosystem to the user base.}{-10mm}{}{}{}

This augmentation would work by providing the Agents with the opportunity to interact inside the Agent-pools, outside of the evolutionary optimisation of the evolving Agent Populations, to determine if they are functionally similar based on their \emph{semantic descriptions}. Similar Agents will compare their \emph{migration histories} to determine Habitats where they could find a niche (i.e. be valuable). This would lead to additional highly targeted migration of Agents throughout the Habitat network, optimising the set of Agents and Agent-sequences at the Agent-pools, and therefore indirectly optimising and accelerating the evolving Agent Populations within the Digital Ecosystem. As each evolving Agent Population within the Digital Ecosystem would be accelerated, the entire ecosystem would operate more efficiently.

This augmentation would strengthen the Agent concept within the \acl{EOA} of Digital Ecosystems, endowing the individual Agents with some intelligence and control over their behaviour. Interestingly, the effectiveness of this augmentation relies on the local interactions of the Agents, producing an emergent global optimising effect on the evolving Agent Populations to accelerate the \emph{ecological succession} of a Digital Ecosystem.

\subsection{Choice of Augmentation}

\tfigure{scale=0.9}{choiceOfApproach}{graffle}{Effect of The Proposed Augmentations}{The \getCap{ch1ifAp}. \getCap{ch2ifAp}}{-5mm}{!h}{}{}

The question of which of the alternative augmentations are most promising, and therefore which we should pursue to theoretical completion and then experimental confirmation, is not obvious. The \emph{evolving Agent Populations} are the embodiment of \emph{evolutionary computing} \cite{eiben2003iec} in Digital Ecosystems, while the Habitats are the embodiment of the \emph{ecology-based computing} we have developed for Digital Ecosystems. So, we will start by considering the \setCap{effect of the different augmentations on the \emph{evolutionary dynamics} (the evolving Agent Populations) and the \emph{ecological dynamics} (the Habitats)}{ch1ifAp}, as shown in Figure \ref{choiceOfApproach}. \setCap{This separation of concerns is an artificial construct, but useful in summarising the potential of the different augmentations, before we decide upon which to pursue.}{ch2ifAp}

The \emph{clustering catalyst} has potential to accelerate evolving Agent Populations with clusters, and therefore the process of \emph{ecological succession} \cite{begon96}. However, a computationally negligible technique is needed to determine the required clustering, else the overall effect will be counterproductive. Nevertheless, we will pursue this augmentation further; first to theoretical completion, and then to experimental simulations for confirmation.

The \emph{replacement aggregator} could prove effective for using a range of techniques when finding the optimal aggregation of the Agents into Agent-sequences in response to user requests. However, it would weaken the \acl{EOA} of Digital Ecosystems and presume more effective techniques than \emph{evolutionary computing} \cite{eiben2003iec} can be found for the \emph{combinatorial optimisation} \cite{papadimitriou1998coa} of the Agent aggregation. The first point would obviously be undesirable, potentially risking the loss of valuable behaviour, such as emergent self-organisation, scalability and sustainability, imbibed from creating the digital counterpart of \emph{biological ecosystems}. Regarding the second point, it has been shown \cite{D8.1} that when considering the combinatorial optimisation of Agent aggregation as the weighted set-cover problem that \emph{evolutionary computing} and \emph{simulated annealing} are more effective than \emph{steepest descent}, \emph{Tabu search}, and \emph{random search}. Also that \emph{evolutionary computing} is more widely applicable (without performance degradation) than \emph{simulated annealing} \cite{D8.1}. So, while there may be other applicable techniques that have not been evaluated, the certainty of success with this augmentation is considerably reduced, and therefore will not be pursued further.

The \emph{Agent-pool aggregation} could prove very effective in optimising and accelerating the process of \emph{ecological succession} \cite{begon96}, but the computational cost would be considerable, most likely making it impractical. So, while we are optimistic regarding the potential success of this augmentation theoretically, the experimental impracticality leads us not to pursue it any further.

The \emph{targeted migration} also has considerable potential to optimise and accelerate the process of \emph{ecological succession} \cite{begon96}, by improving the migration of Agents through the Habitat network of the Digital Ecosystem. Also, it would directly address the scope for optimisation identified from the self-organised diversity experiment of section \ref{divmodexp}. Furthermore, it is the only augmentation to effect the ecological dynamics directly, which makes it desirable to pursue, because while evolution may be well understood in computer science under the auspices of \emph{evolutionary computing} \cite{eiben2003iec}, ecology until our efforts had not been widely explored. So, there is inherently more potential to improve the ecological dynamics than the evolutionary dynamics, and therefore more potential in this augmentation than the others. However, creating a mechanism for the Agents to determine if they are functionally similar with one another based on their \emph{semantic descriptions} will be a challenge. Nevertheless, we will pursue this augmentation further; first to theoretical completion, and then to experimental simulations for confirmation.

\section{Clustering Catalyst}
\label{ccfull}

The \emph{clustering catalyst} will directly optimise the evolutionary self-organisation of evolving Agent Populations with clusters, by encouraging intra-cluster crossover. Crossover involves the crossing of two Agent-sequences in the creation of new Agent-sequences, during the replication stage of the evolutionary cycle \cite{back1996eat}. Theoretical completion of the \emph{clustering catalyst} requires consideration of how best to determine the clusters within an evolving Agent Population, so we will now consider suitable clustering algorithms.

\subsection{Clustering}
\label{haacspec}

We have understood clustering, within the context of evolving Agent Populations, as the amassing of same or similar sequences around an optimum genome \cite{begon96}, but more generally clustering is the classification of objects into different groups, or more precisely, the partitioning of a data set into subsets (clusters), so that the data in each subset share some common trait, often proximity according to a distance measure \cite{jain1988acd}. If the number of clusters $k$ is not apparent from prior knowledge, several methods are available for its determination \cite{milligan1985epd}. For our simulations we will make use of prior knowledge to determine the number of clusters $k$, because if available it is obviously the most effective method, and because our focus is on determining the effectiveness of the \emph{clustering catalyst}. Naturally, the most suitable method for determining the number of clusters can be investigated if the \emph{clustering catalyst} proves to be effective.

An important step in any clustering is to select a distance measure, which calculates the similarity of two elements \cite{jain1988acd}. We will use a distance measure based on our simulated \emph{fitness function} from section \ref{fitnessFunction}, because it is itself based on a distance metric. So, given two Agent-sequences $A$ and $B$, consisting of a set of attributes ${a_1, a_2, ...}$ and ${b_1, b_2, ...}$ respectively, the distance between them will be
\begin{equation}
distance(A,B) = \sum_{b \in B}{|b-a|},
\label{df}
\end{equation}
where $a$ is the member of $A$ such that the difference to the required attribute $b$ is minimised.

A range of clustering algorithms are available, to the extent that taxonomies having been proposed \cite{grabmeier2002tca, jain1999dcr} for their classification. The top-level classification being between hierarchical and partitional algorithms, both of which we shall now explore.

\subsubsection{Hierarchical Clustering}

Hierarchical clustering builds (agglomerative), or breaks up (divisive), a hierarchy of clusters \cite{jain1999dcr}. The traditional representation of this hierarchy is a tree (called a dendrogram), with individual elements at one end and a single cluster containing every element at the other \cite{jain1988acd}. Agglomerative algorithms begin at the leaves of the tree, whereas divisive algorithms begin at the root \cite{jain1999dcr}. 

Agglomerative clustering starts with all the objects as individual clusters, which are then merged according to their similarities until all are fused into a single cluster, with the most similar objects being grouped first \cite{jain1988acd}. The similarity criterion is determined by the linkage analysis, for which there are three common forms \cite{olson1995pah}: 
\vspace{-4mm}
\begin{itemize}
\item Single-link (nearest neighbour or minimum distance) \cite{sneath1973nt}: is obtained by fusing clusters according to the distance between their nearest members.
\item Complete-link (farthest neighbour or maximum distance) \cite{king1967swc}: is obtained by fusing clusters according to the distance between their farthest members.
\item Average-link (average distance) \cite{olson1995pah}: is obtained by fusing clusters according to the 
average distance between pairs of members in the respective sets. 
\end{itemize}
\vspace{-4mm}
Most hierarchical agglomerative clustering algorithms are variants of the single-link, complete-link and average-link algorithms \cite{jain1999dcr}. Most notably, the minimum-variance algorithm \cite{ward1963hgo}, which is a variant of the average-link algorithm, except instead of minimising an average distance it minimises a squared distance weighted by cluster size \cite{murtagh1984sra}. The single-link and complete-link algorithms are the most popular \cite{jain1999dcr}. However, the single-link algorithm suffers from a chaining effect \cite{nagy1968sap}, which tends to produce clusters that are straggly or elongated \cite{jain1999dcr}. In contrast, the complete-link algorithm produces tightly bound or compact clusters \cite{baezayates1992ids}. While the average-link algorithm \cite{olson1995pah} is designed to reduce the dependence of the cluster-linkage criterion on extreme values, such as the most similar or dissimilar of the single-link and complete-link algorithms \cite{olson1995pah}, and results in clusters that tend to have approximately equal within-cluster variability \cite{larose2005dkd}. The minimum-variance algorithm tends to join clusters with a small number of observations first, being strongly biased to producing clusters with the same number of observations, and therefore is very sensitive to outliers \cite{milligan1980ees}.

Divisive clustering methods start with one cluster containing all the objects, which are successively separated into smaller subgroups until the number of clusters equals the number of objects \cite{jain1999dcr}. There are two forms: monothetic, which divides the data by the possession of a single specified attribute, and polythetic, where divisions are based on several attributes \cite{jain1999dcr}. Agglomerative algorithms make clustering decisions based on local patterns without initially considering the global distribution, and these early decisions cannot be undone; while divisive clustering benefits from complete information about the global distribution when making top-level partitioning decisions \cite{manning2008iir}. It also has the advantage of being more efficient if we do not generate a complete hierarchy all the way down to the individual objects \cite{manning2008iir}. However, divisive clustering is conceptually more complex than agglomerative clustering, since a second flat clustering algorithm is required as a subroutine \cite{manning2008iir}. There are also graph-theoretic divisive clustering algorithms, with the best-known based on the construction of the minimal spanning tree of the data, deleting the edges with the largest lengths to generate the clusters \cite{zahn1971gtm}.

\subsubsection{Partitional Clustering}

A partitional clustering algorithm determines a single partition of the data \cite{jain1999dcr}, instead of a clustering structure such as the dendrogram produced by a hierarchical algorithm \cite{jain1988acd}. Partitional algorithms usually produce clusters by optimising a criterion function defined either locally (on a subset of the patterns) or globally (defined over all the patterns) \cite{jain1999dcr}. A combinatorial search for the set of possible labellings, to determine the optimum value of a criterion, is clearly computationally prohibitive, and so in practise the algorithm is run multiple times with different starting states, with the best configuration being used as the output of the clustering \cite{jain1999dcr}. Partitional algorithms have advantages in applications involving large data sets for which the construction of a dendrogram is computationally prohibitive \cite{jain1999dcr}. A problem accompanying partitional algorithms is choosing the number of desired output clusters beforehand \cite{dubes1987mcb}, which is not required for hierarchical clustering. There are many forms of partitional clustering algorithms, including mixture-resolving \cite{jain1988acd}, mode-seeking \cite{jain1988acd}, nearest neighbour \cite{lu1978ssc}, fuzzy clustering \cite{zadeh1996fs}, artificial neural networks \cite{sethi1991ann}, and others \cite{jain1999dcr}.

The most intuitive and frequently used criterion function in partitional clustering algorithms is the \emph{squared error criterion}, which tends to work well with isolated and compact clusters \cite{jain1999dcr}. The most commonly used algorithm employing a squared error criterion is the k-means algorithm \cite{macqueen1965smc}, popular because it is easy to implement \cite{kanungo2002ekm}. It starts with a random initial partition and keeps reassigning the patterns to clusters, based on the similarity between a pattern and the cluster centres, until a convergence criterion is met \cite{macqueen1965smc}. However, a major problem with this algorithm is that it is sensitive to the selection of the initial partition, and may converge to a local optimum of the criterion function value if the initial partition is not properly chosen \cite{galindo2008hrf}.

\subsubsection{Choice of Algorithm}

The choice of the optimal clustering algorithm very much depends on the structure of the data, because clustering is subjective, such that the same data can be partitioned differently for different purposes \cite{jain1999dcr}. We also require a clustering algorithm with a negligible computational cost for the \emph{clustering catalyst} to be effective. So, we choose a hierarchical clustering algorithm over a partitional one, as it is more appropriate for small data sets \cite{jain1999dcr}, such as expected from an evolving Agent Population with clusters. We choose a hierarchical agglomerative clustering algorithm over a divisive one, because it is conceptually simpler \cite{manning2008iir}, and because the efficiency advantage \cite{manning2008iir} of a divisive algorithm would be negligible, given the small size of the expected data set. Finally, we choose a hierarchical agglomerative average-link clustering algorithm of time complexity $O(n^2 \log n)$ \cite{manning2008iir}, over a single-link one of $O(n^2)$ time complexity, a complete-link one of $O(n^2 \log n)$ time complexity, or a minimum-variance one of $O(n^2)$ time complexity \cite{day1984eaa}, because the average-link algorithm is designed to reduce the dependence of the cluster-linkage criterion on extreme values, such as the most similar or dissimilar of the single-link and complete-link algorithms \cite{olson1995pah}; and because the minimum-variance algorithm is biased to producing clusters with the same number of objects \cite{milligan1980ees}, which would be problematic for clusters emerging over the generations. Any difference in execution time of the algorithms would be minimal, despite the different algorithmic time complexities, because of the small size of the expected data set $n$, the Population size. So, we choose a hierarchical agglomerative average-link clustering algorithm \cite{olson1995pah} for our \emph{clustering catalyst}.

\subsection{Physical Complexity Clustering}
\label{pccspec}

We also considered a clustering algorithm based on our extended Physical Complexity from Chapter \ref{ch:investigation}, because clustering is subjective in nature \cite{jain1999dcr}, and our extended Physical Complexity was developed, in section \ref{cluster123}, to understand the clustering of evolving Agent Populations. In our algorithm, with the number of clusters $k$ determined also from prior knowledge, we first sort the evolving Agent Population, then process its Agent-sequences linearly, adding each to the cluster that maximises the Efficiency $E_{c}$ (\ref{efficiencyMultiple}) of the Population. The pre-assignment sorting ensures that the cores of the clusters are established for the Efficiency $E_{c}$ to have the necessary sensitivity when assigning the Agent-sequences of greater uniqueness. The pseudocode for our algorithm is shown in Figure \ref{phyComClus}.

\setlength{\fboxsep}{5mm}
\mfigure{
\framebox{\begin{minipage}[]{160mm}
\begin{center}
\begin{minipage}[]{155mm}
\normallinespacing
\bf clusters[k];\\
group duplicate Agent-sequences within the Population\\
order the groups within Population by greatest size\\
for each Agent-sequence in the ordered Population\\
\white{XX}if a duplicate and not first instance\\
\white{XXXX} then: assign to same cluster as first instance\\
\white{XXXX} else: the cluster that maximises the Efficiency $E_{c}$ of the Population\\
\white{XX}end if\\
end for
\end{minipage}
\end{center}
\end{minipage}}
}{Pseudo-Code for Physical Complexity Clustering}{With the number of clusters $k$ determined from prior knowledge, we first sort the evolving Agent Population, then process its Agent-sequences linearly, adding each to the cluster that maximises the Efficiency $E_{c}$ (\ref{efficiencyMultiple}) of the Population. The pre-assignment sorting is for the Efficiency $E_{c}$ to have the necessary sensitivity when assigning the Agent-sequences of greater uniqueness.}{phyComClus}{4mm}{-4mm}{!h}

Our algorithm will be computationally negligible, because, based on the pseudo-code, it will have a time complexity of $O(n^2)$, where $n$ is the Population size. It will also be as accurate as an exhaustive search, which would have and exponential time complexity, because of the sensitivity of the Efficiency $E_{c}$ when assigning the Agent-sequences to the clusters.

Now that the \emph{clustering catalyst} is theoretically complete, with two alternative clustering algorithms, we can confirm its effect experimentally through simulations.

\section{Targeted Migration}

\label{tm2section}

The \emph{targeted migration} will directly optimise the ecological migration, and therefore indirectly complement the evolutionary self-organisation of the evolving Agent Populations, through the highly \emph{targeted migration} of the Agents to their niche Habitats. The \emph{migration probabilities} between the Habitats produces the existing passive Agent migration, allowing the Agents to spread in the correct general direction within the Habitat network, based primarily upon success at their current location. The \emph{targeted migration} will work in a more active manner, allowing the Agents highly targeted migration to specific Habitats, based upon their interaction with one another to discover Habitats where they could be valuable (i.e. find a niche). Theoretical completion of the \emph{targeted migration} requires further consideration of how it will operate, including its effect on the Agent life-cycle, and a suitable \emph{pattern recognition} \cite{jain2000spr} technique for the required \emph{similarity recognition}.

The \emph{targeted migration} will occur when users deploy their services, specifically when deploying their representative Agents to their Habitats within the Digital Ecosystem, and upon the execution of applications (groups of services), specifically the resulting passive migration of their representative Agent-sequences between the Habitats. The Agent-sequences arriving at Habitats, with respect to the \emph{targeted migration}, will be treated as individual Agents arriving at the Habitats. So, an Agent arriving at a Habitat interacts one-on-one with Agents already present within the Agent-pool of the Habitat, and upon determining functional similarity, based upon comparing their \emph{semantic descriptions}, will share other Habitats successfully visited from their respective \emph{migration histories}. An Agent \emph{migration history}, as defined in section \ref{migrationHistory}, is the migratory path of the Agent through the Habitat network, including its use at the Habitats visited. So, similar Agents can share their \emph{migration histories} to discover new Habitats where they could be valuable, and then use \emph{targeted migration} (via a copy, and not a move) to explore the most promising of the recently acquired Habitats. This will allow successfully interacting Agents to target specific Habitats where they will potentially be useful, but risks potentially infinite \emph{targeted migration}, because \emph{targeted migration} itself can lead to further \emph{targeted migration}. So, each Agent will require a dynamic \emph{targeted migrations counter}, which defines the number of permitted targeted migrations of the Agent. This counter will be incremented upon an Agent's execution in response to a user request, and decremented upon performing a \emph{targeted migration}.

\tfigure{scale=1.0}{newLifeCycle}{graffle}{Agent Life-Cycle With Targeted Migration}{\getCap{nlccap} \getCap{nlc2cap}.}{-2mm}{}{}{}

\setCap{The Agent life-cycle, defined in section \ref{agentLifeCycle}, will change to support the \emph{targeted migration}, as shown}{nlccap} in Figure \ref{newLifeCycle} \setCap{by the blue circle. Specifically, there will be more opportunities for Agent migration, but more importantly these opportunities will be for \emph{targeted migration}, which will help to optimise the set of Agents found at the Habitats}{nlc2cap}, and therefore support the evolving Agent Populations created in response to user requests for applications. The \emph{targeted migration} will essentially short-circuit the hierarchical topology of the Habitat network, which is what allows it to specialise and localise to communities, providing specific solutions to specific requests from specific users. However, the \emph{targeted migration} will also reinforce the hierarchical topology of the Habitat network, because \emph{targeted migration} between connected Habitats will accelerate the existing migration of Agents, while between unconnected Habitats will assist the Digital Ecosystem in supporting emerging communities. So, the \emph{targeted migration} will help strengthen and catalyse the formation of clusters within the Habitat network, and will also assist in locating Habitats within the correct clusters. Therefore, the optimisation of the Digital Ecosystem will be a global emergent effect resulting from the local interactions of the Agents, allowing for niches to be fulfilled faster and so accelerating the process of \emph{ecological succession} \cite{begon96}. Also, the Digital Ecosystem will adapt faster to changing environmental conditions (e.g. changes in the request behaviour of user communities). In biological terms the \emph{targeted migration} endows the Agents with a form of \emph{reciprocal altruistic behaviour} \cite{trivers1971era}, consistent with the Agent paradigm of the \acl{EOA}.

\subsection{Similarity Recognition}

For the \emph{targeted migration} to work successfully an effective technique will be required for the \emph{similarity recognition} between the \emph{semantic descriptions} of two Agents. Each Agent will have an embedded \emph{similarity recognition} component to maintain the consistency of the Agent paradigm of \aclp{EOA}. So, the Agents will interact one-on-one to determine functional similarity based upon their \emph{semantic descriptions}, using their embedded \emph{similarity recognition} components, with each of the two interacting Agents determining similarity for themselves. Again, this is to maintain the consistency of the Agent paradigm. \emph{Similarity recognition} between the \emph{semantic descriptions} of two Agents will require some form of \emph{pattern recognition}, because there is no single standard for the \emph{semantic description} of services \cite{cabral2004asw}, and adopting one over the others would be inconsistent with the inclusive nature of Digital Ecosystems. So, we will now consider the field of \emph{pattern recognition} to determine suitable techniques for the \emph{similarity recognition} components to be embedded within the Agents.

\emph{Pattern recognition} aims to classify data (patterns) based on \emph{priori knowledge} or on statistical information extracted from the data \cite{ripley1996pra}. \emph{Pattern recognition} requires a sensor or sensors for data acquisition, a pre-processing technique, a data representation scheme, and a decision making model \cite{jain2000spr}. Also, learning from a set of examples (training set) is an important and desirable feature of most pattern recognition systems \cite{ripley1996pra}. The four best known approaches for pattern recognition are \cite{jain2000spr}: 

\vspace{-2mm}

\narrowlinespacing
\begin{itemize}
\item Template Matching
\item Statistical Classification
\item Structural Matching
\item Neural Networks
\end{itemize}
\normallinespacing

\vspace{-5mm}

These approaches are not necessarily independent, and sometimes the same pattern recognition method exists with different interpretations \cite{jain2000spr}. For example, attempts have been made to design hybrid systems involving multiple approaches, such as the notion of \emph{attributed grammars} which unifies structural and statistical pattern recognition \cite{fu1982spr}.

\subsubsection{Template Matching}

Template Matching is the simplest and earliest approach to pattern recognition, and involves a generic operation to determine the similarity between two entities (points, curves, or shapes) of the same type \cite{jain2000spr}. A template (typically, a 2D shape) is available, with the pattern to be recognised being matched against the stored template, while considering all allowable changes translation, rotation and scale \cite{jain2000spr}. The similarity measure, often a correlation, may be optimised based on a training set, and often the template itself is defined from a training set \cite{jain2000spr}.

Template Matching is computationally demanding, but the availability of ever faster processors has made it more feasible \cite{jain2000spr}. While effective for some application domains, it has several disadvantages \cite{nixon2007fei}. For instance, poor performance if the patterns are distorted from the imaging process or a viewpoint change, or if there are large intra-class variations among the patterns \cite{jain2000spr}. Deformable template models \cite{grenander2007ptr} or rubber sheet deformations \cite{bajcsy1989mem} can help to compensate when the deformation cannot be easily explained or modelled directly.

\subsubsection{Statistical Classification}

In Statistical Classification each pattern is represented in terms of $d$ features or measurements, and is viewed as a point in a $d$-dimensional feature space, with the goal being to choose those features that allow pattern vectors belonging to different categories to occupy compact and disjoint regions in the $d$-dimensional feature space \cite{webb1999spr}. The effectiveness of which is determined by how well patterns from different classes can be separated \cite{jain2000spr}. The decision boundaries of the $d$-dimensional feature space can be determined from probability distributions, of the patterns belonging to each class, which must be specified or learnt \cite{devroye1996ptp, duda1973pca}. 

One can also take a discrimination analysis based approach to classification, in which a parametric form of a decision boundary is specified, and then the best decision boundary of the specified form is found based on the classification of training patterns \cite{jain2000spr}. Such boundaries can be constructed using, for example, a mean squared error criterion \cite{webb1999spr}. These direct boundary construction approaches are supported by the philosophy \cite{vapnik2006edb} that if you possess a restricted amount of information for solving some problem, try to solve the problem directly and never solve a more general problem as an intermediate step, because it is possible that the available information is sufficient for a direct solution but insufficient for solving a more general intermediate problem.

Statistical approaches are generally characterised by having an explicit underlying probability model, which provides a probability of being in each class and not just a classification, and hence some human intervention is assumed regarding variable selection and transformation, and the overall structuring of the problem \cite{michie1995mln}.

\subsubsection{Structural Matching}

In Structural Matching a hierarchical perspective is adopted, where a pattern is viewed as being composed of simple sub-patterns, which are built from yet simpler sub-patterns \cite{pavlidis1977spr}, and therefore it is applicable to many recognition problems involving complex patterns \cite{jain2000spr}. The elementary (simplest) sub-patterns to be recognised are called primitives, with the given complex pattern to be represented in terms of the interrelationships between these primitives \cite{pavlidis1977spr}. Where the structure is syntactic, a formal analogy is drawn between the structure of patterns and the syntax of language. So, the primitives are viewed as the alphabet of the language, and the patterns are viewed as sentences generated according to the grammar of the language \cite{fu1982spr}. Thus, a large collection of complex patterns can be described by a small number of primitives and grammatical rules, which must be inferred from the available training samples \cite{jain2000spr}.

Structural pattern recognition is intuitively appealing because, in addition to classification, it also provides a description of how the given pattern is constructed from the primitives \cite{jain2000spr}. It is used in situations where the patterns have a definite structure that can be captured by a set of rules, such as electrocardiogram waveforms \cite{trahanias1990spr}, textured images \cite{haralick1979sas}, and the shape analysis of contours \cite{loncaric1998ssa}. However, the implementation of structural approaches leads to many difficulties, including the segmentation of noisy patterns (to detect primitives) and the inference of grammar from training data \cite{jain2000spr}. There can also be a combinatorial explosion of possibilities to be investigated, demanding large training sets and significant computational effort \cite{perlovsky1998ccc}.

\subsubsection{Neural Networks}

\acp{NN} can be viewed as massively parallel computing systems consisting of an extremely large number of simple processors with many interconnections \cite{ripley1996pra}. \ac{NN} models attempt to use certain organisational principles (such as learning, generalisation, adaptivity, fault tolerance, distributed representation, and computation) in a network of weighted directed graphs, in which the nodes are artificial neurons, and the directed edges (with weights) are connections between the neuron outputs and inputs \cite{ripley1996pra}. The main characteristics of \acp{NN} are their ability to learn complex nonlinear input-output relationships, use sequential training procedures, and adapt themselves to the data \cite{jain2000spr}. 

The most commonly used family of \acp{NN} for pattern classification tasks is the feed-forward network, including multilayer perceptrons, which are organised into layers and has unidirectional connections between the layers \cite{jain2000spr}. Another popular network is the \acl{SOM}, or Kohonen-Network \cite{kohonen}, which is often used for feature mapping \cite{jain2000spr}. The increasing popularity of NN models to solve pattern recognition problems has been primarily because of their low dependence on domain-specific knowledge (relative to model-based and rule-based approaches) and the availability of efficient learning algorithms \cite{jain2000spr}. The learning process involves updating the network architecture and connection weights so that a network can efficiently perform a specific classification \cite{ripley1996pra}.

\acp{NN} provide a suite of nonlinear algorithms for feature extraction (using hidden layers) and classification (e.g. multilayer perceptrons) \cite{jain2000spr}. In addition, existing feature extraction and classification algorithms can be mapped onto \ac{NN} architectures for efficient (hardware) implementation \cite{table1994hia}. Despite the seemingly different underlying principles, most of the well-known \ac{NN} models are implicitly equivalent or similar to classical statistical pattern recognition methods \cite{jain2000spr}. However, \acp{NN} offer several advantages, such as unified approaches for feature extraction and classification, and flexible procedures for finding good, moderately nonlinear solutions \cite{jain2000spr}.

\subsubsection{Support Vector Machines}

One of the most interesting recent developments in classifier design is the introduction of the \ac{SVM} \cite{vapnik1998slt}, which is primarily a two-class classifier, and therefore highly suitable for the required \emph{similarity recognition} component of our \emph{targeted migration}. It uses an optimisation criterion that is the width of the margin between the classes, i.e. the empty area around the decision boundary defined by the distance to the nearest training patterns \cite{burges98tutorial}. These patterns, called \emph{support vectors}, define the classification function, and their number is minimised by maximising the margin \cite{burges98tutorial}. This is achieved through a kernel function $K$, which transposes the data into a higher-dimensional space where a hyperplane performs the separation \cite{burges98tutorial}. In its simplest form the kernel function is just a dot product between the input pattern and a member of the support set, resulting in a linear classifier, while nonlinear kernel functions lead to a polynomial classifier \cite{jain2000spr}.

\acp{SVM} are closely related to \aclp{NN}, being a close cousin to classical multilayer perceptrons, with the use of a sigmoid kernel function making them equivalent to two-layer perceptrons \cite{abe2005svm}. However, in the training of \acp{NN}, such as multi-layer perceptrons, the weights of the network are found by solving a non-convex unconstrained minimisation problem, while the use of a kernel function in \acp{SVM} solves a quadratic programming problem with linear constraints \cite{scholkopf1997csv}.

An important advantage of \acp{SVM} is that they offer the possibility to train generalisable nonlinear classifiers in high-dimensional spaces using a small training set \cite{valafar2002s}. Furthermore, for large training sets a small support set is typically selected for designing the classifier, thereby minimising the computational requirements during training \cite{valafar2002s}.

\subsubsection{Choice of Technique}

Template Matching is not suitable for the required \emph{pattern recognition} of our \emph{targeted migration}, because its effective use is domain specific \cite{jain2000spr} and the \emph{similarity recognition} between the \emph{semantic descriptions} of Agents is very different to the domains that it is typically applied \cite{nixon2007fei}. Statistical Classification is also not suitable, because the embedded \emph{similarity recognition} component of each Agent would require human intervention for variable selection and transformation \cite{michie1995mln}. Structural Matching is suitable theoretically, but implementations lead to many difficulties \cite{jain2000spr}, including the segmentation of noisy patterns (to detect primitives) and the inference of grammar from training data \cite{jain2000spr}. There can also be a combinatorial explosion of possibilities to be investigated, demanding large training sets and significant computational effort \cite{perlovsky1998ccc}, neither of which is available. \aclp{NN} are suitable, given their low dependence on domain-specific knowledge and the availability of efficient learning algorithms \cite{jain2000spr}. \aclp{SVM}, albeit a recent development \cite{vapnik1998slt}, are also suitable \cite{joachims1997tcs}, being primarily a binary classifier \cite{jain2000spr} for training generalisable nonlinear classifiers in high-dimensional spaces using small training sets \cite{valafar2002s}. So, we will make use of both \aclp{NN} and \aclp{SVM} for the theoretical completion and implementation of our \emph{targeted migration}.

\subsection{Neural Networks}
\label{NNsubsection}

So, in the first instance, we will leverage the \emph{pattern recognition} capabilities of \acfp{NN} for the embedded \emph{similarity recognition} components of the Agents, allowing them to determine similarity to one another based on the similarity of their \emph{semantic descriptions}. We will use \emph{multilayer perceptrons} (feed-forward artificial \acp{NN}) with \emph{backpropagation} \cite{haykin1998nnc} to provide the required \emph{pattern recognition} behaviour, because of their ability to solve problems stochastically, which allows for approximate solutions to extremely complex problems \cite{haykin1998nnc}. They are a modification of the \emph{standard linear perceptron} \cite{rojas1996nns}, using three or more layers of neurons (nodes) with nonlinear activation functions to distinguish data that is not linearly separable, or separable by a hyperplane \cite{bishop1995nnp}. The power of the multilayer perceptron comes from its similarity to certain biological neural networks in the human brain, and because of their wide applicability has become the standard algorithm for any supervised-learning \emph{pattern recognition} process \cite{haykin1998nnc}.

A pre-processing \cite{bishop1995nnp} of Agent \emph{semantic descriptions} will be required that is consistent across the entire Digital Ecosystem, requiring an alphabetical ordering of the attribute tuples within a \emph{semantic description}, a standardisation of the length of the attributes, before finally making use of a binary encoding for processing by a \ac{NN} \cite{bishop1995nnp}. The assumption of information structured as tuples, including an attribute name and attribute value, is accurate for our simulated \emph{semantic descriptions}, but is also a reasonable assumption for any \emph{semantic description} of web services \cite{dustdar2005sws, medjahed2003cws, debruijn2006wsm, cabral2004asw}. To standardise the length of the attributes, after removing any white-space\footnote{A white-space is any single character or series of characters that represent horizontal or vertical space in typography \cite{friedl2006mre}.}, an average word length of six characters will be used, because 5.39 is the average word length for business English \cite{fox1999sim}. For the binary encoding we propose using Unicode (UTF-8), which is based on extending ASCII to provide multilingual support \cite{gillam2002udp}. However, ASCII's support of only English \cite{gillam2002udp} will be sufficient for our simulations. The size (number of neurons) of the \emph{input layer} \cite{haykin1998nnc} will be proportional to the \emph{semantic description} of the Agent in which the \ac{NN} is embedded, taking advantage of the variation in length of different \emph{semantic descriptions}, which will assist the \ac{NN}-based \emph{pattern recognition} in determining dissimilarity.

We will use a single \emph{hidden layer}, which is usually sufficient for most tasks \cite{bishop1995nnp}. The size of which will be determined through \emph{exploratory programming} \cite{sommerville2006se} in our simulations, because of the difficulty in determining the optimal size without training several networks and estimating the generalisation error \cite{sarle2008}, evident by the range of inconsistent \emph{rules of thumb} \cite{blum1992nnc, swingler1996, berry1997dmt, boger1997kea} available to define the optimal size. The \emph{output layer} \cite{bishop1995nnp} will consist of a single neurone to provide a binary (true or false) response to the question of whether another Agent's \emph{semantic description} is similar to the Agent's own \emph{semantic description}. We will use a threshold of $0.90$ on its output for the determination of similarity. The overall structure of the \acl{NN} is visualised in Figure \ref{NNstructure}.

\tfigure{scale=1.0}{NNstructure}{graffle}{Neural Network for the Similarity Recognition Component}{of Agents in Targeted Migration: Consisting of an {input layer} proportional to the {semantic description} of the Agent in which it is embedded. A single {hidden layer}, and an {output layer} consisting of a single neurone to provide a binary response to the question of whether another Agent's {semantic description} is similar.}{-5mm}{}{}{}

\emph{Multilayer perceptrons} use nonlinear activation functions, which were developed to model the frequency of action potentials (firing) of biological neurons in the brain \cite{haykin1998nnc}. The main activation function used in current applications is the sigmoid function \cite{haykin1998nnc}, a hyperbolic tangent that is normalised and in which the output $y$ of a neurone is the sum of the weighted input values $x$ \cite{bishop1995nnp}, 
\begin{equation}
y = \frac{1}{(1 + e^{-x})}.
\label{sigmoid}
\end{equation}
The weights $x$ between the neurons will be randomly initialised, then trained to the real numbers that provide the desired functionality, because learning occurs in the perceptron by changing the connection (synaptic) weights after each piece of data is processed, based on the error of the output compared to the expected result \cite{bishop1995nnp}. This is an example of \emph{supervised learning} and is carried out through backpropagation, a generalisation of the least mean squares algorithm \cite{haykin1998nnc}. The network is therefore trained by providing it with input and corresponding output patterns \cite{bishop1995nnp}.

\newcommand{\vspacesvm}{\vspace{5mm}}
\vspacesvm

The \ac{NN}-based embedded \emph{similarity recognition} component of an Agent will be trained when the Agent is deployed to a Habitat of the Digital Ecosystem. The initial \emph{training set} will consist of the \emph{semantic description} of the Agent as a positive match, and variants created from its own \emph{semantic description}. If the variant is less than 10\% different it will be processed as a positive match, else it will be processed as a negative match. The \emph{training set} can be extended based on experience, making use of when an Agent visits a Habitat through \emph{targeted migration} (i.e. one acquired from an inter-Agent interaction); if visiting the Habitat proves successful the \emph{semantic description} of the interacting Agent can be appended to the \emph{training set} as a positive match, else as a negative match.

\vspacesvm

\subsection{Support Vector Machines}
\label{SVMsubsection}

\vspacesvm

In the second instance, we will leverage the \emph{pattern recognition} capabilities of \acfp{SVM} for the embedded \emph{similarity recognition} components of the Agents, allowing them to determine similarity to one another based on the similarity of their \emph{semantic descriptions}. As \acp{SVM} are closely related to \aclp{NN}, being a close cousin to classical multilayer perceptrons \cite{abe2005svm}, we will make use of the \emph{pre-processing} and the \emph{training sets} defined in the previous subsection, which will also ensure a fair comparison of the \emph{pattern recognition} techniques in empowering the \emph{similarity recognition} components of the Agents.

\vspacesvm

The selection of a suitable kernel function is important, since it defines the \emph{feature space} in which the \emph{training set} is classified \cite{svm6}, operating as shown in Figure \ref{svm}. \setCap{A \acf{RBF} is recommended for text categorisation \cite{joachims1997tcs}, with the most common form of the \ac{RBF} being Gaussian \cite{gunn1998svm}.}{svmCap} 

\vspacesvm

Training a \ac{SVM} requires solving a large \ac{QP} optimisation problem, which \acf{SMO} breaks into a series of the smallest possible \ac{QP} problems. \ac{SMO} solves these small \ac{QP} problems analytically, which avoids using a time-consuming numerical \ac{QP} optimisation. \ac{SMO} scales between linear and quadratic time complexity, relative to the size of the \emph{training set}, because it avoids matrix computation \cite{platt1999fts}. The alternative, a standard \ac{PCG} chunking algorithm scales between linear and cubic time complexity, relative to the size of the \emph{training set} \cite{platt1999fts}. So \ac{SMO} is faster, up to a thousand times on real-world sparse data sets \cite{platt1999fts}. 

\tfigure{width=170mm}{svm}{graffle}{\acl{SVM}}{(modified from \cite{takahashi2008}): Visualisation showing the training set in the Input Space, and its binary classification by a hyperplane in the higher dimensional Feature Space, achieved through the kernel function. \getCap{svmCap}}{-5mm}{}{}{}

The issue of the learnt behaviour of the embedded \emph{similarity recognition} component of an Agent, whether \ac{SVM} or \ac{NN} based, being inherited when the Agent reproduces is known as the Baldwin effect \cite{baldwin1896nfe}. The Baldwin effect has always been controversial within biological ecosystems \cite{weber2003eal}, primarily because of the problem of confirming it experimentally \cite{sterelny2004rea}. Also, offspring in \emph{biological ecosystems} can be genetically different to their parents \cite{begon96}, such that any learnt behaviour could potentially be inappropriate. However, the offspring in our Digital Ecosystem are genetically identical to their parents (in terms of the individual Agents), and so it makes little sense to force the loss of learnt behaviour. Therefore, we doubt that the Baldwin effect will adversely affect our Digital Ecosystem, which we will confirm through our simulations.

Now that the \emph{targeted migration} is theoretically complete, with two alternative \emph{pattern recognition} techniques, we can confirm its effect experimentally through simulations.

\section{Simulation and Results}

We simulated the Digital Ecosystem using our simulation from section \ref{simRes} (unless otherwise specified), adding the classes and methods necessary to implement the proposed \emph{clustering catalyst} and \emph{targeted migration} augmentations. Each experimental scenario was run ten thousand times for statistical significance of the means and standard deviations calculated.

\subsection{Clustering Catalyst}

We implemented the \emph{clustering catalyst} as defined in sections \ref{ccsection} and \ref{ccfull}, with the clusters determined by hierarchical agglomerative average-link clustering or our Physical Complexity clustering. We also made use of our simulations from Chapter \ref{ch:investigation} to simulate evolving Agent Populations with between two and six clusters, varied according to a Gaussian distribution, with the crossover rate increased from 10\% to 25\% to provide a greater opportunity for the \emph{clustering catalyst} to operate.

\subsubsection{Control}

The \emph{clustering catalyst} benefited from the additional crossover, which alone could have been responsible for any observed optimisation, because it increased recombination \cite{lawrence1989hsd} in the evolving Agent Populations. So, increasing variation in the exploration of solutions, potentially reducing the number of generations required to evolve the optimal solution. Therefore, our experimental simulations included a \emph{crossover control} for the additional crossover, which excluded the \emph{clustering catalyst}.

\subsubsection{Hierarchical Clustering}

\tfigure{}{selectiveBreedingFit}{graph}{Graph of the Hierarchical Clustering Based Clustering Catalyst}{\getCap{sbcf} generations, \getCap{sbc2f} generations. \getCap{sbc3f} generations.}{-7mm}{}{}{5mm}

We started with the hierarchical agglomerative average-link clustering \cite{olson1995pah} based \emph{clustering catalyst}, as defined in section \ref{haacspec}, making use of RapidMiner \cite{rapidMiner} to perform the required clustering. In Figure \ref{selectiveBreedingFit} we graphed for the simulation runs the average number of generations required to evolve the optimal solution, for the evolving Agent Populations with the \emph{hierarchical clustering based clustering catalyst}, compared to the evolving Agent Populations with the \emph{crossover control}, and the evolving Agent Populations alone. \setCap{The evolving Agent Populations alone averaged 296 (3 s.f.)}{sbcf} generations with a standard deviation of 12.76 (2 d.p.), \setCap{while the evolving Agent Populations with the \emph{crossover control} showed a 9\% reduction, averaging 267 (3 s.f.)}{sbc2f} generations with a standard deviation of 8.91 (2 d.p.). \setCap{The evolving Agent Populations with the \emph{hierarchical clustering based clustering catalyst} failed to provide any further optimisation, averaging 281 (3 s.f.)}{sbc3f} generations with a standard deviation of 20.25 (2 d.p.).

\subsubsection{Physical Complexity Clustering}

\tfigure{}{selectiveBreedingHist}{graph}{Graph of the Physical Complexity Based Clustering Catalyst}{\getCap{sbch}, \getCap{sb2ch}, \getCap{sb3ch}.}{-7mm}{}{}{5mm}

Next we considered the \emph{Physical Complexity based clustering catalyst}, as defined in section \ref{pccspec}, making use of our simulations from section \ref{specs}. In Figure \ref{selectiveBreedingHist} we graphed for the simulation runs the average number of generations required to evolve the optimal solution, for the evolving Agent Populations with the \emph{Physical Complexity based clustering catalyst}, compared to the evolving Agent Populations with the \emph{hierarchical clustering based clustering catalyst}, the evolving Agent Populations with the \emph{crossover control}, and the evolving Agent Populations alone. \setCap{The evolving Agent Populations with the \emph{Physical Complexity based clustering catalyst} averaged 274 (3 s.f.)}{sbch} generations with a standard deviation of 17.53 (2 d.p.), \setCap{better than the evolving Agent Populations with the \emph{hierarchical clustering based clustering catalyst} which averaged 281 (3 s.f.) generations}{sb2ch} with a standard deviation of 20.25 (2 d.p.), \setCap{but still worse than the evolving Agent Populations with the \emph{crossover control} which averaged 267 (3 s.f.) generations}{sb3ch} with a standard deviation of 8.91 (2 d.p.).

\vspace{5mm}

Additional experiments were conducted in which we varied several different parameters to determine if there were any conditions under which the \emph{clustering catalyst} was effective. The varied parameters included the population size, the mutation rate and the crossover rate. However, extensive testing through multiple scenarios failed to show any significant reduction in the number of generations required to evolve the optimal solution. So, confirming that the evolving Agent Populations with the \emph{clustering catalyst} were less effective than the evolving Agent Populations with the \emph{crossover control}. \setCap{As the \emph{clustering catalyst} was unsuccessful,}{sbrc} in Figure \ref{selectiveBreedingRun} \setCap{we graphed a typical run of each scenario to observe its behaviour and so better understand why it failed. However, there was no unexpected behaviour, confirming that the evolving Agent Populations with the \emph{clustering catalyst} were simply less efficient than the evolving Agent Populations with the \emph{crossover control}.}{sbr2c}

\tfigure{}{selectiveBreedingRun}{graph}{Graph of Typical Runs for the Clustering Catalysts}{\getCap{sbrc} \getCap{sbr2c}}{-7mm}{}{}{}

The results showed that the \emph{clustering catalyst}, using hierarchical clustering \cite{olson1995pah} or Physical Complexity clustering, failed to optimise the evolutionary processes (with clusters) of the Digital Ecosystem. Additionally, typical runs of each showed no adverse behaviour, which might have explained the failure to optimise the evolutionary processes. Therefore, the results collectively confirm that the intra-cluster crossover assignment of the \emph{clustering catalyst} was less efficient than the random crossover assignment of the \emph{crossover control}. The \emph{clustering catalyst} intuitively had potential, but most likely failed because the individuals within the evolving Agent Populations lacked sufficient complexity (relative to biological populations \cite{begon96}) for the mechanism to be effective, leading to the crossing of very similar individuals, producing offspring that were very similar to their parents, and therefore not actually achieving valuable change.

\subsection{Targeted Migration}

We implemented the \emph{targeted migration} as defined in sections \ref{tmsection} and \ref{tm2section}, using both \acf{NN} and \acf{SVM} based \emph{similarity recognition} components embedded within the Agents. We also made use of our Xgrid \cite{kramer2004ulg} modifications from Chapter \ref{ch:investigation} to take advantage of the \emph{grids} mentioned in the acknowledgements.

\subsubsection{Controls}

The \emph{targeted migration} was dependent on additional Agent migration, which alone could have been responsible for any observed optimisation, because it led to greater distribution of the Agents within the Digital Ecosystem, potentially improving responsiveness for the user base. So, we included a \emph{migration control} in our experimental simulations for the additional Agent migration, being random instead of targeted. Furthermore, to determine the contribution of the \acp{NN} and \acp{SVM} on the \emph{targeted migration} we created a \emph{pattern recognition control}, using a rudimentary \emph{distance function} adapted from the \emph{fitness function} defined in section \ref{fitnessFunction}.

In Figure \ref{graphTarMigContHist} we graphed for the simulation runs the average of the percentage response rate after a thousand time steps (user request events), for the Digital Ecosystem with the \emph{migration control}, and the Digital Ecosystem with the \emph{pattern recognition control}, compared to the Digital Ecosystem alone. \setCap{The Digital Ecosystem alone averaged a 68.0\% (3 s.f.)}{tmcf} response rate with a standard deviation of 2.61 (2 d.p.), \setCap{while the Digital Ecosystem with the \emph{migration control} showed a significant degradation to 49.6\% (3 s.f.)}{tmc2f} with a standard deviation of 1.96 (2 d.p.), \setCap{and the Digital Ecosystem with the \emph{pattern recognition control} showed only a small increase to 70.5\% (3 s.f.)}{tmc3f} with a standard deviation of 2.60 (2 d.p.). Therefore, any observed improvement from the \emph{targeted migration} was not from the additional migration but its targeting, and that the effectiveness of the \emph{pattern recognition} functionality will be significant if the \emph{targeted migration} is to be effective.

\tfigure{}{graphTarMigContHist}{graph}{Graph of the Targeted Migration Controls and the Digital Ecosystem}{\getCap{tmcf} response rate, \getCap{tmc2f}, \getCap{tmc3f}.}{-8mm}{}{}{}

\tfigure{}{graphTarMigControl}{graph}{Graph of Typical Runs for the Targeted Migration Controls}{and the Digital Ecosystem: \getCap{boohoo}. The migration control with additional random migration ultimately decreased the responsiveness, while the pattern recognition control performed only slightly better.}{-8mm}{!b}{-15mm}{}

In Figure \ref{graphTarMigControl} we graphed a typical run of the Digital Ecosystem with the \emph{migration control}, and the Digital Ecosystem with the \emph{pattern recognition control}, compared to the Digital Ecosystem alone (taken from Figure \ref{DigEcoSuc}). \setCap{The Digital Ecosystem alone performed as expected, adapting and improving over time to reach a mature state}{boohoo} through the process of \emph{ecological succession} \cite{begon96}. The Digital Ecosystem with the \emph{migration control}, which included additional random migration, while initially beneficial, ultimately decreased the responsiveness of the Digital Ecosystem. Finally, the Digital Ecosystem with the \emph{pattern recognition control} performed only marginally better than the Digital Ecosystem alone.

\subsubsection{Neural Networks}

We started with the \ac{NN}-based \emph{targeted migration}, as defined in section \ref{NNsubsection}. We made use of Joone (Java Object Oriented Neural Engine) \cite{joone} to implement the required \acp{NN}, and \emph{exploratory programming} \cite{sommerville2006se} to determine that a \emph{hidden layer} 1.5 times the size of the \emph{input layer} was effective for the \ac{NN}-based \emph{similarity recognition} components.

\tfigure{}{graphTarMigHistNN}{graph}{Graph of Neural Networks Based Targeted Migration}{\getCap{nnbtmCap}}{-7mm}{!b}{5mm}{}

In Figure \ref{graphTarMigHistNN} we graphed for the simulation runs the average of the percentage response rate after a thousand time steps (user request events), for the Digital Ecosystem with the \ac{NN}-based \emph{targeted migration}, compared to the Digital Ecosystem alone. \setCap{The Digital Ecosystem alone averaged a 68.0\% (3 s.f.) response rate with a standard deviation of 2.61 (2 d.p.), while the Digital Ecosystem with the \ac{NN}-based \emph{targeted migration} showed a significant improvement to a 92.1\% (3 s.f.) response rate with a standard deviation of 2.22 (2 d.p.).}{nnbtmCap}

\subsubsection{Support Vector Machines}

\tfigure{}{graphTarMigHistSVM}{graph}{Graph of \acl{SVM} Based Targeted Migration}{\getCap{tmsf}, \getCap{tms2f}, \getCap{tms3f}.}{-8mm}{!b}{5mm}{}

Next we considered the \ac{SVM}-based \emph{targeted migration}, as defined in section \ref{SVMsubsection}, making use of LIBSVM (Library for Support Vector Machines) \cite{libsvm} to implement the required \acp{SVM}. In Figure \ref{graphTarMigHistSVM} we graphed for the simulation runs the average of the percentage response rate after a thousand time steps (user request events), for the Digital Ecosystem with the \ac{SVM}-based \emph{targeted migration}, compared to the Digital Ecosystem with the \ac{NN}-based \emph{targeted migration}, and the Digital Ecosystem alone. \setCap{The Digital Ecosystem with the \ac{SVM}-based \emph{targeted migration} averaged a 92.8\% (3 s.f.) response rate}{tmsf} with a standard deviation of 2.09 (2 d.p.), \setCap{slightly better than the \ac{NN}-based \emph{targeted migration} at 92.1\% (3 s.f.)}{tms2f} with a standard deviation of 2.22 (2 d.p.), \setCap{and so significantly better than the Digital Ecosystem alone at 68.0\% (3 s.f.)}{tms3f} with a standard deviation of 2.61 (2 d.p.).

\vspace{4mm}

In Figure \ref{graphTarMigSVM} we graphed typical runs of the \setCap{Digital Ecosystem with the \ac{SVM}-based \emph{targeted migration}, the Digital Ecosystem with the \ac{NN}-based \emph{targeted migration}, and the Digital Ecosystem alone.}{gSVM2cap} \setCap{The Digital Ecosystem alone performed as expected, adapting and improving over time to reach a mature state through the process of \emph{ecological succession} \cite{begon96}}{gNN2cap}, approaching 70\% effectiveness for the user base. The \setCap{Digital Ecosystem with the \emph{targeted migration}, \ac{NN} or \ac{SVM}-based, showed a significant improvement}{gNN3cap} in the \emph{ecological succession}, reaching the same performance in less than a fifth of the time, before reaching over 90\% effectiveness for the user base. To show more clearly \setCap{the greater effectiveness of the SVM-based \emph{targeted migration}, compared to the NN-based \emph{targeted migration}}{histogram2Cap}, we graphed in Figure \ref{NNhistogram} the \setCap{frequency of poor matches ($<$50\%) every one hundred time steps, for the Digital Ecosystem with the SVM-based \emph{targeted migration}, compared to the Digital Ecosystem with the NN-based \emph{targeted migration}, and the Digital Ecosystem alone}{histogramCap}.

\tfigure{}{graphTarMigSVM}{graph}{Graph of Typical Runs for the Digital Ecosystem and Targeted Migration}{\getCap{gNN2cap} In comparison, the \getCap{gNN3cap}.}{-8mm}{}{}{}

\tfigure{}{NNhistogram}{graph}{Graph of Frequencies for the Targeted Migration}{The \getCap{histogramCap}. It shows \getCap{histogram2Cap} from the seven hundredth generation onwards.}{-7mm}{}{}{}

The results showed that the \emph{targeted migration} optimised and accelerated the \emph{ecological succession} \cite{begon96} of our Digital Ecosystem, constructively interacting with its ecological and evolutionary dynamics. The results also showed that it was not the additional migration, but its targeting that created the improvement in the Digital Ecosystem, and that an effective \emph{pattern recognition} technique was required for the \emph{targeted migration} to operate effectively. Both \acp{NN} and \acp{SVM} proved to be effective, \acp{SVM} marginally more than \acp{NN}. The results also showed that there were no adverse side-effects from the Baldwin effect \cite{baldwin1896nfe}, the inheritance of learnt behaviour in the Agents from the embedded \emph{similarity recognition} components, whether \ac{SVM} or \ac{NN} based. Finally, based on the experimental results, and our theoretical understanding, we would recommend \acp{SVM} for the \emph{pattern recognition} functionality of the \emph{targeted migration}.

\section{Summary and Discussion}

We started by reviewing the scope for optimisation and acceleration resulting from the evolutionary self-organisation of \emph{ecological succession} (the formation of a mature ecosystem) being a slow process \cite{begon96}, even the accelerated form present in our Digital Ecosystem, which was identified and confirmed in the previous chapters. In the results of Chapter \ref{ch:creation}, specifically the \emph{ecological succession} experiment, the Digital Ecosystem reached only 70\% responsiveness, identifying potential for improvement, which was confirmed by the results of Chapter \ref{ch:investigation}, specifically the \emph{self-organised diversity} experiment for which the Digital Ecosystem responded more slowly in some scenarios than others, confirming the potential for improvement. Furthermore, the optimisation of Digital Ecosystems sought was not that of parameter optimisation, which is achievable through \emph{exploratory programming} \cite{sommerville2006se}, but an augmentation to the \acl{EOA} providing a significant improvement in performance, i.e. better solutions for the users than the Digital Ecosystem alone could achieve. We then discovered that we would be unlikely to find optimising or accelerating augmentations for our Digital Ecosystem from biological ecosystems research, because ecological optimisation is concerned with the maintenance of diversity and stability \cite{walters1978eoa, abrams1984fto, naeem1997bee}, and ecological acceleration is similarly concerned with the re-establishment of diversity and stability \cite{mcclanahan1993afs, lugo1997apr, wunderle1997ras}. This was unsurprising, because one of the fundamental differences between biological and digital ecosystems lie in the motivation and approach of their researchers; given that biological ecosystems are ubiquitous natural phenomena whose maintenance is crucial to our survival \cite{balmford2002erc}, whereas Digital Ecosystems are a technology engineered to serve specific human purposes.

So, we therefore proposed, constructed, and explored several alternative augmentations to accelerate or optimise the evolutionary and ecological self-organising dynamics of our Digital Ecosystem, based on the understanding and results from the previous chapters, and our general knowledge and intuition. The \emph{clustering catalyst} aimed to optimise the evolutionary self-organisation of evolving Agent Populations with clusters, by encouraging intra-cluster crossover to directly accelerate the evolving of applications (Agent-sequences) in response to user requests, and therefore the responsiveness of the Digital Ecosystem to the user base. The \emph{replacement aggregator} would have replaced the use of \emph{evolutionary computing} \cite{eiben2003iec}, for the aggregation of the Agents into optimal applications (Agent-sequences) in response to user requests, with an alternative technique, directly optimising the responsiveness of the Digital Ecosystem to the user base. The \emph{Agent-pool aggregation} aimed to allow for the creation of potentially useful applications (Agent-sequences) or partial applications inside the Agent-pools, optimising the Agent-sequences found at the Agent-pools of the Habitats, which would in turn optimise the evolving Agent Populations, and therefore the responsiveness of the Digital Ecosystem to the user base. The \emph{targeted migration} aimed to complement the evolving Agent Populations indirectly, with additional highly targeted migration to support the existing Agent migration between the Habitats, optimising the Agents found at the Agent-pools of the Habitats, which would in turn optimise the evolving Agent Populations, and therefore the responsiveness of the Digital Ecosystem to the user base. The \emph{replacement aggregator} would have weakened the \acl{EOA} of Digital Ecosystems, and the \emph{Agent-Pool aggregation} would have incurred an impractical computational cost to operate, so we chose the most promising augmentations, the \emph{clustering catalyst} and the \emph{targeted migration}, to be completed theoretically and then investigated experimentally through simulations.

The \emph{clustering catalyst} augmentation aimed to directly optimise the evolutionary self-organisation of evolving Agent Populations with clusters, by encouraging intra-cluster crossover. Crossover involves the crossing of two Agent-sequences in the creation of new Agent-sequences, occurring during the replication stage of the evolutionary cycle \cite{back1996eat}. Theoretical completion of the \emph{clustering catalyst} required an algorithm to perform the clustering of evolving Agent Populations. So, we considered alternative clustering algorithms \cite{jain1999dcr, jain1988acd, olson1995pah, ward1963hgo, lu1978ssc, zadeh1996fs, sethi1991ann} for their suitability to our evolving Agent Populations. We chose a hierarchical agglomerative average-link clustering algorithm \cite{olson1995pah} for our \emph{clustering catalyst}, because a hierarchical algorithm was more appropriate \cite{jain1999dcr} than a partitional one for the small size of the data sets of evolving Agent Populations, and because an agglomerative algorithm was conceptually simpler than a divisive one \cite{manning2008iir}, for which the efficiency advantage \cite{manning2008iir} would have been negligible given the expected size of the data sets. Also, because the average-link algorithm is designed to reduce the dependence of the cluster-linkage criterion on extreme values, such as the most similar or dissimilar of the single-link and complete-link algorithms \cite{olson1995pah}; and because the minimum-variance algorithm is biased to producing clusters with the same number of objects \cite{milligan1980ees}, which would have been problematic for clusters emerging over the generations. We also considered a clustering algorithm based on our extended Physical Complexity from Chapter \ref{ch:investigation}, because clustering is subjective by nature \cite{jain1999dcr}, and our extended Physical Complexity was augmented in section \ref{cluster123} to understand the clustering of evolving Agent Populations. So, we implemented the \emph{clustering catalyst} using the hierarchical agglomerative average-link clustering \cite{olson1995pah} and our Physical Complexity clustering. The results showed that the \emph{clustering catalyst}, using either clustering algorithm, failed to optimise the evolutionary processes of the Digital Ecosystem. It intuitively had potential, but most likely failed because the individuals within the evolving Agent Populations lacked sufficient complexity (relative to biological populations \cite{begon96}) for the mechanism to be effective, leading to the crossing of very similar individuals, producing offspring that were very similar to their parents, and therefore not actually achieving valuable change.

The \emph{targeted migration} augmentation aimed to directly optimise the ecological migration, and therefore indirectly complement the evolutionary self-organisation of the evolving Agent Populations, through the highly \emph{targeted migration} of the Agents to their niche Habitats. The \emph{migration probabilities} between the Habitats produces the passive Agent migration, allowing the Agents to spread in the correct general direction within the Habitat network, based primarily upon success at their current location. The \emph{targeted migration} works in a more active manner, allowing the Agents highly targeted migration to specific Habitats, based upon their interaction with one another to discover Habitats where they could be valuable (i.e. find a niche). Theoretical completion of the \emph{targeted migration} required further consideration of how it would operate, including its effect on the Agent life-cycle, and a suitable \emph{pattern recognition} \cite{jain2000spr} technique for the required \emph{similarity recognition} between the \emph{semantic descriptions} of the Agents. So, we considered alternative \emph{pattern recognition} techniques \cite{jain2000spr, ripley1996pra, fu1982spr, nixon2007fei, webb1999spr, pavlidis1977spr, jain1996ann, gunn1998svm} for \emph{similarity recognition} components to be embedded within the Agents. Template Matching was not suitable because its effective use is domain specific \cite{jain2000spr} and the \emph{similarity recognition} between the \emph{semantic descriptions} of Agents is very different to the domains that it is typically applied \cite{nixon2007fei}. Statistical Classification was also not suitable, because the embedded \emph{similarity recognition} component of each Agent would have required human intervention for variable selection and transformation \cite{michie1995mln}. While Structural Matching was a suitable technique theoretically, implementations have had many difficulties \cite{jain2000spr}, including the segmentation of noisy patterns (to detect primitives) and the inference of grammar from training data \cite{jain2000spr}. There can also be a combinatorial explosion of possibilities to be investigated, demanding large training sets and significant computational effort \cite{perlovsky1998ccc}, neither of which was available. \acfp{NN} were suitable, given their low dependence on domain-specific knowledge and the availability of efficient learning algorithms \cite{jain2000spr}. \acfp{SVM}, albeit a recent development \cite{vapnik1998slt}, were also suitable \cite{joachims1997tcs}, being primarily a binary classifier \cite{jain2000spr} for training generalisable nonlinear classifiers in high-dimensional spaces using small training sets \cite{valafar2002s}. So, we implemented the \emph{targeted migration} using both \ac{NN} and \ac{SVM} based \emph{similarity recognition} components embedded within the Agents. The results showed that the \emph{targeted migration} accelerated and optimised the \emph{ecological succession} of the Digital Ecosystem, constructively interacting with its ecological and evolutionary dynamics, marginally more when powered by \acp{SVM} than \acp{NN}. So, based on the experimental results, and our theoretical understanding, we would recommend \acp{SVM} for the \emph{pattern recognition} functionality of the \emph{targeted migration}.

The \emph{targeted migration} also resulted in the Baldwin effect \cite{baldwin1896nfe}, the inheritance of learnt behaviour in the Agents, from the embedded \emph{similarity recognition} components, whether \ac{SVM} or \ac{NN} based. While the Baldwin effect has always been controversial within \emph{biological ecosystems} \cite{weber2003eal}, primarily because of the problem of confirming it experimentally \cite{sterelny2004rea}, it undoubtedly occurred in our Digital Ecosystem. However, the experimental results and our \emph{exploratory programming} \cite{sommerville2006se} showed no adverse side-effects, therefore supporting the possibility of its presence in \emph{biological ecosystems}.

In this chapter we attempted the acceleration and optimisation of Digital Ecosystems, because \emph{ecological succession} (the formation of mature ecosystems) is a slow process \cite{begon96}, even the accelerated form present in our Digital Ecosystems. While not all our attempts were successful, understandable considering the constructive nature of our efforts in this chapter, we have optimised and accelerated our Digital Ecosystem through the \emph{targeted migration} of its Agents. The \emph{targeted migration} significantly enhanced the \emph{ecological succession}, constructively interacting with the ecological and evolutionary dynamics, helping the Agents to optimise their migration and distribution within our Digital Ecosystem.

\vfill

\pagebreak
\thispagestyle{plain}

\chapter{Conclusions}
\label{ch:conclusions}

\vspace{-5mm}

\section{Achievements}

Substantial parts of our efforts are original contributions in the area of \acl{BIC} \cite{forbes2004ilb} and the emerging field of Digital Ecosystems, with our major research contributions being as follows:

\begin{itemize}

\item We have created the first interpretation of Digital Ecosystems where the word \emph{ecosystem} is more than just a metaphor, which we have confirmed experimentally. They are the digital counterparts of \emph{biological ecosystems}: having their properties of self-organisation, scalability and sustainability \cite{Levin}; created through combining understanding from theoretical ecology \cite{levins1969sda}, evolutionary theory \cite{ec16}, \aclp{MAS} \cite{moaspaper}, distributed evolutionary computing \cite{lin1994cgp}, and \aclp{SOA} \cite{soa1w}. Furthermore, the \acl{EOA} of Digital Ecosystems includes a novel form of distributed evolutionary computing, an optimisation technique working at two levels: a first optimisation, migration of Agents which are distributed in a peer-to-peer network, operating continuously in time; this process feeds a second optimisation, based on evolutionary computing, operating locally on single peers and is aimed at finding solutions that satisfy locally relevant constraints. So, the local search is improved through this twofold process to yield better local optima faster, as the distributed optimisation provides prior sampling of the search space through computations already performed in other peers with similar constraints. We have also defined the interaction of Digital Ecosystems with \emph{business ecosystems} \cite{moore1996}, specifically in supporting and enabling them to create \aclp{DBE}.

\item We have investigated the emergent self-organising properties of Digital Ecosystems, because a primary motivation for our research is the desire to exploit the self-organising properties \cite{Levin} of \emph{biological ecosystems}. We started with the evolutionary self-organisation of \emph{ecological succession} \cite{begon96}, which conformed to expectations \cite{Hubbell}. Next we considered the self-organisation of the order constructing processes (the evolving Agent Populations). We extended Physical Complexity \cite{adami20002} to include evolving Agent Populations, which required extending definitions for populations of variable length sequences, creating a measure for the \emph{efficiency} of information storage, and an understanding of clustering within Populations to support the non-atomicity of Agents. We then extended Chli-DeWilde stability \cite{chli2} to include the evolutionary dynamics of evolving Agent Populations, building upon this to construct an entropy-based definition for the \emph{degree of instability}, which was used to study the stability of evolving Agent Populations. Finally, the unique hybrid nature of Digital Ecosystems resulted in us creating our own definition for the self-organised \emph{diversity}, based on the global distribution of the Agents in the Populations relative to the request behaviour of the user base. Overall an insight has been achieved into where and how self-organisation occurs in Digital Ecosystems, including what forms it can take and how it can be quantified.

\item We have optimised and accelerated Digital Ecosystems, because the evolutionary self-organisation of \emph{ecological succession} \cite{begon96} (the formation of a mature ecosystem) is a slow process, even the accelerated form present in Digital Ecosystems. So, we considered alternative augmentations, including the accelerating effect of a \emph{clustering catalyst} on the evolutionary dynamics, through the acceleration of the evolutionary processes; and the optimising effect of \emph{targeted migration} on the ecological dynamics, through the emergent optimisation of the Agent migration patterns. The experimental results showed that the \emph{clustering catalyst} failed, despite intuitively having potential, most likely because the individuals within the evolving Agent Populations lacked sufficient complexity (relative to biological populations \cite{begon96}) for the augmentation to be effective. However, the experimental results also showed that the \emph{targeted migration} optimised and accelerated the \emph{ecological succession} of Digital Ecosystems, constructively interacting with the ecological and evolutionary dynamics. We also discovered that there were no adverse side-effects from the Baldwin effect \cite{baldwin1896nfe}, the inheritance of learnt behaviour, in Digital Ecosystems. Therefore, supporting the possibility of the Baldwin effect in \emph{biological ecosystems}, which has always been controversial \cite{weber2003eal}.

\end{itemize}

\section{Future Directions} 

Our efforts offer considerable scope for the future, with there being several interesting avenues to pursue, some of which we discuss below.

\subsection{Ecosystems Conceptualisation}

Conceptualising ecosystems has been an inherent part of this work, which presents us with an opportunity to formalise our current and future efforts to improve the cross-disciplinary knowledge transfer required.

\subsubsection{Biology of Digital Ecosystems}

In creating Digital Ecosystems, the digital counterpart of \emph{biological ecosystems}, we naturally asked their likeness to the \emph{biological ecosystems} from which they came. Further to this, we could consider the applicability of other aspects of ecosystems theory in understanding and analysing the dynamics of Digital Ecosystems. For example, \emph{energy pyramids}\footnote{\emph{Energy pyramids} show the dissipation of energy at trophic levels, positions that organisms occupy in a food chain, e.g. producers or consumers \cite{odum1968efe}.} of \emph{biological ecosystems}, what is their equivalent in Digital Ecosystems? Given that Digital Ecosystems are information-centric, whereas \emph{biological ecosystems} are energy-centric \cite{begon96}, they would undoubtedly be \emph{information pyramids}, but further definition would naturally require more research.

\subsubsection{Biological Design Patterns}

A \emph{design pattern} is a general reusable solution to a commonly occurring problem in software design \cite{gamma1995dpe}. It is not a finished design that can be transformed directly into code, but a description or template for how to solve a problem that can be used in many different situations \cite{gamma1995dpe}. For example, object-oriented design patterns typically show relationships and interactions between classes or objects, without specifying the final application classes or objects that are involved \cite{gamma1995dpe}. \acfp{BDP} would extend this concept to catalogue common interactions between biological structures using a pattern-oriented modelling approach \cite{grimm2005pom}, which when applied would endow software systems with the desirable properties of biological systems, such as self-organisation, self-management, scalability and sustainability.

\subsubsection{Classes of Ecosystems}

While \emph{evolutionary theory} \cite{ec16} was well understood within computer science, under the auspices of \emph{evolutionary computing} \cite{eiben2003iec}, \emph{ecosystems theory} \cite{begon96}, until our efforts, was not. Similarly, while \emph{evolutionary theory} is well understood within linguistics \cite{croft2000elc} and economics \cite{nelson1982ete}, equally \emph{ecosystems theory} is not \cite{mufwene2001ele}. So, using our efforts as a \emph{case study}, we could follow the same process to create Language Ecosystems and Economic Ecosystems. For example, there are many separate efforts within linguistics using evolution to model language change \cite{christiansen2003le}, but there is no unifying framework, which has resulted from different linguists independently adopting elements of evolutionary theory \cite{christiansen2003le}. So, we could provide a wide-ranging and encompassing definition of Language Ecosystems, which would unify the many disparate efforts in linguistics aimed at understanding language evolution.

\subsubsection{Generic Ecosystem Definition}

In the creation of Digital Ecosystems we considered aspects of \emph{biological ecosystems}, including \acl{ABM} \cite{Greenetal2006} and \acf{CAS} \cite{Levin}, and then constructed their counterparts in Digital Ecosystems. After which we considered the possibility of a Generic Ecosystem definition, as we suggested at the end of Chapter \ref{ch:creation}, without which some of the counterparts we constructed appeared to be compromised, when they were actually the realisation of generic abstract concepts in Digital Ecosystems. Most notably the network structure, which is energy-centric in \emph{biological ecosystems} \cite{begon96}, while information-centric in Digital Ecosystems. So, there is potential to create a Generic Ecosystem definition, using a suitable modelling technique such as \ac{CAS} \cite{waldrop1992ces}, which would abstractly define the key properties of an ecosystem, and would theoretically be applicable to any domain where the modelling technique has been applied. Therefore, the Generic Ecosystem definition would provide a framework for the application of ideas, concepts, and models from \emph{biological ecosystems} to other classes of ecosystems, including Digital Ecosystems, Language Ecosystems and Economic Ecosystems.

\subsection{Simulation Framework}

An open-source simulation framework for Digital Ecosystems \cite{eveSim} was created by the \acf{DBE} project \cite{DBE}, and is currently supported by the \acf{OPAALS} project \cite {OPAALS} to assist further research into Digital Ecosystems, including the wider implications of interacting with social systems, such as \emph{business ecosystems} of \acfp{SME}.

\subsection{Digital Business Ecosystems}

In an old market-based economy, made up of sellers and buyers, the parties exchange property \cite{delcloque2001dii}. While in a new network-based economy, made up of servers and clients in a \emph{business ecosystem} \cite{moore1996}, the parties share access to services and experiences \cite{delcloque2001dii}. Digital Ecosystems are a platform for the network-based economy of \emph{business ecosystems}, providing mechanisms for the creation of \aclp{DBE}.

\subsubsection{Service Futures Market}

One such mechanism the Digital Ecosystem could provide to the network-based economy of \emph{business ecosystems} \cite{moore1996}, would be a \emph{futures market}\footnote{An auction market in which participants buy and sell commodities for an agreed price, that the sellers have yet to produce \cite{hull2005ffa}.} for services. As each service (Agent) consists of an \emph{executable component} and a \emph{semantic description}, the later acting as a guarantee of behaviour, and the evolving Agent Populations only requiring the guarantees (\emph{semantic descriptions}) to operate, the actual \emph{executable component} of a service (Agent) is only required once an application (Agent-sequence) has been assembled. Therefore, service (Agent) evolution could operate entirely on the \emph{semantic descriptions}, with business users only needing to supply the \emph{executable component} of a service (Agent) once there is a demand, i.e. when the \emph{semantic description} of one of their services has been used in the construction of an application which meets the request of another business user. Therefore, creating a \emph{futures market} for evolving services within Digital Business Ecosystems.

\subsubsection{Regional Deployment}

A partial reference implementation \cite{eveNet} for our Digital Ecosystem, which includes an implementation of the \emph{targeted migration}, was created by the \aclp{DBE} project \cite{DBE}, and we expect that once completed will be deployed as part of the software platform intended for the regional deployment of their \emph{Digital Ecosystems} \cite{den4dek, OPAALS}. \emph{Digital Ecosystems} (distributed adaptive open socio-technical systems, with properties of self-organisation, scalability and sustainability, inspired by natural ecosystems \cite{OPAALS}) are emerging as a novel approach to the catalysis of sustainable regional development driven by \acfp{SME} \cite{OPAALS}. The community focused on the deployment of \emph{Digital Ecosystems}, REgions for Digital Ecosystems Network (REDEN) \cite{reden}, is supported by projects such as the Digital Ecosystems Network of regions for (4) DissEmination and Knowledge Deployment (DEN4DEK) \cite{den4dek}, a thematic network that aims to share experiences and disseminate all the necessary knowledge that will allow regions to plan an effective deployment of \emph{Digital Ecosystems} at all levels (economic, social, technical and political) to produce real impacts in the economic activities of European regions through the improvement of \ac{SME} business environments. So, the next major step in our research will be to collect real world data, confirming that Digital Ecosystems operate effectively with \emph{business ecosystems} in creating Digital Business Ecosystems.

\section{Concluding Remarks}

The ever-increasing challenge of software complexity in creating progressively more sophisticated and distributed applications, makes the design and maintenance of efficient and flexible systems a growing challenge \cite{newsArticle1, slashdot, newsArticle3}, for which current software development techniques have hit a \emph{complexity wall} \cite{lyytinen2001nwn}. In response we have created Digital Ecosystems, the digital counterparts of biological ecosystems, possessing their properties of self-organisation, scalability and sustainability \cite{Levin}; \aclp{EOA} that overcome the challenge by automating the search for new algorithms in a scalable architecture, through the evolution of software services in a distributed network.


\bShell
./urlfix.sh thesis.bbl
\eShell

\chapter*{Bibliography}
\addcontentsline{toc}{chapter}{Bibliography}
\narrowlinespacing
\bibliographystyle{plainurl}
\begin{spacing}{1}
\bibliography{references}
\end{spacing}
\vfill

\ifthenelse{\boolean{final}}{\execute{cd pdfbookmark; ./pdfbookmk.sh Abstract Acknowledgements "List of Figures" Bibliography &}}{}

\bShell
./mkthesis2aux.sh &
\eShell

\end{document}